\documentclass[12pt]{qmulphd}  

\usepackage{amsmath}
\usepackage{amssymb}
\usepackage{graphicx}
\usepackage{verbatim}
\usepackage{listings}
\usepackage{algpseudocode}
\usepackage{multirow}
\usepackage{array}
\usepackage{url}
\usepackage{rotating}
\usepackage{booktabs}
\usepackage{hyperref}
\usepackage{enumitem}
\usepackage{color}
\usepackage{parskip}
\usepackage{tikz}

\usepackage[T1,T5]{fontenc}
\hypersetup{
    pdftitle={Model Counting Modulo Theories},
    pdfauthor={Quoc-Sang Phan},
    pdfsubject={Formal Verification},
    pdfkeywords={Model Counting Modulo Theories} {Quantitative Information Flow} {Satisifiability Modulo Theories} {Symbolic Execution},
    citecolor=blue,
    linkcolor=black,
    urlcolor=blue,
}
\newcommand{\stmt}[1]{{\small\texttt{#1}}}
\newcommand{\api}[1]{{\normalfont\textbf{#1}}}
\newcommand{\mt}[1]{\mathtt{#1}} 
\newcommand{\code}[1]{{\texttt{#1}}}
\newcommand{\dpllt}{DPLL($\mathcal{T}$)}
\newcommand{\todo}[1]{\relax}
\newtheorem{theorem}{Theorem}
\newtheorem{corollary}{Corollary}
\newtheorem{lemma}{Lemma}
\newtheorem{proposition}{Proposition}
\newtheorem{definition}{Definition}

\newtheorem{proofsketch}{Proof sketch}
\algdef{SE}[IF]{If}{EndIf}[1]{\algorithmicif\ #1\ }{\algorithmicend\ \algorithmicif}
\algdef{C}[IF]{IF}{ElsIf}[1]{\algorithmicelse\ \algorithmicif\ #1\ }
\algdef{SE}[WHILE]{While}{EndWhile}[1]{\algorithmicwhile\ #1\ }{\algorithmicend\ \algorithmicwhile}
\algtext*{EndWhile}
\algtext*{EndIf}
\algtext*{EndFunction}
\newsavebox{\leftlisting}
\newsavebox{\rightlisting}

\lstnewenvironment{qsplisting}[1][] 
 {\lstset{
    mathescape, 
    language=C,
    morekeywords={assert},
    frame=single,
    basicstyle=\footnotesize\ttfamily,
    numbers=left,
    stepnumber=0,
    #1
  }}{}

\title{Model Counting Modulo Theories}

\author{Quoc-Sang Phan}             

\degree{Doctor of Philosophy}     
\degreedate{London 2015}       

\newtheorem{proof}{Proof}

\begin{document}

\baselineskip=18pt plus1pt

\setcounter{secnumdepth}{2}
\setcounter{tocdepth}{2} 


\maketitle                  

\begin{originalitylong}
I, Quoc-Sang Phan, confirm that the research included within this thesis is my own work or that where it has been carried out in collaboration with, or supported by others, that this is duly acknowledged and my contribution indicated. Previously published material is also acknowledged in the next page.

I attest that I have exercised reasonable care to ensure that the work is original, and does not to the best of my knowledge break any UK law, infringe any third party's copyright or other Intellectual Property Right, or contain any confidential material.

I accept that the College has the right to use plagiarism detection software to check the electronic version of the thesis.

I confirm that this thesis has not been previously submitted for the award of a degree by this or any other university.

The copyright of this thesis rests with the author and no quotation from it or information derived from it may be published without the prior written consent of the author.

\vspace{2cm}
Quoc-Sang Phan\\
April 9, 2015



\end{originalitylong}

\begin{publicationslong}
\label{publications}
Parts of this thesis have been published; Chapters \ref{chap:SQIF}, \ref{chap:SymExDPLL}, \ref{chap:jpf-bmc}, and \ref{chap:qilura} have been published as
the papers listed below, in reverse chronological order. Chapter \ref{chap:allsmt} is currently under submission.


\begin{description}[font=\normalfont]
  \item[\cite{Phan:2015:CBM}] \textbf{Quoc-Sang Phan}, Pasquale Malacaria, and Corina~S. P\u{a}s\u{a}reanu. Concurrent Bounded Model Checking. {\em SIGSOFT Software Engineering Notes}, 40(1):1--5, February 2015.
  \item[\cite{Phan:OASIcs:2014:4774}]\textbf{Quoc-Sang Phan}. Symbolic Execution as DPLL Modulo Theories. In {\em 2014 Imperial College Computing Student Workshop}, volume 43 of {\em OpenAccess Series in Informatics (OASIcs)}, pages 58-65, Dagstuhl, Germany, 2014. Schloss Dagstuhl-Leibniz-Zentrum fuer Informatik.
  \item[\cite{Phan:2014:QILURA}] \textbf{Quoc-Sang Phan}, Pasquale Malacaria, Corina~S. P\u{a}s\u{a}reanu, and
  Marcelo d'Amorim. Quantifying Information Leaks Using Reliability Analysis. In {\em Proceedings of the 2014 International SPIN Symposium on Model Checking of Software}, SPIN 2014, pages 105-108, New York, NY, USA, 2014. ACM.

  \item[\cite{Phan:2014:AMC}] \textbf{Quoc-Sang Phan} and Pasquale Malacaria. Abstract Model Counting: a novel approach for Quantification of Information Leaks. In {\em Proceedings of the 9th ACM Symposium on Information, Computer
  and Communications Security}, ASIACCS '14, pages 283--292, New York, NY, USA, 2014. ACM.
  \item[\cite{Phan:OASIcs:2013:4277}] \textbf{Quoc-Sang Phan}. Self-composition by Symbolic Execution. In {\em 2013 Imperial College Computing Student Workshop}, volume 35 of {\em OpenAccess Series in Informatics (OASIcs)}, pages 95-102, Dagstuhl, Germany, 2013. Schloss Dagstuhl-Leibniz-Zentrum fuer Informatik.
  \item[\cite{Phan:2012:SQI:2382756.2382791}] \textbf{Quoc-Sang Phan}, Pasquale Malacaria, Oksana Tkachuk, and Corina~S. P\u{a}s\u{a}reanu. Symbolic Quantitative Information Flow. {\em SIGSOFT Software Engineering Notes}, 37(6):1--5, November 2012.
\end{description}

\end{publicationslong}

\begin{abstractlong}
This thesis is concerned with the quantitative assessment of security in software. More specifically, it tackles the problem of efficient computation of \emph{channel capacity}, the maximum amount of confidential information leaked by software, measured in Shannon entropy or R\'{e}nyi's min-entropy.  

Most approaches to computing channel capacity are either efficient and return only (possibly very loose) upper bounds, or alternatively are inefficient but precise; few target realistic programs. In this thesis, we present a novel approach to the problem by reducing it to a model counting problem on first-order logic, which we name \emph{Model Counting Modulo Theories} or \#SMT for brevity. 

For quantitative security, our contribution is twofold. First, on the theoretical side we establish the connections between measuring confidentiality leaks and fundamental verification algorithms like Symbolic Execution, SMT solvers and DPLL. Second, exploiting these connections, we develop novel \#SMT-based techniques to compute channel capacity, which achieve both accuracy and efficiency. These techniques are scalable to real-world programs, and illustrative case studies include C programs from Linux kernel, a Java program from a European project and anonymity protocols.

For formal verification, our contribution is also twofold. First, we introduce and study a new research problem, namely \#SMT, which has other potential applications beyond computing channel capacity, such as returning multiple-counterexamples for Bounded Model Checking or automated test generation. Second, we propose an alternative approach for Bounded Model Checking using classical Symbolic Execution, which can be parallelised to leverage modern multi-core and distributed architecture.

For software engineering, our first contribution is to demonstrate the correspondence between the algorithm of Symbolic Execution and the DPLL($\mathcal{T}$) algorithm 
used in state-of-the-art SMT solvers. This correspondence could be leveraged to improve Symbolic Execution for automated test generation. Finally, we show the relation between computing channel capacity and reliability analysis in software.
\end{abstractlong}

\begin{dedication}
To my parents {\fontencoding{T5}\selectfont Phan Qu\'\ocircumflex{}c K\'inh and Nguy\~\ecircumflex{}n Th\d i H\uhorn{}\ohorn{}ng}
\end{dedication}

\begin{acknowledgementslong}
I feel extremely lucky to have Pasquale Malacaria as my supervisor, and I would like to thank him for his guidance, encouragement and support for the last three years. Pasquale was always available for discussions and advice, which was important for the development of the ideas in this thesis. I also truly appreciate the opportunities he has given me to present my work, and network with the research community.
I would also thank him for establishing the foundation of 
Quantitative Information Flow, the main topic of this thesis.  

I am also deeply grateful to Corina P\u{a}s\u{a}reanu for her advice, encouragement and kindness. Indeed, Corina has been my \emph{de facto} co-supervisor, and a large part of this thesis would not have been existed without the Symbolic PathFinder platform that she developed. I also want to thank Corina for sharing her office with me, and escorting me almost every day during my internship at NASA Ames.

I wish to thank my mentors, Johann Schumann and Kristin Rozier, for giving me a chance to work at NASA, and for sharing with me parts of their vast knowledge and experience. Particular thanks to Johann for teaching me Unmanned Aerial Systems, and for allowing me to stay in his house during my internship. 

I am grateful to Oksana Tkachuk and Marcelo d'Amorim for mentoring me during the Google Summer of Code programs in 2012 and 2013. I am also thankful to Nikos Tzevelekos, Michael Tautschnig and Dino Distefano for fruitful discussions and for their careful comments on early versions of my papers.

I would like to thank my family, their love has helped me overcome all difficulties. In particular, my wife Huyen Do has loved and cared for me throughout my PhD. My son Voi, born at the beginning of my third year of PhD, has brought me endless happiness.

I wish to thank my office-mate and fellow PhD student Nhat Anh Dang for his helps and friendship. 
Although we work in totally different areas, this has not prevented us from having countless discussions, and Nhat Anh even managed to teach me several mathematical concepts. I am also thankful 
for his kind helps in my personal life.

I am grateful for the scholarship of the School of Electronic Engineering and Computer Science, which covered my tuition fees and stipends to live in London. I also benefited greatly from the research environment of the school, in particular the weekly research seminars of the Theory group have helped me to broaden my knowledge. Thank Nikos Tzevelekos for having done an excellent job organizing these seminars.

\end{acknowledgementslong}

\tableofcontents            


\chapter[Introduction]{Introduction}
\section{Motivation}
The year 2014 has witnesses several high-profile software security incidents, e.g. the disclosure of heartbleed bug~\cite{heartbleed}, the leaks of celebrity photos in iCloud~\cite{photoleaks}, the hack of Sony Pictures~\cite{sony}. The damage caused by leaking confidential information varies, from personal embarrassment to the lost of dozens of millions of dollars. 

These incidents show the importance of protecting confidential data from being leaked by software. Access control systems~\cite{Sandhu:312842} can limit access to information, but cannot control internal information propagation once accessed. This motivates the research on \emph{information flow security}~\cite{DBLP:journals/jsac/SabelfeldM03}, which aims to track the flows of information in software, and forbid illegal flows that leaks information to public observers.

However, leakage of information is hardly avoidable in computer programs. Even ``secure'' programs do leak some information about the secret data being processed. A popular example is the password checking program, which can be considered as secure with a reasonably strong password. Every time it rejects an input string, it reveals to the adversary that the password is different from that string.
This amount of information is small, but if the adversary is allowed to make enough attempts, i.e. brute force attack, the program will eventually leak all information to the attacker. 
As leakage of information is unavoidable, it is important to assess the leaks to decide if they are acceptable. This leads to the question \emph{how much} information a program could leak to an adversary.

The research area of \emph{Quantitative Information Flow} analysis (QIF~\cite{Clark:2007:SAQ:1370628.1370629,Malacaria:2007:AST:1190216.1190251}) has been developed to provide a rigorous framework to answer the question above. Intuitively, after observing an execution of the program, the adversary gains more information and has less uncertainty about the confidential data. The difference of uncertainties before and after his observation is the amount of information leaked by the program. QIF has based its foundation on the entropy concept of Information Theory~\cite{Cover:1991:EIT:129837}: the uncertainties about the confidential data are measured in bits by, for example, Shannon entropy; leakage is then computed by numerical subtraction. 

Manual computation of QIF, e.g. in \cite{Malacaria:2007:AST:1190216.1190251}, is tedious, expensive and infeasible for complex programs. In order to apply the theory of QIF to practice, it is crucial to have automated techniques that can efficiently quantify leakage in software systems. However, most approaches to automation of QIF are either efficient and return only (possibly very loose) upper bounds~\cite{McCamant:2008:QIF:1375581.1375606,Newsome:2009:MCC:1554339.1554349,Meng:2011:CBI:2166956.2166957}, or alternatively are inefficient but
precise~\cite{Backes:2009:ADQ:1607723.1608130,Heusser:2010:QIL:1920261.1920300,Klebanov12a}; few target realistic programs.
This thesis is a significant step towards efficient and accurate computation of QIF by means of formal methods. 

\section{State of the art 
}
This section outlines some of the key advances that have led to the current state of the art for automation of QIF.
Emphasis is given to work on deterministic programs.

The concept of information flow was first introduced in the 1977 paper of the the Dennings~\cite{Denning:1977:CPS:359636.359712}, who described a lattice model where variables are partitioned into security labels: \code{H}, standing for ``high'', for variables containing sensitive data and \code{L}, standing for ``low", for variables containing public information. The partial order $\code{L} \leq \code{H}$ in the lattice indicates that information flows from variables with label \code{H} to variables with label \code{L} are not allowed.

In 1982, Goguen and Meseguer~\cite{DBLP:conf/sp/GoguenM82} described \emph{non-interference}, a security policy for a general automaton framework where there are a set of state changing commands and a set of users. A group of users \code{G} have non-interference on another group of users \code{G}' if and only if any user in the group \code{G} cannot observe the effects of commands used by any user in the group \code{G}'.

In 1996, Volpano, Smith, and Irvine~\cite{Volpano:1996:STS:353629.353648} achieved an impressive milestone by proving the soundness of the Dennings' analysis using a type system, which coincides with the idea of non-interference. Since then, research in information flow has evolved in different directions. In 2003, Sabelfeld and Myers~\cite{DBLP:journals/jsac/SabelfeldM03} published an excellent survey of the field up to that time, citing 147 papers.

The first complete quantitative analysis of information flow had been developed by Clark, Hunt and Malacaria through a series of papers~\cite{Clark2002238,Clark:2005:QIW:1705540.1705938,Clark:2005:QIF:1094472.1094516,Clark:2007:SAQ:1370628.1370629} from 2002 to 2007. The authors illustrated their analysis on a simple deterministic While language, and provided each command of the language a lower bound and an upper bound on the amount of leakage. Their treatment for loops is very pessimistic, assuming that all information in the looping conditions will be leaked. Malacaria~\cite{Malacaria:2007:AST:1190216.1190251} then showed a more precise treatment for looping constructs by using the partition property of entropy. This work is labour intensive and impossible to automate.

In 2009, Smith~\cite{Smith:2009:FQI:1532848.1532876} proposed to use R\'{e}nyi min-entropy as a metric for QIF, and showed that the \emph{channel capacity} of information flow, i.e. the maximum leakage, measured by this metric is equal to the logarithm of the number of possible outputs.
This result agrees with previous observations of Malacaria and Chen~\cite{Malacaria:2008:LMM:1375696.1375713} that channel capacity measured by Shannon entropy also has the same tight upper bound.

Meanwhile, in the same year (2009), 
Backes, K\"{o}pf and Rybalchenko~\cite{Backes:2009:ADQ:1607723.1608130} reached an important milestone by introducing the first automatic technique for QIF analysis using model checking~\cite{clarke1999model} and Barvinok algorithm~\cite{Barvinok:1994:PTA:187096.187093} for model counting. This analysis is very expensive, and only applicable to a limited class of programs.

In 2010, Heusser and Malacaria~\cite{Heusser:2010:QIL:1920261.1920300} published a paper demonstrating QIF analysis for programs from Linux kernel. Prior to this paper, all the work on QIF analysis were demonstrated with small ``toy" examples. Heusser and Malacaria made assumptions that the program had \code{N} different possible outputs, and asserted that there did not exist an output different from those \code{N} outputs. 
Since these assumptions and assertion required the composition of \code{N}+1 copies of the program, their analysis suffers severely from the state space explosion problem.

A year later, in 2011, Meng and Smith~\cite{Meng:2011:CBI:2166956.2166957} described a fast approximation technique to compute an upper bound on channel capacity. They infer the relations between all pairs of bits of the output as a propositional formula, then using a \#SAT solver~\cite{mathematica} to count the model. The analysis was largely manual and was demonstrated with toy examples. The upper bound can be very loose if outputs are sparse and scattered.

In summary, the problem of an automated QIF analysis is still very challenging. A simpler problem, called bounding QIF, is considered in~\cite{DBLP:journals/jcs/YasuokaT11,Heusser:2010:QIL:1920261.1920300}: deciding if a program $P$ leaks less than a constant $ q$.
In previous work, Yasuoka and Terauchi have proved that bounding QIF is not a \emph{k-safety} problem for any $k$~\cite{DBLP:journals/jcs/YasuokaT11}. Cerny et al. then proved that
in the case of Shannon entropy, bounding QIF is PSPACE-complete~\cite{Cerny:2011:CQI:2056311.2056572}. Therefore,
QIF and bounding QIF remain a huge challenge.
Available techniques are either inefficient or imprecise. This thesis investigates techniques that improves both precision and efficiency for QIF analysis.

\section{Contributions}
This thesis is about solving a problem, QIF, using a set of tools referred to as Formal Methods. Through the process of solving the problem, we also gain the insights to improve the tools. As a result, this thesis makes contributions both to approaches for automated QIF analysis using Formal Methods as well as techniques for improving Formal Methods.
\subsection*{Contributions to Quantitative Information Flow}
This thesis makes a theoretical advance by casting the QIF problem into a variant of the Satisfiability Modulo Theories (SMT) problem~\cite{DeMoura:2011:SMT:1995376.1995394}. This results in a general algorithm for QIF analysis based on the \dpllt{} algorithm~\cite{Nieuwenhuis:2006:SSS:1217856.1217859} for SMT.
This theoretical advance leads to practical advance: compared to the work of Heusser and Malacaria~\cite{Heusser:2010:QIL:1920261.1920300}, the technique proposed in this thesis dramatically reduces the time of analysing programs from Linux kernel, from some hours to a few seconds.

The second theoretical contribution of this thesis is to show that classical Symbolic Execution~\cite{King:1976:SEP:360248.360252} can be understood as a variant of the \dpllt{} algorithm. In other words, Symbolic Executors are SMT solvers. This view enables us to develop the first QIF analysis tool for Java bytecode, built on top of the Symbolic PathFinder symbolic execution platform. 

Our third contribution is to show the relation between Quantitative Information Flow and Reliability analysis~\cite{Cheung:1980:USR:1313319.1313511}, which are two totally separate research areas prior to this thesis. This relation leads to the development of an efficient QIF technique based on an available Reliability analyser.

The practical contribution of this thesis is the development of several QIF analysis tools for programs in both C and Java, and the experiments of these tools on real-world programs: C programs from the Linux kernel, a Java tax program from the European project HATS, and anonymity protocols.
 
\subsection*{Contributions to Formal Methods}
The insights we have learned from solving the QIF problem lead us to investigate techniques that could improve Formal Methods. 
Our first contribution in this thesis is the introduction of the Model Counting Modulo Theories (or \#SMT) problem, which has several potential applications beyond QIF.

Our second contribution is: from the view of Symbolic Executors as SMT solvers, we propose a new methodology for Bounded Model Checking based on Symbolic Execution. Our methodology is naturally parallelizable, thus it can exploit modern multi-core machines and distributed architecture. Experimental results show that it outperforms the state-of-the-art Bounded Model Checker CBMC~\cite{ckl2004} in several complex case studies.

The third contribution of this thesis is a light weight method for All-Solution SAT Modulo Theories (All-SMT). Unlike available approaches, our method can be used for any SMT problem with any ground theories, including bit vectors.

Another contribution of this thesis is two applications of All-SMT solvers: the first one is to find multiple-counter examples in Bounded Model Checking; the second one is to combine Bounded Model Checking and All-SMT solver for automated test vector generation.
\section{Thesis structure}
This thesis can be divided in two parts: the first part includes chapters \ref{chap:SQIF}, \ref{chap:qilura}, and \ref{chap:allsmt} focusing on automation of QIF using Formal Methods; the second part includes chapters \ref{chap:SymExDPLL} and \ref{chap:jpf-bmc} focusing on improving Formal Methods.

\emph{Chapter \ref{chap:background} } provides necessary preliminaries on the two main concepts in this thesis: QIF and SMT.

\emph{Chapter \ref{chap:SQIF} } introduces the \#SMT problem as a generalization of the SMT problem, and casts the QIF problem into \#SMT. The result is a general algorithm for QIF analysis based on the \dpllt{} algorithm used in SMT solvers.

\emph{Chapter \ref{chap:SymExDPLL} } shows how Symbolic Execution can be viewed as a variant of the \dpllt{} algorithm. This view enables one to turn a classical Symbolic Executor into 
a \#SMT solver for QIF analysis with little effort.

\emph{Chapter \ref{chap:jpf-bmc} } presents a new methodology for Concurrent Bounded Model Checking using classical Symbolic Execution. The effectiveness of the methodology is illustrated with several complex
case studies in both C and Java.

\emph{Chapter \ref{chap:qilura} } describes the relation between QIF and Reliability analysis. This relation is exploited to build an efficient QIF analysis tool
for Java bytecode based on an available Reliability analysis engine.

\emph{Chapter \ref{chap:allsmt} } presents two lightweight algorithms for \#SMT in a pure logic settings, and three applications of a \#SMT/All-SMT solver apart from QIF, namely multiple-counterexample analysis for Bounded Model Checking, automated test generation and reliability analysis.

\emph{Chapter \ref{chap:conclusions} } concludes the thesis with a summary of contributions and discusses possible directions for future work.

\chapter{Preliminaries}
\label{chap:background}
\graphicspath{{chapter2/figs/}}
In order to make the thesis self-contained, this chapter provides some basic notions and terminology about discrete probability theory, information theory and first-order
theories. We also recall the elements and general concepts of formal methods. 

The content of this chapter is synthesized from 
several sources, e.g. \cite{DBLP:journals/jsac/SabelfeldM03,MSC:9408698,Smith:2009:FQI:1532848.1532876,Shannon:2001:MTC:584091.584093,Cover:1991:EIT:129837,Biere:2009:HSV:1550723,deMoura:2007:TSM:1770351.1770358,DBLP:books/daglib/0076838}.
\emph{None of the results in this chapter were discovered by the author of this thesis}; a couple of proofs for well-known results, e.g. maximal entropy, are lifted almost verbatim from text books.

\section{Quantitative Information Flow}
\begin{figure}[htp]
\centering
 \begin{minipage}{.4\linewidth}

    \includegraphics[scale=.8]{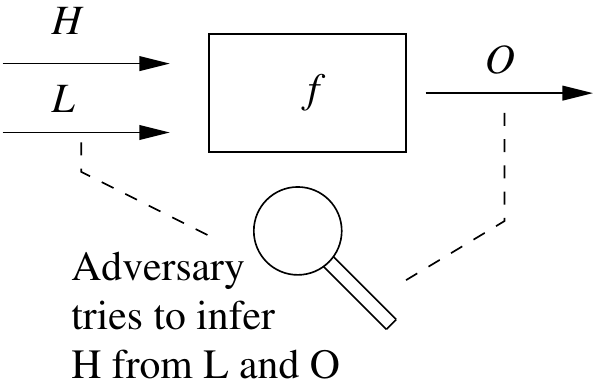}
    \caption{Attacker Model}
    \label{attackermodel}
  \end{minipage}
\end{figure}

Our attacker model is depicted in Figure \ref{attackermodel}. The program \code{P}, characterized by a function $f$, takes confidential input \code{H}, public 
input \code{L}, and produces output \code{O}. \code{L} may or may not be controlled by an adversary.
The adversary tries to infer information from \code{H} by observing \code{L} and \code{O}.

\subsection{Information Flow and Non-interference}
\label{sec:selfcomp}
Information flow is the (illegal) transmission of information from a variable \code{H} to a variable \code{O} in a given process. The 
simplest case of information flow is \emph{explicit flow} (or direct flow) where the whole or partial value of \code{H} is copied directly to \code{O}, for example:

\begin{figure}[htp]
\centering
\begin{minipage}[b]{0.17\linewidth}
\begin{lstlisting}
O = H + 3;
\end{lstlisting}
\end{minipage}
\end{figure}

There are more subtle cases, which are categorized as \emph{implicit flow} (or indirect flow). Consider, for example, the program below, which simulates a common password checking 
procedure:

\begin{figure}[htp]
\centering
\begin{minipage}[b]{0.25\linewidth}
\begin{lstlisting}
if (H == L)
    O = true;
else
    O = false;
\end{lstlisting}
\end{minipage}
\caption{A password checking program}
\label{passcheck}
\end{figure}

\code{H} is the password, i.e. the confidential data; \code{L} is the public input provided by the user; \code{O} is the observable output, ``\code{O = true;}'' means the password is
accepted. Although \code{H} is not directly copied to \code{O}, there is still information flow leaked from $\code{H} \rightarrow \code{O}$. This information is ``small'', but one can reveal all information
about \code{H} if he is allowed to make enough attempts.

Obviously, information flow from confidential data to observable output is not desirable, which is the motivation of research in \emph{secure information
flow}. Dating back to the pioneering work of the Dennings in the 1970s \cite{Denning:1977:CPS:359636.359712}, secure information flow analysis has been an active research topic for the last
four decades.

A popular security policy that guarantees the absence of information flow leaks is non-interference~\cite{Cohen:InformationB,DBLP:conf/sp/GoguenM82}. It was introduced by Goguen and Meseguer as a security policy for a general automaton framework,
here we give a definition of non-interference in language-based settings:

\begin{definition}[Non-interference]\label{DEF:NI}
Suppose a program \code{P} takes secret input \code{H}, public input \code{L} and produces public output \code{O}. \code{P} has the non-interference property if and only if: for all possible pairs of input vector $[ \code{H}_1,\code{L}_1 ]$,
and $[ \code{H}_2,\code{L}_2 ]$:
$$ \code{L}_1 = \code{L}_2 \wedge \code{O}_1 = P ([\code{H}_1,\code{L}_1 ]) \wedge \code{O}_2 = P ([\code{H}_2,\code{L}_2 ]) \Rightarrow \code{O}_1 = \code{O}_2
$$
\end{definition}
where $\code{O}_1 = P ([\code{H}_1,\code{L}_1 ])$ denotes that $\code{O}_1$ is the result of executing the program \code{P} with the 
input vector $[\code{H}_1,\code{L}_1 ]$.

We use the two terms "\emph{information flow}" and "\emph{interference}" interchangeably, since information flow from \code{H} to \code{O} means \code{O} is interfered by \code{H}. The non-interference policy guarantees the absence of interference or information flow.
\subsubsection{Type system approach to non-interference}
There has been a large body of work that has used type systems for validating non-interference, following the idea of Volpano, Irvine and Smith~\cite{Volpano:1996:STS:353629.353648}.
Type systems are fast and they support automated, compositional verification. Moreover, the analysis is \emph{safe}, which means if a program is classified as ``\emph{secure}'', then it is actually secure, there are no false negatives. 

However, this approach is not extensible. Even a small modification in the information flow policy or in the programming language, e.g. adding a new feature,
requires a non-trivial extension of the type system and its soundness proof.
This approach also returns too many false positives, which means secure programs can be classified as ``\emph{insecure}''. For example, consider again the two examples with a small
modification to make them satisfy non-interference: 

\begin{figure}[htp]
\centering
\begin{minipage}[b]{0.25\linewidth}
\begin{lstlisting}
O = H - H + 3; 
\end{lstlisting}
\end{minipage}
\hspace{1cm}
\begin{minipage}[b]{0.25\linewidth}
\begin{lstlisting}
if (H == L)
    O = false;
else
    O = false;
\end{lstlisting}
\end{minipage}
\label{secpass}
\end{figure}

Typing rules would always classify programs like the above as insecure. Another tricky case is of programs that leak information in the intermediate states, but sanitize
information at the end, for example:

\begin{figure}[htp]
\centering
\begin{minipage}[b]{0.20\linewidth}
\begin{lstlisting}
O = H + 3;
O = 3;
\end{lstlisting}
\end{minipage}
\end{figure}

Given that the attacker can only observe the final value of the output \code{O}, the program is secure. However, it would be classified as insecure by type systems.

\subsubsection{Theorem proving approach to non-interference} 
Another prominent approach for secure information flow is to use theorem proving, in which non-interference is logically formulated by \emph{self-composition} \cite{Darvas:2005:TPA:2154040.2154072,Barthe:2004:SIF:1009380.1009669}, as non-interference itself is not
a logical property. 

We assume a similar setting as in the case of non-interference: given a program \code{P} that takes secret input H, public input \code{L} and producing public output \code{O}, we denote by $\code{P}_1$ the same program as \code{P}, with all variables renamed: \code{H} as $\code{H}_1$, \code{L} as $\code{L}_1$ and \code{O} as $\code{O}_1$.
Self-composition is expressed in Hoare-style framework as \cite{Barthe:2004:SIF:1009380.1009669}:
\begin{equation}
\{ \code{L} = \code{L}_1 \} \code{P};\code{P}_1 \{ \code{O} = \code{O}_1 \} 
\label{scHoare}
\end{equation}
The Hoare triple states that if the precondition $\code{L} = \code{L}_1$ holds, then after the execution of $P; P_1$, the postcondition $\code{O} = \code{O}_1$ also holds.
The purpose of having the copy $\code{P}_1$ with all variables renamed is to have \emph{another} \code{P} to compare with \code{P}, so self-composition is
logical formulation of non-interference. Compared to type system approach, the theorem proving approach is much more precise, returning no false positives.

For example, consider again the password checking program \code{P} in Figure \ref{passcheck}, the composition of \code{P} and its copy $\code{P}_1$ is shown in Figure \ref{fig:selfcomp-pass}.
\begin{figure}[htp]
\centering
\begin{minipage}[b]{0.28\linewidth}
\begin{lstlisting}
if (H == L)
    O = true;
else
    O = false;
if (H1 == L1)
    O1 = true;
else
    O1 = false;
\end{lstlisting}
\end{minipage}
\caption{Self-composition of the password checking program}
\label{fig:selfcomp-pass}
\end{figure}
By choosing $\code{H} = \code{L} \wedge \code{H}_1 \neq \code{L}_1$, it is easy to find a counter-example for the Hoare triple in (\ref{scHoare}), such that $\code{L} = \code{L}_1$ holds and $\code{O}  = \code{O}_1$ does not hold. 
Therefore, the password checking program violates non-interference.

Compared to type system approach, the theorem proving approach is much more precise, returning no false positives. However, it is impractical in reality, as elegantly put in
\cite{Terauchi:2005:SIF:2156802.2156828} by Terauchi and Aiken: 

\emph{``When we actually applied the self-composition approach, we found that
not only are the existing automatic safety analysis tools not powerful enough to
verify many realistic problem instances efficiently (or at all), but also that there
are strong reasons to believe that it is unlikely to expect any future advance.''}

Terauchi and Aiken also pointed out that the limitations of self-composition come from the symmetry and redundancy of the self-composed program, which
lead to some partial-correctness conditions that hold between \code{P} and \code{P}$_1$. To find these conditions is crucial for the effectiveness of the analysis, 
however, finding them is in general impractical. 
Moreover, 
the use of an interactive theorem prover requires considerable user interaction and verification expertise.

\subsection{Information Theoretical Measurement of Interference}
The fact that the password checking program, Figure \ref{passcheck}, leaks information indicates that non-interference is over-pessimistic, and often unachievable.
Moreover, there are situations that one needs to decide, between two programs, which one is more secure? Consider the following program:

\begin{figure}[htp]
\centering
\begin{minipage}[b]{0.25\linewidth}
\begin{lstlisting}
if (H >= L)
    O = true;
else
    O = false;
\end{lstlisting}
\end{minipage}
\end{figure}

Obviously this program is much less secure than the one in Figure \ref{passcheck}, a fact that cannot be proved with non-interference.
These examples show the need of a more quantitative assessment of information flow: instead of asking ``\emph{does the program leak information}?'', we would like to know ``\emph{how much does it leak}?''
Information Theory~\cite{Shannon:2001:MTC:584091.584093} provides the tools to answer this question.
\subsubsection{Discrete Probability Theory}
We introduce the concepts that are to be used in the construction of an information theoretical analysis of information flow. A thorough background and description can be found
on standard textbook~\cite{probabilitybook}.
\begin{definition}[Sample space] The sample space of an experiment, denoted by $\Omega$, is a countable set of all possible outcomes of that experiment.
\end{definition}
Each outcome is a complete description of the state of the real world as a result of the experiment. It needs not be a number, e.g. the outcome of 
tossing a coin is either ``head'' or ``tail''. Therefore, probability theory needs the following definition:
\begin{definition}[Random variable]
A random variable is a function $X: \Omega \rightarrow N$ ($N \subseteq \mathbb{R}$) that associates a real number with each possible outcome in $\Omega$. 
\end{definition}
Each element $x \in N$ is assigned a ``\emph{probability}'' value, denoted by $p(x)$ or $p(X=x)$, which is the measure of the likeliness that $X$ takes the value $x$.
\begin{definition}[Probability distribution] The probability distribution of a random variable $X :\Omega \rightarrow N$ is
a function $p: N \rightarrow [0, 1]$ such that:
$$\sum_{x \in N} p(x) = 1$$
\end{definition}
For discrete probability, the function $p$ is also known as \emph{probability mass function}. In this thesis, we are interested in the \emph{uniform distribution}, in which all the elements of $N$ have the same probability. This means every
outcome of the sample space is equally likely to occur.

Any subset $E \subseteq  \Omega$ is refereed to as an \emph{event}. The probability of an event $E$ is defined as follows:
$$p(E) = \sum_{\omega \in E} p(X(\omega))$$
It is straightforward from the definition of probability distribution that $p(\Omega) = 1$, which means $\Omega$ is an event that always occurs.
At other extreme, it is easy to show that $p(\emptyset)$ = 0, i.e. the empty set is an event that never occurs.

\begin{definition}[Conditional Probability] The probability of an event $A$ given that another event $B$ has occurred is defined as follows:
$$p(A|B) = \dfrac{p(A \cap B)}{p(B)}$$
\end{definition}
\begin{definition}[Joint Probability] The joint probability mass function of two discrete random variables $X$, $Y$ is defined as:
$$p(x,y) = p(X=x\ \mathrm{and}\ Y=y) = p(y | x) \ p(x) = p(x | y) \ p(y)$$
\end{definition}
\subsubsection{Information Theory}
Information Theory~\cite{Shannon:2001:MTC:584091.584093,Cover:1991:EIT:129837} provides the mathematical foundation to reason about ``\emph{information}'' in a quantitative sense.
Its main concept is ``entropy'', which measures the uncertainty about a random variable.
\begin{definition}[Entropy] Given a random variable $X: \Omega \rightarrow N$ with the probability mass function $p(x)$. Shannon entropy is defined as:
%
$$F(X) = - \sum_{x \in N}  p(x)\ \log p(x) $$
\end{definition}
The logarithm is to the base 2 by convention, hence the units are \emph{bits}. We also write $F(p)$ for the above quantity.

\begin{lemma} \label{maxe} Given a random variable $X: \Omega \rightarrow N$ with the probability mass function $p(x)$, the entropy of $X$ is bounded by:
 $$ 0 \leq F(X) \leq \log |N|$$
 where $|N|$ is the cardinality of the set $N$.
\end{lemma}
\textbf{Proof:}
By definition $0 \leq p(x) \leq 1$ for all $x \in N$, which leads to $p(x)\log p(x) \leq 0$. As a result, the first
part of the lemma $F(X) \geq 0$ holds.

The second part of the lemma is proved by using Lagrange multiplier to find the maximal entropy. For simplicity, we define a system of index for 
all elements in $N$ as follows: $x_1, x_2,\ldots ,x_n$ (which also means $|N| = n$). 

We write $p_i$ for $p(X = x_i)$, and maximize the function: $$f(p_1,p_2,\ldots,p_n) = - \sum_{i=1}^{n} p_i \log p_i$$
subject to the condition of probability: $$g(p_1,p_2,\ldots,p_n) = \sum_{i=1}^{n} p_i = 1$$

Lagrange multipliers is used to find the maximum entropy $\vec{p}^{\,*} = \{p_1^{\,*}, p_2^{\,*},\ldots, p_n^{\,*}\}$ across all probability distributions $\vec{p} = \{p_1, p_2, \ldots, p_n\}$ of $X$.
It is required that:
$$    \left.\frac{\partial}{\partial \vec{p}}(f+\lambda (g-1))\right|_{\vec{p}=\vec{p}^{\,*}}=0$$
This equation is expanded into a system of $n$ equations ($i ={1, 2,\ldots,n}$):
$$\left.\frac{\partial}{\partial p_i}\left\{-\left (\sum_{j=1}^n p_j \log_2 p_j \right ) + \lambda \left(\sum_{j=1}^n p_j - 1\right) \right\}\right|_{p_i=p^*_i} = 0$$
Carrying out the differentiation of these $n$ equations results in:
$$    -\left(\frac{1}{\ln 2}+\log_2 p^*_i \right) + \lambda = 0$$
Since the value of $p^*_i$ only depends on $\lambda$, they are all equal:
$$p_1^{\,*} = p_2^{\,*} = \ldots = p_n^{\,*}$$
Moreover, by definition of probability distribution:
$$\sum_{i=1}^{n} p_i^{\,*} = 1$$
This leads to:
$$p_1^{\,*} = p_2^{\,*} = \ldots = p_n^{\,*} = \frac{1}{n}$$
The maximal entropy is:
$$F(p^{\,*}) = - \sum_{1}^{n} p_i \log p_i = - \sum_{1}^{n} \frac{1}{n} \log \frac{1}{n} = n = |N|$$
%
This proves the second part Lemma \ref{maxe}: $F(X) \leq \log |N|$, equality holds when $X$ has uniform distribution. 
\begin{definition}[Joint Entropy] Given two random variables $X: \Omega_X \rightarrow N_X$ and $Y: \Omega_Y \rightarrow N_Y$, the joint Shannon entropy is defined as follows:
    $$F(X,Y) = -\sum_{x \in N_X} \sum_{y \in N_Y} p(x,y) \log{p(x,y)}$$
where $p(x,y)$ is the joint probability mass function of $X$ and $Y$.
\end{definition}
The joint entropy measures how much uncertainty there is in the two random variables $X$ and $Y$
taken together.
\begin{definition}[Conditional Entropy] \label{conde} Given two random variables $X :\Omega_X \rightarrow N_X$ and $Y :\Omega_Y \rightarrow N_Y$
, the conditional entropy of $X$ given the knowledge of $Y$ is defined as follows:
$$F(Y|X) = \sum_{x\in N_X}\,p(x)\,F(Y|X=x)$$
\end{definition}
The definition of conditional entropy can be expanded as:
\begin{align*}
F(Y|X) &= - \sum_{x\in N_X}\,p(x)\, \sum_{y\in N_Y}\,p(y|x)\,\log p(y|x) \\
       &= - \sum_{x\in N_X} \sum_{y\in N_Y} \, p(x,y) \,\log p(y|x) \\
       &= - \sum_{x\in N_X} \sum_{y\in N_Y} \, p(x,y) \,\log \left( \dfrac{p(x,y)}{p(x)} \right) \\
       &= -\sum_{x\in N_X} \sum_{y\in N_Y} \, p(x,y) \,\log p(x,y) + \sum_{x\in N_X} \sum_{y\in N_Y} \, p(x,y) \,\log p(x)\\
       &= F(X,Y) + \sum_{x\in N_X} \, p(x) \,\log p(x) \\
       &= F(X,Y) - F(X)
\end{align*}
If $X$ and $Y$ are independent, then knowing the value of $X$ does not change uncertainty about $Y$, which means $F(Y|X) = F(Y)$. At the other
extreme, if the value of $Y$ is completely determined by the value of $X$, then there is no uncertainty about $Y$ when already knowing $X$, and $F(Y|X) = 0$ holds.
\begin{lemma} \label{jointprop} Given two random variables $X$ and $Y$, the following inequalities hold:
$$F(X,Y) \geq \max\{F(X),F(Y)\}$$
\end{lemma}
\textbf{Proof:} Without loss of generality, assume that $\max\{F(X),F(Y)\} = F(X)$, we need to prove:
$$F(X,Y) \geq F(X) \iff F(X) + F(Y|X) \geq F(X) \iff F(Y|X) \geq 0 $$
By definition of conditional entropy, we have $F(Y|X) = \sum\,p(x)\,F(Y|X=x)$. Similar to the proof of the first part of Lemma \ref{maxe}, we can show that $F(Y|X=x) \geq 0$ holds.
As a result, $F(Y|X) \geq 0$ also holds. This proves the inequality in Lemma \ref{jointprop}. 
Equality holds if and only if $F(Y|X) = 0$ holds, which means
the value of $Y$ is completely determined by the value of $X$.
%
%
\begin{definition}[Mutual Information] Given two random variables $X :\Omega_X \rightarrow N_X$ and $Y :\Omega_Y \rightarrow N_Y$, the mutual information or mutual dependence between $X$ and $Y$ is defined as:
$$I(X;Y) = \sum_{y \in N_Y} \sum_{x \in N_X} p(x,y) \log{ \left(\frac{p(x,y)}{p(x)\,p(y)} \right) } $$
\end{definition}
The mutual information can be rewritten as:
\begin{align*}
I(X;Y) &=  \sum_{y \in N_Y} \sum_{x \in N_X} p(x,y) \log{ \left(\frac{p(x|y)}{p(x)} \right) } \\
       &=  - \sum_{y \in N_Y} \sum_{x \in N_X} p(x,y) \log{ p(x) } + \sum_{y \in N_Y} \sum_{x \in N_X} p(x,y) \log{ p(x|y) } \\
       &= F(X) - F(X|Y)
\end{align*}
%
%
\subsubsection{Measurement of leakage}
In the context of information flow analysis, we are interested in the value of variables \code{H}, \code{L}, and \code{O} as in our attacker model in Figure \ref{attackermodel}. 
For each variable, we denote its sample space and random variable as follows:
$X_H: \Omega_H \rightarrow N_H$, $X_L:\Omega_L \rightarrow N_L$, and $X_O: \Omega_O \rightarrow N_O$.

The random variable $X_H$ represents the \emph{a priori} knowledge of the adversary about the secret data \code{H}. Consider, for example, the password
checking program in Figure \ref{passcheck}: if the adversary already knows that the password \code{H} is ``abc'', this means $\Omega_H = \{\text{``abc''}\}$,
and $p(X_H(\text{``abc''})) = 1$.
Or suppose that $\Omega_H$ contains all possible strings of length up to 30, but the adversary knows in advance that the victim has the habit to use his 
daughter's name, ``Alice'', as password. In this case, in the probability distribution of $X_H$, the value of $p(X_H(\text{``Alice''}))$ is close to 1.
The most general case is the adversary does not have any information about the password, which means $X_H$ has a uniform distribution. 

The entropy $F(X_H)$ measures the initial uncertainty of the adversary about 
$H$. When the adversary already knows that the password 
is ``abc'', there is no uncertainty at all. $p(X_H(\text{``abc''})) = 1$ leads to $F(X_H) = 0$. On the other
hand, if the adversary has no information about the password in advance, his uncertainty is maximal.

After an execution of the program \code{P}, there is some information about \code{H} leaked via information flow from \code{H} to \code{O}.
As a result, the adversary learns some information, and reduces his uncertainty about \code{H}.
The difference in his uncertainties before and after the observation is the amount of leakage:
\begin{equation}
\text{{\ttfamily leakage}} = \text{{\ttfamily initial uncertainty}} - \text{{\ttfamily remaining uncertainty}}
\end{equation}
The remaining uncertainty about \code{H} given the knowledge of $O$ is, by definition, the conditional entropy $F(X_H | X_O)$. As a 
result, leakage is calculated as follows:
$$\Delta_F(X_H) = F(X_H) - F(X_H | X_O) = I(X_H;X_O)$$
In the case of non-interference, \code{O} is independent from \code{H}, or in other words \code{O} is not interfered by \code{H}, $F(X_H | X_O) = F(X_H)$ holds.
As a result, the leakage is $\Delta_F(X_H) = 0$.

$\Delta_F(X_H) = I(X_H;X_O)$ is always positive. This property can be proved by using Jensen's inequality. The interested reader 
is referred to the textbook by Cover and Thomas~\cite{Cover:1991:EIT:129837} for a proof in details.
\subsection{Problem Statement}
Programs such as the one in our attacker model in Figure \ref{attackermodel} can be viewed as a channel, in which (confidential) information flows from the input \code{H}
to the output \code{O}. The \emph{channel capacity} $C$ is defined as the maximum amount of information can be transmitted in this channel. More formally,
$C$ is maximum mutual information of $X_H$ and $X_O$ over all possible input distributions $p$ of $X_H$.
\begin{theorem}[Channel Capacity]\label{theo:capcity} Given a program \code{P} as a discrete channel from \code{H} to \code{O}, the channel capacity $C$ is computed by:
 $$C = \log |N_O|$$
\end{theorem}
\textbf{Proof:} By definition of conditional entropy, the leakage can be rewritten as:
 \begin{align*} 
\Delta_F(X_H) &=  F(X_H) - F(X_H | X_O) = F(X_H) - (F(X_H,X_O) - F(X_O)) \\
	      &= F(X_O) - (F(X_H) - F(X_H,X_O))
 \end{align*}
From Lemma \ref{jointprop}, we have $F(X_H,X_O) \geq F(X_H)$. This leads to:
$$\Delta_F(X_H) \leq F(X_O)$$
From Lemma \ref{maxe}: $F(X_O) \leq \log |N_O|$, which means $\Delta_F(X_H) \leq \log |N_O|$ holds.
Equality holds when both of the equalities in Lemma \ref{maxe} and Lemma \ref{jointprop} hold. Which means the value of \code{O} is completely determined
by the value of \code{H}, and $X_O$ has uniform distribution.

As such, counting the number of observables is the basis
of state-of-the-art QIF analysis, e.g.~\cite{Heusser:2010:QIL:1920261.1920300,Meng:2011:CBI:2166956.2166957,Kopf:2012:AQC:2362216.2362268,kmm:qest:2013}, and also
the basis for this thesis.
The channel capacity theorem also justifies the following:
\begin{definition}[The QIF problem]\label{DEF:QIF}
Given a program \code{P}, QIF is the problem of counting $N$, 
the number of possible outputs of \code{P}.
\end{definition}
\section{Logical Satisfiability Problems}
This section provides some logical concepts used throughout in this thesis. The interested readers are referred to~\cite{Biere:2009:HSV:1550723} for more details.
\subsection{Propositional Satisfiability}
\begin{definition}[Propositional atom]
A propositional atom or Boolean atom is a statement or assertion that must be true or false.
\end{definition}
Examples of Boolean atoms are: ``all humans are mortal'' and ``program \code{P} leaks $k$ bits''.
Boolean atoms are the most basic building blocks of \emph{propositional formulas}, each Boolean atoms $A_i$ is also a formula.

Propositional formulas are constructed from Boolean atoms using \emph{logical connectives}: \code{not} ($\neg$), \code{and} ($\wedge$), \code{or} ($\vee$), and \code{imply} ($\rightarrow$). That means if $\varphi_1$ are $\varphi_2$ are formulas, 
then $\neg \varphi_1$, $\varphi_1 \wedge \varphi_2$, $\varphi_1 \vee \varphi_2$, and $\varphi_1 \rightarrow \varphi_2$ are also formulas. 
For example, $(\neg A_1 \wedge A_2) \rightarrow A_3$ is a propositional formula.

A Boolean atom $A_i$ or its negation $\neg A_i$ is called a \emph{literal}.
We denote by \emph{Atom}($\varphi$) the set $\{A_1, A_2, \dots A_n\}$ of Boolean atoms that occur in $\varphi$. The truth of a propositional
formula $\varphi$ is a function of the truth values of the Boolean atoms it contains.

We denote by $\top$ and $\bot$ the truth values of true and false, respectively.
\begin{definition}
Given a propositional formula $\varphi$, a truth assignment $\mu$ of $\varphi$ is defined as a function
which assigns each Boolean atom of $\varphi$ a truth value:
$$\mu: Atom( \varphi ) \rightarrow \{\top, \bot\} $$
\end{definition}
A partial truth assignment of a formula $\varphi$ is a function $\mu: \mathcal{A} \rightarrow \{\top, \bot\} $ where $\mathcal{A}$ is any subset of \emph{Atom}($\varphi$).
A (partial) truth assignment $\mu$ satisfies a propositional formula $\varphi$, denoted by $\mu \models \varphi$ , if $\varphi$ is evaluated to $\top$
under $\mu$. For example $\mu: A_3 \mapsto \bot$ satisfies the formula $(\neg A_1 \wedge A_2) \rightarrow A_3$.

A formula $\varphi$ is \emph{satisfiable} if there exists a (partial) truth assignment such that $\mu \models \varphi$.
If $\mu \models \varphi$ for every truth assignment $\mu$, then $\varphi$ is \emph{valid}. Either a formula is valid or its negation is satisfiable.
\begin{definition}
A propositional formula $\varphi$ is in Conjunctive Normal Form (CNF) if and only if it is a conjunction of
disjunctions of literals:
$$\varphi = \bigwedge_{i = 1}^N \bigvee_{j = 1}^{M_i} l_{ij}$$
\end{definition}
Any propositional formula can be converted to CNF by an algorithm with worst-case linear time~\cite{Plaisted:1986:SCF:7240.7244,BoydelaTour:1992:ORC:147064.147065}.
\begin{definition} [The SAT problem] Given a propositional formula $\varphi$ in CNF, the Boolean Satisfiability Problem (SAT) is the problem of
 finding an assignment $\mu$ that satisfies $\varphi$.
\end{definition}
\todo{DPLL}
As the SAT problem is NP-complete, there is no algorithm that efficiently works on all instances of the problem. However, there are two main families of algorithms for state-of-the-art SAT solvers: DPLL~\cite{Davis:1960:CPQ:321033.321034,Davis:1962:MPT:368273.368557} and Stochastic Local Search~\cite{morgan2005}. This thesis focuses on the DPLL algorithm, which will be described in details later in chapter~\ref{chap:SymExDPLL}.
\subsection{Model Counting}

\begin{definition} [The \#SAT problem]
Given a propositional formula $\varphi$, the Model Counting problem (\#SAT) is the problem of counting all the solutions of the SAT problem.
\end{definition}

\todo{The blocking clause approach}

\subsection{Satisfiability Modulo Theories}
We assume countable sets of variable $\mathcal{V}$, function symbols $\mathcal{F}$ and predicate symbols $\mathcal{P}$. A first-order logic \emph{signature} is defined
as a partial function $ \Sigma: \mathcal{F} \cup \mathcal{P} \rightarrow A $ ($A \subset \mathbb{N}$). Each $a \in A$ corresponds to the \emph{arity} of an symbol.
Obviously, a 0-ary predicate is a Boolean atom, and a 0-ary function symbol is called a \emph{constant}.

A $\Sigma$-term $\tau$ is either a variable $x \in \mathcal{V}$ or it is built by applying function symbols in $\mathcal{F}$ to $\Sigma$-terms, e.g.
$f(\tau_1,\dots ,\tau_n)$ where $f \in \mathcal{F}$ and $\Sigma(f) = n$. For example, $f(x,g(x))$ is a $\Sigma$-term if $\Sigma(f) = 2$
and $\Sigma(g) = 1$.
\begin{definition}
If $\tau_1,\dots ,\tau_n$ are $\Sigma$-terms, and $p \in \mathcal{P}$ is a predicate symbol 
such that $\Sigma(p)= n $, then $p(\tau_1,\dots ,\tau_n)$ is a $\Sigma$-atom.
\end{definition}
A $\Sigma$-atom or its negation is called $\Sigma$-literal. We use the infix equality sign ``$=$'' as a shorthand for the equality predicate. If $\tau_1$ and $\tau_2$ are $\Sigma$-terms, then the $\Sigma$-atom 
$\tau_1 = \tau_2$ is called $\Sigma$-equality. $\neg(\tau_1 = \tau_2)$ or $\tau_1 \neq \tau_2$ is called $\Sigma$-disequality.

$\Sigma$-atoms are the most basic building blocks of $\Sigma$-formulas, each $\Sigma$-atom $p(\tau_1,\dots ,\tau_n)$ is also a $\Sigma$-formula.
Similar to the construction of propositional formulas, $\Sigma$-formulas are constructed from $\Sigma$-terms which are glued together by \code{universal quantifiers} ($\forall$), \code{existential quantifiers} ($\exists$), and logical connectives. That means if $\varphi_1$ and $\varphi_2$ are 
$\Sigma$-formulas, then $\forall x: \varphi_1$, $\exists x: \varphi_1$, $\neg \varphi_1$, $\varphi_1 \wedge \varphi_2$, $\varphi_1 \vee \varphi_2$, and $\varphi_1 \rightarrow \varphi_2$ are also $\Sigma$-formulas.

A \emph{quantifier-free} $\Sigma$-formula does not contain quantifiers; a \emph{sentence} is a $\Sigma$-formula without free variables.
A first-order theory is defined as follows:
\begin{definition}[First-order theory]
A first-order theory $\mathcal{T}$ is is a set of first-order sentences with signature $\Sigma$.
\end{definition}
A $\Sigma$-structure $M$ is a triple ($|M|,\Sigma,\mathcal{I}$) consisting of a non-empty domain $|M|$, a signature $\Sigma$, and an interpretation $\mathcal{I}$.
The interpretation $\mathcal{I}$ assigns meanings to symbols of $\Sigma$: for each function symbol $f \in \mathcal{F}$ such that $\Sigma(f) = n$, $f$ 
is assigned a $n$-ary function $\mathcal{I}(f)$ on the domain $|M|$; for each predicate symbol $p \in \mathcal{P}$ such that $\Sigma(p) = n$, $p$ is assigned a $n$-ary
predicate $\mathcal{I}(p)$, represented by a subset of $|M|^n$. For each variable $x \in \mathcal{V}$, $\mathcal{I}(x) \in |M|$.

A $\Sigma$-structure $M$ is a model of the $\Sigma$-theory $\mathcal{T}$ if it satisfies all sentences in $\mathcal{T}$. If a $\Sigma$-formula is 
 satisfiable in a model of $\mathcal{T}$, then it is called $\mathcal{T}$-\emph{satisfiable}.

Henceforth, for simplicity we will omit the prefix ``$\Sigma-$'' from term, atom, formula, etc. Instead, we will often use the prefix ``$\mathcal{T}$-''
to denote ``in the theory $\mathcal{T}$''.

We define a bijective function $\mathcal{BA}$ (\emph{Boolean abstraction}) which maps Boolean atoms into themselves
and $\mathcal{T}$-atoms into fresh Boolean atoms. 
The \emph{Boolean refinement} function $\mathcal{BR}$ is then defined as the inverse of $\mathcal{BA}$, which means $\mathcal{BR}$ = $\mathcal{BA}^{-1}$.

\begin{definition} [The SMT problem] Given a theory or a combination of theories $\mathcal{T}$ and a $\Sigma$-formula $\varphi$, the 
Satisfiability Modulo Theories problem (SMT) is the problem of deciding $\mathcal{T}$-satisfiability of $\varphi$.
\end{definition}

Most of the work on SMT focus on quantifier-free formulas, and \emph{decidable} first-order theories. The SMT problem is NP-hard, as it 
subsumes the SAT problem.

Most state-of-the-art SMT solvers implement the DPLL($\mathcal{T}$) algorithm, which is the integration of two components: (i) an \emph{enumerator} integrating a DPLL-based SAT solver enumerates truth
assignments satisfying the Boolean abstraction of the input formula; (ii) $\mathcal{T}$-solvers validate the consistency w.r.t. theories
$\mathcal{T}$ of the (partial) assignment produced by the SAT solver.
The DPLL($\mathcal{T}$) algorithm will be described in more details later in chapter \ref{chap:SymExDPLL}.
\todo{TODO: definition of theory of arithmetic, fixed-width bit vectors, see the handbook}
\section{The programming language and the program}
\label{sec:transit}
For simplicity, we illustrate our methodologies using the \emph{guarded command}  
language \cite{Flanagan:2001:AEE:360204.360220} instead of C or Java.
The grammar of the 
language is depicted  in Figure~\ref{fig:lang}.

\begin{figure}[htp]
$$
\begin{array}{llll}
\text{\emph{program} }  &\equiv         &stmt*\\
\text{\emph{stmt s} }   &\equiv         &\text{ \textbf{assume} \emph{e} } | \text{ \textbf{assert} \emph{e} } |\text{ \emph{ v = e }}  
                        &               |\text{ \textbf{if} \emph{e} \textbf{then goto} \emph{s} \textbf{else goto} \emph{s}}\\
\text{\emph{v} }        &\in            & \text{\emph{Var}} & \text{\emph{(variables)}}\\
\text{\emph{e} }        &\in            & \text{\emph{Exp}} & \text{\emph{(expressions)}}\\
\end{array}
$$
\label{fig:lang}
\caption{The guarded command language}
\end{figure}

In this thesis, we focus on safety properties: note that the two commands $assume (c)$ and $assert (c)$ are powerful enough to encode expressive temporal properties~\cite{Clarke:2003:BCC:775832.775928}, and also support assume-guarantee style compositional reasoning.
Any program can be viewed as a system that transits between states. There are many ways to describe this system depending on how much detail
of the program that needs to be captured. Apart from chapter \ref{chap:SymExDPLL}, in this thesis a program \code{P} is modelled as a transition system
as follows:
\begin{equation}
P = (S, I, F, T)
\label{equa:transystem}
\end{equation}
where $S$ is the set of program states;
$I \subseteq S$ is the set of initial states; $F \subseteq S$ is the set of final states; and
$T \subseteq S \times S$ is the transition relation.

Under this setting, a trace of (a concrete) execution of the program \code{P} is represented by a sequence of states: $\rho = s_0s_1..s_k$ such that $s_0 \in I,
s_k \in F$ and $\langle s_i,s_{i+1}\rangle\in T$ for all $i \in \{0,..,k-1\}$.

We define two functions \emph{init} and \emph{fin} to get the initial state and final state of $\rho$:
$\text{\emph{init}}(\rho)=s_0 \text{ and } \text{\emph{fin}}(\rho)=s_k$
The semantics of $P$ is then defined as the set $\mathcal{R}$ of all possible traces.

In the context of the information flow problem, we assume that each initial state $s\in I$ is a pair $\langle H,L \rangle$, i.e. 
$I=I_H \times I_L$, in which $H$ is the confidential component to be protected and $L$ is the public component that may be controlled by an attacker.
\section{Formal Methods}
Formal methods refer to a set of mathematical-based techniques used in Computer Science for the specification and
verification of software and hardware systems. These techniques base their foundations on several conceptual frameworks: automata theory, 
logic calculi, formal languages, program semantics and so on.

The act of using formal methods to prove or to disprove the correctness of a system, with respect to a certain property, is called 
\emph{formal verification}. Compared to testing, formal verification is much more expensive. However, it is crucial in the development of systems whose failure can cause huge financial lost or even cost human lives. 
History has witnessed several computer-related disasters that could have been prevented if formal verification had been used~\cite{clarke1999model}.
\todo{TODO: ariel missile, etc}

There are three main components involving in the formal verification of a hardware or software system:
\begin{itemize}
 \item \emph{A formal model} of the system. Models used in formal verification vary in the level of abstraction, from an automaton describing status 
 changes of the system, to source code or machine code of the system.
 \item \emph{A formal specification}, often described in a formal languages. These formal languages also have different power of expressiveness. 
 \item \emph{A formal method}, implemented in a fully or partially automated tool, to prove or disprove the conformance of the formal model to the formal specification.
\end{itemize}

There are three possible cases for the result: the first case is the program conforms to the specification;
the second case is the system violates the specification, in which a counterexample might be returned; the final case is 
the tool fails to prove or disprove within a period of time.

Naturally, there is a trade-off between the level of abstraction of the model and the expressiveness of the specification. Typically, formal methods-based tools can check
complicated specification on highly abstract models, and simple specification in detailed models.
In this thesis, the formal model of a program is C source code or Java bytecode. The formal specification is the reachability of some assertions in
the source code or bytecode. The formal methods we used are Bounded Model Checking and Symbolic Execution.

\subsection{Bounded Model Checking and CBMC}\label{Sec:BMC-CBMC}
The aim of Bounded Model Checking \cite{Biere:1999:SMC:646483.691738} (BMC) is to find bugs or to prove their absence up to some bounded \emph{k} number of transitions. 
Recall that a program \emph{P} is modelled as a transition system $P = (S, I, F, T)$, and a trace is represented by a sequence of states $\rho = s_0s_1..s_k$.

A trace can be also seen in logical form:
the set \emph{I} and the relation \emph{T} can be written as their characteristic functions: $s_0 \in I$ iff $I(s_0)$ holds; $\langle s_i,s_{i+1}\rangle\in T$ iff
$T(s_i, s_{i+1})$ holds. In this way, a trace $\rho$ is represented by the formula:
\begin{equation}
 I(s_0) \wedge \bigwedge_{i=0}^{k-1} T(s_i, s_{i+1})
\label{equa:constraints}
\end{equation}

Clearly the transition system $P$ is a model for such a formula, i.e. $P$ is a model for  all formulas representing traces of the program. 
As BMC aims to find bugs or prove their absence up to some bounded \emph{k} number of transitions, it explores all traces $\rho = s_0s_1..s_k$ of
the program \emph{P}, in which $s_k$ needs not to be in \emph{F}. 

Notice that because of the bound $k$  there are only a finite number of traces to explore. Hence we can represent the bounded program as a formula $\mathcal{C}$ which is a conjunction of formulas, whose conjoints are possible traces.
Notice formulas can also represent symbolic traces, for example if in a formula the value of a program variable is left unspecified then there can be several concrete traces satisfying that formula. Formulas satisfied by set of concrete traces can be referred to as symbolic traces.

CBMC translates a C program into a logical formula $\mathcal{C}$ which is then used as a model for the property  $\mathcal{P}$ to be verified. The property is   verified by the C program 
 iff $\mathcal{C} \wedge \mathcal{P}$ is valid. This can be checked by a satisfiability solver on $\mathcal{C} \wedge\neg \mathcal{P}$.
 In fact if $\mathcal{C} \wedge\neg \mathcal{P}$ is true in the model then one trace will satisfy $\neg \mathcal{P}$  hence the property is not valid. On the other hand if  $\mathcal{C} \wedge\neg \mathcal{P}$ is false in the model then no trace will satisfy $\neg \mathcal{P}$ hence $\mathcal{P}$ is valid.

\subsection{Symbolic Execution and Symbolic PathFinder}
\label{sec:SPF}
Symbolic Execution~\cite{King:1976:SEP:360248.360252} (SE) is a programming analysis technique which executes programs 
on unspecified inputs, by using symbolic inputs instead of concrete data. For each executed program path, SE builds a \emph{path condition} which 
is the condition on the inputs for the execution to follow that path, according to the branching conditions in the code. 

A path condition $pc$ is initialized as empty, and it doesn't change when executing non-branching instructions. For an \textbf{if} statement with condition $c$, there are three possible cases: (\textbf{i}) $pc \vdash c$: SE chooses the \textbf{then} path; (\textbf{ii}) $pc \vdash \neg c$: SE chooses
the \textbf{else} path; (\textbf{iii}) ($pc \nvdash c$) $\wedge$ ($pc \nvdash \neg c$): SE executes both paths: in the \textbf{then} path, it updates the path condition
$pc_1 = pc \wedge c$, in the \textbf{else} path it updates the path condition $pc_2 = pc \wedge \neg c$.

In classical SE, the satisfiability of the path condition is checked at every branching point, using off-the-shelf constraint solvers. In this way only 
feasible program paths are explored. A symbolic execution tree characterizes the execution paths followed during the symbolic execution of a program. The nodes represent 
program (symbolic) states and the arcs represent transitions between states. 

Symbolic PathFinder (SPF) is a SE framework built on top of the Java PathFinder
(JPF) model checking tool-set for Java bytecode analysis. It implements a bytecode interpreter that
replaces the standard, concrete execution semantics of bytecodes with a non-standard symbolic execution.

SPF implements classical SE, in its default mode SPF only explores feasible symbolic paths.
However, it also has an option to run without constraint solving, which means for an \textbf{if} statement with condition $c$, both $pc \nvdash c$ and $pc \nvdash \neg c$
are assumed to be true. As a result of this option, SPF will explore all the possible paths (feasible and infeasible) through the program, up to the given bound.
This particular option will be used later in chapter~\ref{chap:jpf-bmc} for the design of a concurrent bounded model checker.

\chapter[Model Counting Modulo Theories]{Model Counting Modulo Theories}
\label{chap:SQIF}
\newcommand{\qif}{QIF}
\graphicspath{{chapter3/figs/}}

This chapter introduces the \#SMT problem and a \#SMT-based technique for QIF analysis, which provides a dramatic improvement on state-of-the-art implementations of QIF analysis. On the theoretical side, this work establishes a connection between fundamental verification algorithms and QIF. This connection is
exploited to mitigate the state explosion problem by developing  a novel approach for QIF based on
SMT. More specific contributions are:
\begin{enumerate}
 \item Introduction of a new research problem, Model Counting Modulo Theories or \#SMT, and its applications to QIF.
 \item A framework, called \#DPLL($\mathcal{T}$), to build a solver for \#SMT-based QIF.
 \item A prototyping tools for QIF analysis: \textbf{sqifc} built on top of CBMC~\cite{ckl2004}.
 \item Analysis of complex code, including recent vulnerabilities from the National Vulnerability Database of the US government~\cite{nvd} and anonymity protocols.
\end{enumerate}
\section{Illustrative Example}
\label{chap3:example}
To illustrate our approach, consider the data sanitization program \code{P} from~\cite{Newsome:2009:MCC:1554339.1554349,Meng:2011:CBI:2166956.2166957}, shown 
in Figure \ref{sanitize}. 
\begin{figure}[htp]
\centering
\begin{minipage}[b]{0.22\linewidth}
\begin{lstlisting}
L = 8;
if (H < 16)
  O = H + L;
else
  O = L;
\end{lstlisting}
\end{minipage}
\caption{Data sanitization}
\label{sanitize}
\end{figure}
It is straightforward to show that only integer values from 8 to 23 are possible outputs of this program. 
An attacker has hence available 16 possible output observations: observing outputs  9 .. 23 will know the secret \code{H} is 1 .. 15 and observing 8 will know 
the secret is 0 or greater than 15. Assuming the attacker has no prior knowledge of the secret \code{H} apart that is a 32 bits variable his \emph{a-priori} probability 
of guessing the value of \code{H} in one try is $\frac{1}{2^{32}}$, and the expected probability of guessing the secret in one try after observing the outputs is:
\[ \frac{15}{2^{32}}+\frac{{2^{32}}-15}{2^{32}}\frac{1}{2^{32}-15}= \frac{15}{2^{32}}+ \frac{1}{2^{32}} =\frac{16}{2^{32}}\]
We can measure the leakage of the program as the difference (of the $-\log$ base 2) between the probability of guessing the secret before and after observing the outputs of the program; in this case:
\[ -\log(\frac{1}{2^{32}}) - (-\log\frac{16}{2^{32}}) = \log(16) = 4\]
The result of this measurement, $\log(16)=$ log (number of output observations), is an alternative explanation for theorem \ref{theo:capcity}, i.e. theorem of channel capacity,
that we have proved in the previous chapter. Our goal is to develop an efficient automated technique to compute this number of output observations.

Notice that the output \code{O} is stored in the computer memory as a string of 32 bits $b_1b_2\dots b_{32}$, which can be represented by a set of Boolean 
variables $V_I = \{p_1, p_2, \dots p_{32}\}$ such that $p_i = \top$ if and only if $b_i$ is 1, and $p_i = \bot$ if and only if $b_i = 0$.
Thus, each possible value of \code{O} corresponds to a truth assignment for $V_I$. For example, \code{O} = \code{1000b} corresponds to:
$p_1 \mapsto \bot, p_2 \mapsto \bot, p_3 \mapsto \bot, p_4 \mapsto \top, p_5 \mapsto \bot, \dots , p_{32} \mapsto \bot$.

Since there are 16 possible values of \code{O} from 8 to 23, there are 16 possible truth assignments for $V_I$. We can view these truth assignments 
as partial models of a logical satisfiability problem on a logical formula $\varphi_P$. Obivously, this formula $\varphi_P$ characterizes the 
behaviour of the program \code{P} because of the correspondence between a possible output of \code{P} and a partial model of $\varphi_P$.

Our goal is to count all possible values of \code{O}, which corresponds to counting all models of $\varphi_P$ with respect to the set $V_I$.
In other words, we cast the problem of measuring information leaks of \code{P} to a model counting problem on $\varphi_P$. In the next sections,
we will give a formal definition for this problem, which we name Model Counting Modulo Theories, and develop an algorithm for QIF analysis based on it.
\section{Model Counting Modulo Theories}
\label{method}
\begin{figure}[htp]
\centering
\begin{tikzpicture}
\node at (0,0) {SAT};
\draw [->] (0,-0.5) -- (0,-3.5);
\node [above] at (4,.2) {generalize to first-order theories};
\node at (0,-4) {\#SAT};
\draw [->] (.7,0) -- (7.5,0);
\node at (8.2,0) {SMT};
\node[align=left, below] at (1.7,-1.2)
{generalize to\\counting models};

\draw [->] (8.2,-0.5) -- (8.2,-3.5);
\draw [->] (.7,-4) -- (7.5,-4);
\node at (8.2,-4) {?};
\node [above] at (4,-3.8) {generalize to first-order theories};
\node[align=right, below] at (6.5,-1.2)
{generalize to\\counting models};
\end{tikzpicture}
\caption{Logical satisfiability problems}
\label{fig:satprob}
\end{figure}
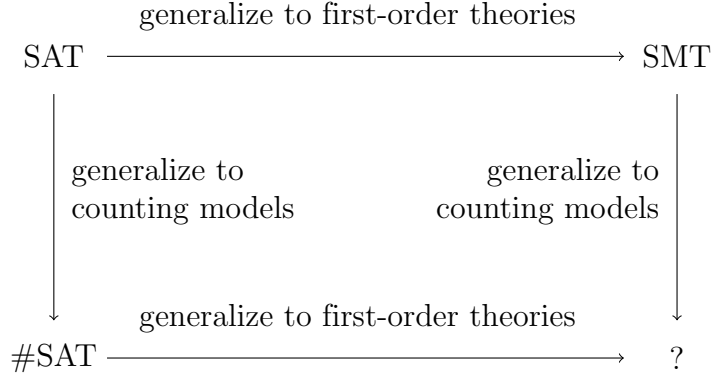

In the previous chapter, we have recalled three logical satisfiability problems that have been studied extensively in recent years, namely SAT, \#SAT, and SMT.
Their relations with each other are depicted in Figure \ref{fig:satprob}. Since propositional logic is a special case of first-order theories whose signature contains 
only 0-ary predicates, the SAT problem is a simple case of the SMT problem. Moreover, the \#SAT problem is a generalization of the SAT problem from finding one model to counting all models. 
What is lacking in this big picture is a satisfiability problem that is a generalization of \#SAT to first-order theories, and is a generalization
of SMT from finding one model to counting all models. 

Note that it is not always possible to count models in SMT as most SMT theories permit an infinite number of models. For example, there are uncountably 
infinite models for the formula:
$\varphi = A \wedge (x > 1)$,
where the background theory $\mathcal{T}$ is $\mathcal{LA}(\mathbb{Q})$, the theory of linear arithmetic over the rationals, which means $x \in \mathbb{Q}$.

Here we restrict the problem to counting all models with respect to a set of Boolean variables. This restriction guarantees that there are always
finite number of models regardless of the background theories.
\begin{definition}[The \#SMT problem]
Given a theory or a combination of theories $\mathcal{T}$ and a $\Sigma$-formula $\varphi$, the Model Counting Modulo Theories problem 
(\#SMT) is the problem of counting all models $M$ of $\mathcal{T}$ with respect to a set of Boolean variables $V_I$ such that $\varphi$ is
$\mathcal{T}$-satisfiable in $M$.
\end{definition}

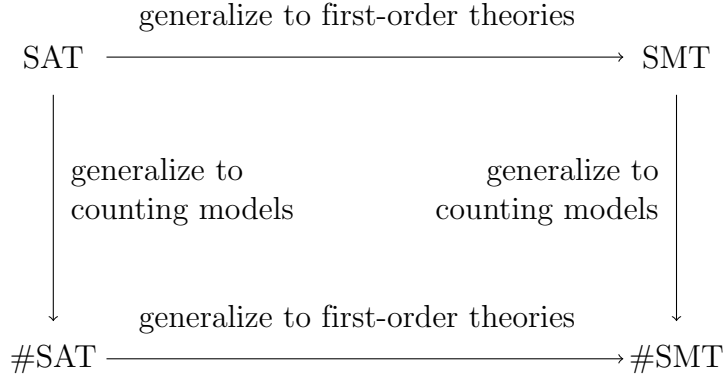
\begin{figure}[htp]
\centering
\begin{tikzpicture}
\node at (0,0) {SAT};
\draw [->] (0,-0.5) -- (0,-3.5);
\node [above] at (4,.2) {generalize to first-order theories};
\node at (0,-4) {\#SAT};
\draw [->] (.7,0) -- (7.5,0);
\node at (8.2,0) {SMT};
\node[align=left, below] at (1.7,-1.2)
{generalize to\\counting models};

\draw [->] (8.2,-0.5) -- (8.2,-3.5);
\draw [->] (.7,-4) -- (7.5,-4);
\node at (8.2,-4) {\#SMT};
\node [above] at (4,-3.8) {generalize to first-order theories};
\node[align=right, below] at (6.5,-1.2)
{generalize to\\counting models};
\end{tikzpicture}
\caption{Logical satisfiability problems}
\label{fig:satprob2}
\end{figure}

With this new \#SMT problem, our picture of logical satisfiability problems in Figure \ref{fig:satprob2} is now complete. 
The \#SMT problem is a generalization of both the \#SAT problem and the SMT problem.

Recall that most state-of-the-art SMT solvers are the integration of two components: (i) an \emph{enumerator} integrating a SAT solver enumerates truth
assignments satisfying the Boolean abstraction of the input formula; (ii) $\mathcal{T}$-solvers validate the consistency w.r.t. theories
$\mathcal{T}$ of the (partial) assignment produced by the SAT solver.

Naturally, an SMT solver can be extended into a \#SMT solver by replacing the SAT solver with a \#SAT solver that can explicitly enumerate all
models.

\section{Quantitative Information Flow as \#SMT}
\label{chap3:method}
We assume the setting in our attacker model in Figure \ref{attackermodel}: a program \code{P}
that takes secret input \code{H}, public input \code{L} and producing public output \code{O}. As per definition \ref{DEF:QIF}, the quantitative information flow problem is to
count the number $N$ of possible values of the output \code{O}, which is an $M$-bit data $b_1b_2\dots b_M$.

Assuming that we have a (first-order) formula $\varphi_P$ with the following properties: (i) $\varphi_P$ contains a set of Boolean variables 
$V_I := \{p_1,p_2,..,p_M\}$; (ii) $p_i = \top$ if and only if $b_i$ is 1, and $p_i = \bot$ if and only if $b_i = 0$. Under these settings, 
the QIF problem of counting $N$ can be viewed as a \#SMT problem on the formula $\varphi_P$ and the set $V_I$ of Boolean variables.

Under this view, on one hand we have a program \code{P} to perform QIF analysis, and a tool box of formal methods. On the other hand, we have 
a logical formula $\varphi_P$ to compute \#SMT and the DPLL($\mathcal{T}$) algorithm. Hence, there are two possible approaches to our QIF problem: the first one
is to construct such a formula $\varphi_P$ from the program \code{P}, then solving it with an \#SMT solver; the second one is to use formal methods in a way
that mimics the DPLL($\mathcal{T}$) algorithm.
\begin{center}
\begin{tikzpicture}
\node at (0,0) {\code{P}};
\draw [<->] (.7,0) -- (5.2,0);
\node at (5.9,0) {$\varphi_P$};
\node at (0,-.8) {QIF};
\node at (5.9,-.8) {\#SMT};
\node at (0,-1.6) {Formal methods};
\node at (5.9,-1.6) {DPLL($\mathcal{T}$)};
\end{tikzpicture}
\end{center}
The first approach is more intuitive. However, it is also more complicated, since there is no available off-the-shelf \#SMT solver. Therefore, we will leave it 
for chapter~\ref{chap:allsmt}. In this chapter, we will explore the second approach, which is much simpler yet powerful enough to analyse real-world programs, and
outperform dramatically the state-of-the-art technique.

An extremely naive technique for QIF using formal methods is to check each number one by one if it can be a value of the output \code{O}.
This can be done as follows:

\begin{figure}[htp]
\centering
\fbox{
\begin{minipage}[b]{0.5\linewidth}
\begin{algorithmic}
\State $N$ = 0
\ForAll {$v$ from 0 to $2^M$}
  \If {(\textbf{assert} \code{O} != $v$ is violated)} 
	\State $N \leftarrow N + 1$ 
  \EndIf
\EndFor\\
\Return $N$
\end{algorithmic}
\end{minipage}
}
\end{figure}

We make an assertion that the output \code{O} is always different from $v$, then checking the validity of this assertion using formal methods.
If the assertion is valid, then $v$ is not a possible value of \code{O}. On the contrary, if the assertion is violated, then \code{O} can take the value $v$,
and we increase the counter.

Assuming that we have a very powerful tool that can verify each assertion in one second, the procedure would take around $2^{32}$ seconds, which is 
approximately 136 years. Therefore, this technique is impractical. The reason is that it checks one concrete value at a time, and thus it is vulnerable
to the state-space explosion problem. Although this technique is naive, it inspires us to develop a procedure to process multiple values at a time.

Consider again the set of Boolean variables $V_I := \{p_1,p_2,..,p_M\}$, for example the partial truth assignment $\mu = \{p_1 \mapsto \top, 
p_2 \mapsto \bot\}$ represents  $2^{M-2}$ concrete values: all the bit configurations over $M$ bits where the first bit is 1 and the second bit is 0. Thus,
if we can verify that $\varphi_P$ cannot be satisfiable in $\mu$, we can ignore those $2^{M-2}$ values.
Although for this approach we do not construct the formula $\varphi_P$, checking that $\varphi_P$ cannot be satisfiable in $\mu$ can be done by
using formal methods to check the assertion that $\neg (b_1 = 1 \text{ \&\& } b_2 = 0)$ is valid.

So a partial truth assignments of $V_I$ is a symbolic representation for a set of values of the output \code{O}. Based on this, we develop a technique,
called \emph{Symbolic Quantitative Information Flow} that mimics the DPLL($\mathcal{T}$) algorithm and explores the state space.
Recall that DPLL($\mathcal{T}$) algorithm consists of a DPLL-based SAT solver to enumerate (partial) truth assignments, and a $\mathcal{T}$-solver to 
check the consistency of these truth assignment. As we use formal methods, in particular model checking, to check if the formula $\varphi_P$ can
be satisfiable in a partial truth assignment $\mu$, the model checker plays the role of the $\mathcal{T}$-solver.
\begin{center}
\begin{tikzpicture}
\node at (0,0) {\code{P}};
\draw [<->] (.7,0) -- (5.2,0);
\node at (5.9,0) {$\varphi_P$};
\node at (0,-.8) {Model Checker};
\node at (5.9,-.8) {$\mathcal{T}$-solver};
\end{tikzpicture}
\end{center}
\section{Symbolic Quantitative Information Flow}
Our first step is to construct the set of Boolean variables $V_I$ that we have described. In a language that supports bitwise operators such as C/C++ and Java,
this can be done by instrumenting the program \code{P} as follows:
\begin{figure}[htp]
\centering
\fbox{
\begin{minipage}[b]{0.37\linewidth}
\begin{algorithmic}
\ForAll {$i$ from 1 to $M$}
  \State $b_i =$ (\code{O} $>>$ ($i$ - 1)) \& 1
  \If {($b_i$ == $1$)} 
	\State $p_i \leftarrow \top$
  \Else
	\State $p_i \leftarrow \bot$
  \EndIf
\EndFor
\end{algorithmic}
\end{minipage}
}
\caption{Program instrumentation}
\label{fig:instru}
\end{figure}

\begin{figure}
\centering
\fbox{
\begin{minipage}[b]{0.45\linewidth}
\begin{algorithmic}
\Function{SymbolicQIF}{$V_I,\varphi_P$}
\State $\varPsi$ = $\epsilon$, $pc$ = $\epsilon$, $N$ = 0, $i$ = 1
\State EarlyPrunning($V_I$) \label{algsqif:unitpropagate}
\State SymCount($V_I, \varPsi, \varphi_P,N, pc, i$) \label{algsqif:symcount}\\
\Return $\varPsi$, $\log_2(N)$
\EndFunction
\end{algorithmic}
\end{minipage}
}
\caption{Symbolic QIF analysis}
\label{symQIFframe}
\end{figure}

\begin{figure}
\centering
\fbox{
\begin{minipage}[b]{0.6\linewidth}
\begin{algorithmic}[1]
\Function{SymCount}{$V_I, \varPsi, \varphi_P,N, pc, i$}
\If{($N \geq 2^k$)} \label{algsqif:policyStart}
    \Return $Insecure$
\EndIf \label{algsqif:policyEnd}
\State Extract $p_i$ from $V_I$
\State $pc_1 \leftarrow pc \wedge p_i$ \label{algsqif:start1}
\If{($\mathcal{T}$-solver($\varphi_P,pc_1$))} \label{algsqif:sat1}
    \If{($i == M$)}
        \State $\varPsi \leftarrow \varPsi \cup \{pc_1\}$
        \State $N \leftarrow N + 1$
    \Else
	\State SymCount($V_I, \varPsi,\varphi_P, N, pc_1, i+1$)
    \EndIf
\EndIf \label{algsqif:end1}
\State $pc_2 \leftarrow pc \wedge \neg p_i$ \label{algsqif:start2}
\If{($\mathcal{T}$-solver($\varphi_P,pc_2$))} \label{algsqif:sat2}
    \If{($i == M$)}
        \State $\varPsi \leftarrow \varPsi \cup \{pc_2\}$
        \State $N \leftarrow N + 1$
    \Else
	\State SymCount($V_I, \varPsi, \varphi_P, N, pc_2, i+1$)
    \EndIf
\EndIf \label{algsqif:end2}
\EndFunction
\end{algorithmic}
\end{minipage}
}
\caption{Symbolic counting for QIF}
\label{symcount}
\end{figure}

This instrumentation guarantees that the variable $p_i$ corresponds to the $i^{th}$ bit of the output \code{O}.
A high level framework to explore the state-space and quantify the leaks of confidential data is described by the procedures {\tt SymbolicQIF} and ${\tt SymCount}$ in Figure \ref{symQIFframe}
and \ref{symcount}. 

$V_I$, $\varPsi,\varphi_P$ and $N$ are passed by reference, while $pc$ and $i$ are passed by value.
$V_I$ is the symbolic representation of the output described in the previous section,  $\varphi_P$ is the formula
representing the program $P$ and $\varPsi$ is the set of models of $\varphi_P$. $N$ is the cardinality of $\varPsi$,
and the procedure ${\tt SymbolicQIF}$ returns $\log_2(N)$ as the channel capacity. 

$M$ is the size of the output data type, e.g. $M = 32$ if
$O$ is a 32-bit integer, and $i$ is the depth of the recursive call.
The parameter $pc$ is a partial assignment of $V_I$, it is
incrementally updated when the search progresses. In {\tt SymCount}, $\mathcal{T}$-solver($\varphi_P,pc$) means the $\mathcal{T}$-solver 
is called to check if there is a model of $\varphi_P$ where $pc$ is (assigned to) $\top$.

We illustrate the algorithm ${\tt SymCount}$ by running it on a simple example (we ignore temporarily lines 2 and 3 that will be clarified in section \ref{SEC:soundCompl}).
Consider again the case study of the data sanitization program in Figure \ref{sanitize}.
Only integer values from 8 to 23 are possible outputs of this program, which means the number of possible outputs is $N = 16$.

At the beginning, all variables are initialised in the procedure ${\tt SymbolicQIF}$ as in Figure \ref{symQIFframe},
the method {\tt EarlyPrunning} employs a heuristic that will be discussed later in this section. The method ${\tt SymCount}$ is then called to count the number of possible models of $\varphi_P$.

When a variable $p_i\in V_I $ is selected, we systematically explore in the same way for both $p_i$ and $\neg p_i$. Hence, the block of code from line \ref{algsqif:start1} to
line \ref{algsqif:end1}, and the one from line \ref{algsqif:start2} to line \ref{algsqif:end2} in Figure \ref{algsqif:symcount} are symmetric: we only explain the first one.

\begin{figure}
\begin{minipage}[b]{1\linewidth}
\centering
\includegraphics[width=0.8\textwidth]{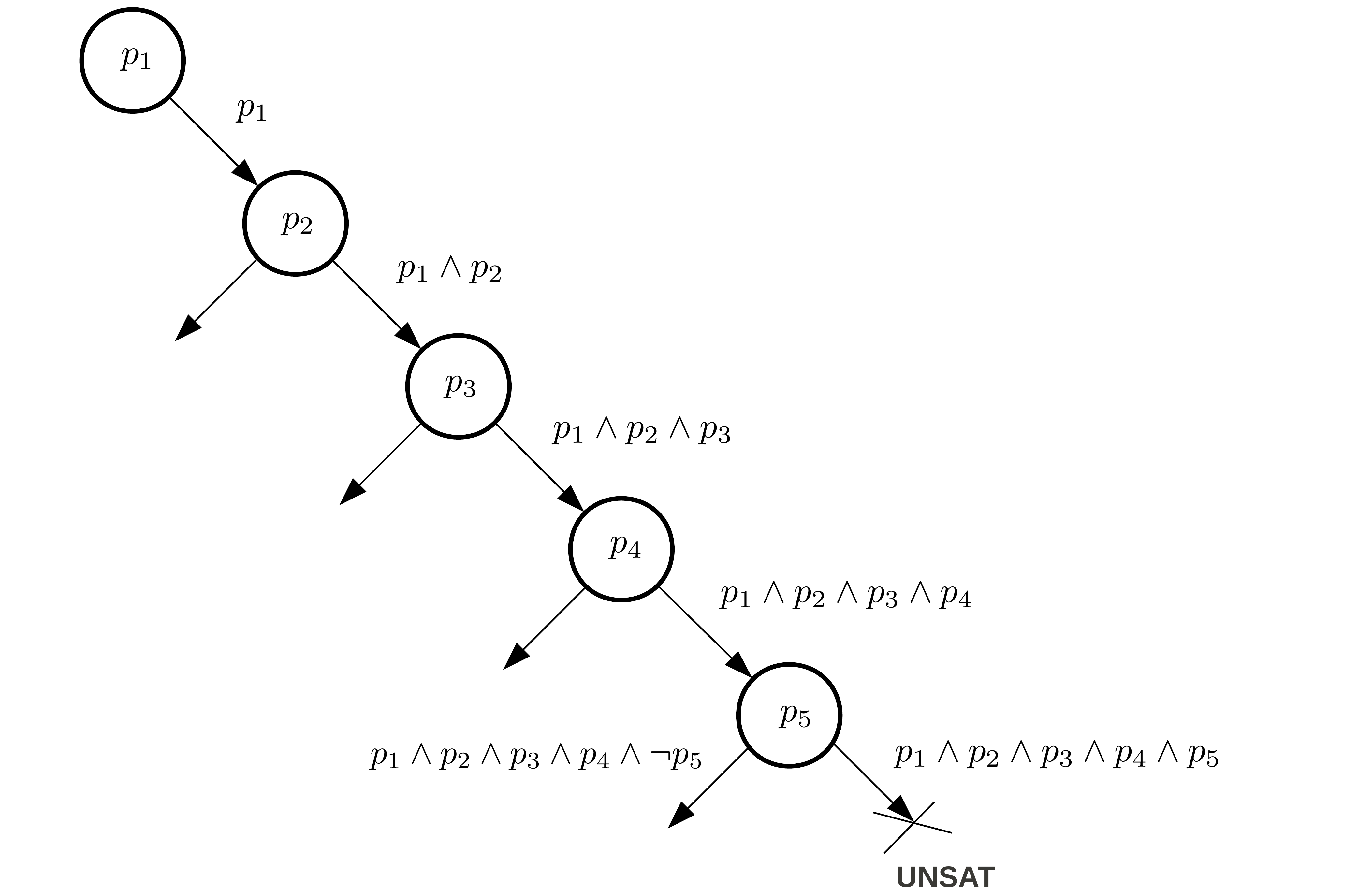}
\caption[Partial exploration path of SQIF]{Partial exploration path of SQIF for the 
program from Figure \ref{sanitize}.}
\label{trace}
\end{minipage}
\end{figure}

A partial run of ${\tt SymCount}$ on the illustrative example is depicted in Figure \ref{trace}.
At the first call of ${\tt SymCount}$: $i = 1$, the variable $p_1$ is in consideration and it is added
to   $pc$  in line \ref{algsqif:start1}. Since $pc$ is initialised to be empty, $pc_1 = p_1$. 
The $\mathcal{T}$-solver is called to check if there is a model of $\varphi_P$ where $p_1$ is (assigned to) $\top$.
This can be done
by using assertion to check the validity of $\neg p_1$ in a program as follows:
$$\text{\textbf{assert} } !p_1;$$

A model checking tool like JPF or CBMC can be used as a $\mathcal{T}$-solver to verify this assertion and it will return $\top$ if the assertion fails, and $False$ otherwise. In this example, the $\mathcal{T}$-solver would return $\top$ since ${p_1}$ stands for  ``first bit is 1'' and all odd values from 9 to 23 are possible outputs
satisfying the condition $p_1$. Hence, SQIF proceeds by calling
${\tt SymCount}$ with $i = 2$. Similarly, the procedure progresses until calling ${\tt SymCount}$ with $i = 5$, which means it needs to verify:
$$\text{\textbf{assert} } !(p_1 \text{ \&\& } p_2 \text{ \&\& } p_3 \text{ \&\& } p_4 \text{ \&\& } p_5);$$
This time the $\mathcal{T}$-solver would return $False$, since $p_1 \wedge p_2..\wedge p_5$ represents a set of outputs of which each element is at least $2^0 + 2^1 + ..+ 2^4 = 31$, while the possible range of $O$ is only
from 8 to 23. For a program with an output of 32-bits, by using {\tt EarlyPrunning}, SQIF trims a set of $2^{27}$ concrete values represented by the family of sets:
$$\{V_I := \{p_1,p_2,..,p_{32}\} : p_1 \wedge p_2 \wedge p_3 \wedge p_4 \wedge p_5\}$$
This is how the state-space explosion problem is mitigated.

At the depth $i = 5$ as above, if SQIF takes the path of $\neg p_5$ from line \ref{algsqif:start2}, then the $\mathcal{T}$-solver returns $\top$ ($O$ = 15 is one of the models). Hence, the procedure
continues with $i = 6$, and from this point until $i = 32$, only the path of $\neg p_i$ is SAT. At $i = 32$, SQIF finds a full path $00..01111$ which represents an output $O = 15$.
This path is added to $\varPsi$, and SQIF increases $N$. Finally, at the end of the method {\tt SymbolicQIF}, we have $\varPsi = \{8, 9,..,23\}$ and $N = 16$, thus we can conclude
that the data sanitization program in Figure \ref{sanitize} leaks at most 4 bits.

The method {\tt EarlyPrunning} implements the idea that if $p_1$ is unsatisfiable, then $p_1 \wedge C$ is also unsatisfiable for any $C$. Therefore, at the beginning of the
{\tt SymbolicQIF}, all $p_i$ are checked for satisfiability, and the results are stored for later use. We note that {\tt EarlyPrunning} speeds up {\tt SymbolicQIF} dramatically when
the number of possible models (outputs of the program) is small.

We have developed a prototyping tool for QIF analysis of C programs, \textbf{sqifc} built on top of CBMC.

\section{Soundness and Completeness}\label{SEC:soundCompl}
By soundness of the SQIF approach we mean that given  $\varPsi$, $\log_2(N)$ returned by {\tt SymbolicQIF}{($V_I,\varphi_P$)},
 each element of $\varPsi$ is a model of $\varphi_P$ i.e. corresponds
to a possible value of the output of the program $P$. By completeness of SQIF, we mean that $\varPsi$ is the set of all models of $\varphi_P$ i.e. all values of the output of $P$.

\begin{theorem}\label{soundness} Given a sound (resp. complete)  $\mathcal{T}$-solver
the SQIF approach is sound (resp. complete) i.e. {\tt SymCount} solves the QIF problem (Definition \ref{DEF:QIF}).
\end{theorem}
\begin{proofsketch}
The SQIF algorithm as described in Figure \ref{symcount} is based on DPLL which itself is a depth-first search procedure. As the search space is a binary tree with bounded depth $M$, the number
of bits of the output, the depth-first search procedure is complete.
The soundness of SQIF is guaranteed by the soundness of the $\mathcal{T}$-solver, i.e. model checker.
\end{proofsketch}
In reality $\mathcal{T}$-solvers are only complete in particular domains. Moreover, even with sound and complete $\mathcal{T}$-solvers, a large leak requires an exponential number of calls to the  $\mathcal{T}$-solver and so in practice SQIF is complete
  only for programs with \emph{small} leaks. 
Since our tools are based on bounded model checker 
, we choose to analyse only bounded programs. Notice however that Theorem \ref{soundness} holds for general $\mathcal{T}$-solvers.

Because of these practical issues about completeness, it has been
 proposed to shift the focus from the question \emph{``How much does it leak?"} to the simpler quantitative question \emph{``Does it leak more than k?"}~\cite{Heusser:2010:QIL:1920261.1920300,DBLP:journals/jcs/YasuokaT11}.
This approach not only makes the problem easier to analysis, but it is also more intuitive in term of security, because the user policy, i.e. \emph{threshold k}, is encoded in the analysis.
The ultimate goal of security analysis is to determine whether a program is \emph{secure} or \emph{insecure}. As discussed in the previous section, the goal of QIF is to relax security policy
from \emph{non-interference} to an acceptable threshold \emph{k bits of interference}, so that we can tolerate ``\emph{small}" leak, and accept more programs as secure.
The SQIF approach can also be used in the same way: with a user policy $k$, if SQIF finds out more than $k = 2^k$ possible outputs, we can stop the procedure and conclude that the
program is \emph{insecure}. This is the meaning of lines \ref{algsqif:policyStart} and \ref{algsqif:policyEnd} of function ${\tt SymCount}$ in Figure \ref{algsqif:symcount}.

A straightforward consequence of the Theorem \ref{soundness} is that, assuming a sound $\mathcal{T}$-solver,  given a user policy $k$, {\tt SymCount} never returns \emph{secure} for a program leaking more than $k$ bits.
This can be formally expressed as:
\begin{corollary}
SQIF is sound w.r.t a user policy $k$.
\end{corollary}

\section{Evaluation}
\label{case}
\lstset{language=C,numbers=left,frame=single,
basicstyle=\footnotesize \ttfamily
}
Only few papers present QIF static code analysis of real-world applications: examples are~\cite{Heusser:2010:QIL:1920261.1920300},~\cite{Kopf:2012:AQC:2362216.2362268} and the more 
recent~\cite{kmm:qest:2013}. Of these three approaches,~\cite{Kopf:2012:AQC:2362216.2362268} uses a different attacker model, namely cache side-channels and so is not directly 
comparable with our approach. The other two,~\cite{Heusser:2010:QIL:1920261.1920300} and~\cite{kmm:qest:2013}, use the same attacker model as we do but are at the moment both 
restricted to C programs, and hence only comparable to \textbf{sqifc}. We will concentrate on~\cite{Heusser:2010:QIL:1920261.1920300}, to which we refer as \textbf{selfcomp}, because it is based on the well-known 
concept of self-composition~\cite{Barthe:2004:SIF:1009380.1009669}.
For the analysis of anonymity protocols, we compare \textbf{sqifc} against \textbf{QUAIL}~\cite{quail,Biondi:2013:CAV}, a state-of-the-art quantitative analyser for probabilistic programs.
The case studies broadly fall  in three categories:
\begin{itemize}
 \item the first category, consisting of case studies from the National Vulnerability Database of the US government \cite{nvd}, is aimed to demonstrate how our analysis is able to deal with complex
C-code,
\item the CRC case study shows the applicability to quantify leakage in applications which leak by design,
\item the case studies Grade and Dining cryptos protocols show how our technique, even if it is unable to analyse probabilistic programs, is able to computing channel capacity for anonymity protocols. 
\end{itemize}

The experiments are conducted on a desktop machine with Intel Core i5 3.3GHz and 8GB of memory.
\subsection{CVE-2011-2208}
This case study is an example of a program that leaks information when the attacker can control the public input. It is taken from the National Vulnerability Database (NVD) of the US government~\cite{nvd}, and it is released on 13/06/2012.

\begin{figure}[htp]
\centering
\begin{minipage}[b]{0.8\linewidth}
\begin{lstlisting}[language=C]
int osf_getdomainname(char __user *name, int namelen)
{
  unsigned len;
  int i, error;
 
  error = verify_area(VERIFY_WRITE, name, namelen);
  if (error)
    goto out;
 
  len = namelen;
  if (namelen > 32)
    len = 32;
 
  down_read(&uts_sem);
  for (i = 0; i < len; ++i) {
    __put_user(system_utsname.domainname[i], 
        name + i);
    if (system_utsname.domainname[i] == '\0')
      break;
  }
  up_read(&uts_sem);
 out:
   return error;
}
\end{lstlisting}
\caption{arch/alpha/kernel/osf\_sys.c}
\label{cve1}
\end{minipage}
\end{figure}

The system call \texttt{osf\_getdomainname}, depicted in Figure \ref{cve1}, 
in the Linux kernel before 2.6.39.4 leaks sensitive information from kernel memory.
This is caused by an integer signedness error: the signed parameter \texttt{namelen} is assigned to the unsigned variable \texttt{len} in line 10,
so a negative value can be transformed into a big positive one.  Therefore, although the condition in line 11 restricts \texttt{namelen} to 32, the number of
characters  returned to the user via the structure \texttt{name} may be much greater including bytes from kernel memory.

In order to quantify the information leakage caused by this vulnerability, we chose the thresholds of security policy $k =64$ and $k = 256$, which means the
program is secure if it leaks less than 6 and 8 bits respectively.
After the times in Figure \ref{chap3:experiment} , \textbf{sqifc} and \textbf{selfcomp} conclude that the program is insecure. We then apply the patch provided for this vulnerability, and
run \textbf{sqifc} again. This time, \textbf{sqifc} found only one possible value for \texttt{name}, which means a leak of zero bit. Hence, we
prove that the patch fixed the leak.

\subsection{CVE-2011-1078}
This case study is also taken from NVD, and it is released on 21/06/2012. The function \texttt{sco\_sock\_getsockopt\_old}  
in the Linux kernel before 2.6.39,
depicted in Figure \ref{cve2}, 
leaks sensitive information from kernel memory.

\begin{figure}[htp]
\centering
\begin{minipage}[b]{0.82\linewidth}
\begin{lstlisting}[language=C]
static int sco_sock_getsockopt_old(
    struct socket *sock, int optname,
    char __user *optval, int __user *optlen)
{
  struct sock *sk = sock->sk;
  struct sco_conninfo cinfo;
  int len, err = 0;
  ...

  lock_sock(sk);

  switch (optname) {
    case SCO_OPTIONS:
      ...

    case SCO_CONNINFO:
      ...

      cinfo.hci_handle = sco_pi(sk)->conn->hcon->handle;
      memcpy(cinfo.dev_class, 
        sco_pi(sk)->conn->hcon->dev_class, 3);

      len = min_t(unsigned int, len, sizeof(cinfo));
      if (copy_to_user(optval, (char *)&cinfo, len))
        err = -EFAULT;
	break;
	...
      }

      release_sock(sk);
      return err;
}
\end{lstlisting}
\caption{net/bluetooth/sco.c}
\label{cve2}
\end{minipage}
\end{figure}

As in line 24, \texttt{cinfo} is copied to the user. Although its total size is 5 bytes, and all bytes are correctly assigned, when compiled it includes an additional padding byte for
alignment purposes. This padding byte is not zeroed out, and hence it contains kernel memory, and is leaked to the user.
Results of the analysis for  $k = 8$ and $k = 64$, are shown in Figure \ref{chap3:experiment}.
\lstset{stepnumber=0
}
\subsection{Cyclic Redundancy Check}
The program in Figure \ref{crc} performs Cyclic Redundancy Check\footnote{\href{http://en.wikipedia.org/wiki/Cyclic\_redundancy\_check}{http://en.wikipedia.org/wiki/Cyclic\_redundancy\_check}} (CRC) and shifts right the result \texttt{sft} bits.
We also have a Java version of the program to test with \textbf{jpf-qif}.

\begin{figure}[htp]
\centering
\begin{minipage}[b]{0.73\linewidth}
\centering
\begin{lstlisting}
unsigned char GetCRC8( 
	unsigned char check , unsigned char ch)
{   
  int i, sft ;
  for ( i = 0 ; i < 8 ; i++ ) {
    if ( check & 0x80 ) {
      check<<=1;
      if ( ch & 0x80 ) { check = check | 0x01;} 
      else { check =check & 0xfe; }
      check = check ^ 0x85;
    } else {
      check <<=1;
      if ( ch & 0x80 ) { check = check | 0x01; }
      else { check = check & 0xfe; }
    }
    ch<<=1;
  }
  check >>= sft;
  return check;
}

\end{lstlisting}
\caption{Cyclic Redundancy Check}
\label{crc}
\end{minipage}
\end{figure}

We quantify the amount of information of the confidential input \texttt{ch} revealed by observing the output of function
\texttt{GetCRC8}. We analyse this program with \textbf{sqifc} and \textbf{selfcomp} for \texttt{sft} values of 3 and 5 giving a maximum leakage for this program of 5 (\textbf{selfcomp} times out on this case) and 3 bits respectively which is consistent with the design of the program.
Results of the analysis are shown in Figure \ref{chap3:experiment}. In the case value of \texttt{sft} is 5, i.e. $k = 8$, \textbf{selfcomp} is faster as the state-space
is still small enough, and \textbf{selfcomp} requires only one call to CBMC. When \texttt{sft} = 3, i.e. $k = 32$, the state-space explosion makes 
\textbf{selfcomp} fail to solve. SQIF requires several calls to the solver, but it is less vulnerable to state-space explosion.

\begin{figure*}[htp]
\centering
\begin{tabular}{|>{\centering}p{3.7cm}<{\centering}|>{\centering}p{1.7cm}<{\centering}|>{\centering}p{1.5cm}<{\centering}|>{\centering}p{2.5cm}<{\centering}|>{\centering}p{3cm}<{\centering}|}
\hline
\textbf{Case Study} & \textbf{Policy} & \textbf{LoC}  & \textbf{sqifc} time & \textbf{selfcomp} time\tabularnewline \hline
Data Sanitization & - &$ <10$  & 11.898 &  timed out\tabularnewline \hline
CVE-2011-2208 &64 & $> 200$  & 22.759 &  119.117 \tabularnewline \hline
CVE-2011-2208 &256 &    & 88.196 &  timed out \tabularnewline \hline
CVE-2011-1078 &8& $>200   $ & 10.380 &  13.853\tabularnewline \hline
CVE-2011-1078 &64&   & 37.899 &  timed out\tabularnewline \hline 
CRC  &8&$ <30 $ &  1.209 &  0.498\tabularnewline \hline
CRC &32 &$ $ &  8.657  & timed out \tabularnewline \hline
\end{tabular}
\caption[Comparing sqifc against selfcomp]{Comparing sqifc against selfcomp. Times are in seconds, timeout is 30 minutes. In the first case study, ``-'' means the policy is not specified.}
\label{chap3:experiment}
\end{figure*}
\lstset{stepnumber=1
}

\subsection{The Grade Protocol}
This case study was used to illustrate protocol analysis in~\cite{Kiefer:2012:AAO:2362216.2362281,Biondi:2013:CAV}.
This anonymity protocol is designed to enable
a group of students to compute the sum of their grades (e.g., to compute the average) without revealing individual grades.
We denote $S_1 , ..., S_k$ be the $k$ students arranged in a ring, each one is given
a secret grade $g_i$ between 0 and $m-1$. To compute the sum of $g_i$ without disclosing them, the students produce $k$ random numbers between 0 and $n =
(m - 1) ∗ k + 1$ such that the number $r_i$ is known only to the students $S_i$ and $S_{(i+1)\%k}$. Each student $s_i$ then outputs a number $d_i = g_i + r_i - r_{(i+1)\%k}$
and the sum of all grades is equivalent to the sum of the outputs modulo $n$.
\begin{figure}[htp]
\centering
\begin{minipage}[b]{0.75\linewidth}
\begin{lstlisting}[language=C]
int func(){
  size_t S = 5, G = 5, i = 0, j = 0;
  size_t n = ((G-1)*S)+1, sum = 0;
  size_t numbers[S], announcements[S], h[S];

  for (i = 0; i < S; i++) h[i] = nondet_int() % G;

  for (i = 0; i < S; i++) 
	numbers[i] = nondet_int() % n;

  while (i<S) {
    j=0;
    while (j<G) {
      if (h[i]==j)
        announcements[i] = 
          (j+numbers[i]-numbers[(i+1)%S])%n;
      j=j+1;
    }
    i=i+1;
  }

  for (i = 0; i < S; i++)	
    sum += announcements[i];

  return sum % n;
}
\end{lstlisting}
\caption{The Grade protocol}
\label{gradec}
\end{minipage}
\end{figure}

This protocol is implemented as a probabilistic program in both~\cite{Kiefer:2012:AAO:2362216.2362281} and~\cite{Biondi:2013:CAV}. Here we implement it
in standard ANSI C with the built-in non-deterministic functions of CBMC. The source code, shown on Figure \ref{gradec},
is based on the one provided in~\cite{Biondi:2013:CAV}. The array \texttt{h[S]} stores the grades of all students, i.e. the secret. The attacker can observe \texttt{sum \% n}.

\begin{figure}[htp]
\centering
\begin{tabular}{|c|c|c|c|c|c|c|c|c|c|}
\cline{1-10}
\multicolumn{2}{|l|}{\textbf{Tool}} & \multicolumn{4}{c|}{\textbf{QUAIL}} & \multicolumn{4}{c|}{\textbf{sqifc}} \\ \cline{1-10}
\multicolumn{2}{|l|}{\textbf{Students}}  & \textbf{2} & \textbf{3} & \textbf{4} & \textbf{5} & \textbf{2} & \textbf{3} & \textbf{4} & \textbf{5} \\ \cline{1-10}
\multicolumn{1}{|c|}{\multirow{4}{*}{\begin{sideways}\textbf{Grades}\end{sideways}} } &
\multicolumn{1}{c|}{\textbf{2}} & 1.500 & 1.811 & 2.030 & 2.198 & 1.585 & 2.000 & 2.322 & 2.585\\ \cline{2-10}
\multicolumn{1}{|c|}{}                        &
\multicolumn{1}{c|}{\textbf{3}} & 2.197 & 2.525 & 2.745 & 2.910 & 2.322 & 2.807 & 3.170 & 3.459 \\ \cline{2-10}
\multicolumn{1}{|c|}{} &
\multicolumn{1}{c|}{\textbf{4}} & 2.655 & 2.984 & 3.201 & 3.365 & 2.807 & 3.322 & 3.700 & 4.000 \\ \cline{2-10}
\multicolumn{1}{|c|}{}                        &
\multicolumn{1}{c|}{\textbf{5}} & 2.999 & 3.325 & 3.541 & timed out & 3.170 & 3.700 & 4.087 & 4.392 \\
\hline
\end{tabular}
\caption[Leakage measured by QUAIL and sqifc]{Leakage measured by \textbf{QUAIL} and \textbf{sqifc}}
\label{sqif-grade}
\end{figure}

To compare \textbf{QUAIL} with our tool, \textbf{sqifc}, we repeat the experiment of the authors for the grade protocol with the tool and examples provided in~\cite{quail}.
However, \textbf{QUAIL} timed out after 1 hours for most of the cases, as showed in Figure \ref{sqif-grade-time} (we had the same results of leakage with the authors in the cases the tool did not time out).
Therefore, we take the result in Figure \ref{sqif-grade} directly from the paper~\cite{Biondi:2013:CAV}. Comparing the results in Figure \ref{sqif-grade},
it is easy to realise that the bounds on the leaks, measured by \textbf{sqifc}, do not exceed the real leaks, measured by \textbf{QUAIL}, by more than 1 bit,
while \textbf{sqifc} required no more than 1 minutes in all cases as showed in Figure \ref{sqif-grade-time}.

\begin{figure}[htp]
\centering
\begin{tabular}{|c|c|c|c|c|c|c|c|c|c|}
\cline{1-10}
\multicolumn{2}{|l|}{\textbf{Tool}} & \multicolumn{4}{c|}{\textbf{QUAIL}} & \multicolumn{4}{c|}{\textbf{sqifc}} \\ \cline{1-10}
\multicolumn{2}{|l|}{\textbf{Students}}  & \textbf{2} & \textbf{3} & \textbf{4} & \textbf{5} & \textbf{2} & \textbf{3} & \textbf{4} & \textbf{5} \\ \cline{1-10}
\multicolumn{1}{|c|}{\multirow{4}{*}{\begin{sideways}\textbf{Grades}\end{sideways}} } &
\multicolumn{1}{c|}{\textbf{2}} & 1.306   & 241.483 & - & - & 5.657 & 7.029 & 10.767 & 9.469\\ \cline{2-10}
\multicolumn{1}{|c|}{}                        &
\multicolumn{1}{c|}{\textbf{3}} & 28.613  & -       & - & - & 9.145 & 11.597 & 17.987 & 20.930 \\ \cline{2-10}
\multicolumn{1}{|c|}{} &
\multicolumn{1}{c|}{\textbf{4}} & 508.313 & -       & - & - & 10.095 & 16.872 & 21.869 & 18.579 \\ \cline{2-10}
\multicolumn{1}{|c|}{}                        &
\multicolumn{1}{c|}{\textbf{5}} & -       & -       & - & - & 14.639 & 20.666 & 33.298 & 40.399 \\
\hline
\end{tabular}
\caption[Elapsed time  in seconds of QUAIL and sqifc]{Elapsed time  in seconds of \textbf{QUAIL} and \textbf{sqifc}}
\label{sqif-grade-time}
\end{figure}

\subsection{The Dining cryptos protocol}
This case study is a variation of the dining cryptographers protocol of Chaum~\cite{Chaum:1988:DCP:54235.54239}, one of the most popular problem in anonymity protocol.
There is a group of cryptographers gathering around a table for dinner. After the meal, they are informed that the bill has been paid by someone, who could be one of them or the
National Security Agency (NSA). Even though the cryptographers respect each other's right to make an anonymous payment, they want to find out whether the NSA paid.
To determine this, they use a protocol as follows: each pair of adjacent cryptographers toss a coin hidden from everybody else, so that each cryptographer only knows the values 
of the coin shared with the one on his left and with the one on his right; then each cryptographer declares aloud the exclusive OR of the two coins he sees, i.e. 0 if they have 
the same value and 1 otherwise. However if the payer is one of the cryptographers,
he declares the opposite. In the end, if the sum of all declared values is even, then it is concluded that the NSA paid the bill. On the other hand, the sum is odd means one of the cryptographers did it.

\begin{figure}[htp]
\centering
\begin{minipage}[b]{0.6\linewidth}
\begin{lstlisting}[language=C]
size_t func(){

  size_t N = 5, output = 0, i = 0;
  size_t coin[N], obscoin[2], decl[N];
  size_t h;

  h = nondet_uchar() % (N+1);
 
  for (i = 0; i < N; i++){
    coin[i] = nondet_uchar() % 2;
  }

  for (i = 0; i < N; i++){
    decl[i] = coin[i] ^ coin[(i+1)%N];
    if (h==i+1){ 
      decl[i] = !decl[i];
    }
    i = i+1;
  }

  for (i = 0; i < N; i++){
    output = output + decl[c];
  }

  return output;
}
\end{lstlisting}
\caption{The Dining cryptos protocol}
\label{dining}
\end{minipage}
\end{figure}

We are interested in knowing how much information about the payer can be leaked by the sum of all declared values (in the dining cryptographers the observation are the declared values instead).
The input code for the protocol is depicted in Figure \ref{dining}. \texttt{h} is the identity of the payer, i.e. the secret, \texttt{output} is the observable.
The coin toss is modelled by a built-in non-deterministic function in line 10.
This model is less precise than implementation in probabilistic programs where it is possible to select random values from a specific distribution. By modelling with non-deterministic
function and computing channel capacity, we can only compute the maximum leakage in all possible distributions.
Figure~\ref{sqif-dining} shows the channel capacity computed by \textbf{sqifc}, and the time to compute them. 

\begin{figure*}[htp]
\centering
\begin{minipage}[b]{0.9\linewidth}
\begin{tabular}{| l | c | c | c | c | c | c | c | c   }
  \hline 
  \textbf{Cryptos}          & 3     & 4      & 5     & 6        & 100        &   300 \\ \hline
  \textbf{Channel capacity} & 2     & 2.32   & 2.59  & 2.81     & 6.658      &  8.234 \\ \hline 
  \textbf{Time in seconds}  & 2.145 & 3.496  & 3.632 &  18.634  & 158.517    & 3326.915 \\ \hline
\end{tabular}
\caption[The dining cryptos protocol analysed by sqifc]{The dining cryptos protocol analysed by \textbf{sqifc}}
\label{sqif-dining}
\end{minipage}
\end{figure*}

\section{Discussion of related work}
\label{related}
Meng and Smith introduce an \emph{approximate} technique to calculate an upper bound on channel capacity in~\cite{Meng:2011:CBI:2166956.2166957}.
The authors' implementation of the method is largely manual, and we proposed an automation for it in~\cite{Phan:2012:SQI:2382756.2382791}.
While the work of Meng and Smith is very inspiring, the technique can be very imprecise, for example when the leaks are sparse in the state space. Moreover, the user policy is not encoded in the analysis which
makes it infeasible when the leaks are not small. Take an example of a program that leaks all 32 bits of integral confidential data, it needs
to make 64 calls to STP solver to determine that all bits are \emph{Non-fixed}. Then, in order to determine two bit patterns of (31*32)/2 = 496 pairs
of \emph{Non-fixed} bits, it needs to make another 496 * 4 = 1984 calls to STP solver, so it is 2048 calls in total.

The first automated method for QIF was proposed by Backes et al.~\cite{Backes:2009:ADQ:1607723.1608130}. The method can be divided into two stages: first,
it employs model checking to compute an equivalence relation $\mathcal{R}$ on the set of confidential inputs w.r.t. observable outputs; secondly, if this relation $\mathcal{R}$ can
be represented by a system of \emph{linear integer inequalities} $A\bar{x} \geqslant \bar{b}$, which means it is a bounded integer \emph{polytopes}, then a variant of Barvinok's algorithm~\cite{Barvinok:1994:PTA:187096.187093}
can be used to count the number of integer solutions of $\mathcal{R}$.
While this work is important as the first effort on automation of QIF analysis, it is not clear however how this approach can be applied to real-world programs because of, for example, bit-wise operators in the CRC case study or non-linear relations and so on.

Closer to our work is the paper of \textbf{selfcomp}~\cite{Heusser:2010:QIL:1920261.1920300} discussed in the previous section. However, as already outlined their approach to address the question \emph{"Does it leak more than k?"} is
quite different from ours.
K\"{o}pf et al.~\cite{Kopf:2012:AQC:2362216.2362268} also apply QIF to real-world applications, i.e. leakage of cache side-channels; their technique is  based on abstract interpretation and hence
not based on bounded models. Because of this however they over-approximate channel capacity.

A preliminary version of this chapter has been presented first in a workshop~\cite{Phan:2012:SQI:2382756.2382791}, and then in a conference~\cite{Phan:2014:AMC}. 
However our definition for \#SMT was a little bit different from the one in this chapter: we required that each of the Boolean variables in the set $V_I$ is 
a Boolean abstraction of some $\mathcal{T}$-atom, hence we named the problem \emph{Propositional Abstract Model Counting}. This requirement makes the
definition more complicated and less general.

A recent paper~\cite{kmm:qest:2013} explores QIF in a pure logical framework. 
The approach is powerful and elegant, however it is more limited when compared to our approach as it relies on the solver to generate models whereas our approach can use any solver instead. For example we can analyse Java by using JPF as a solver for bytecode even if JPF doesn't generate a model in the sense of~\cite{kmm:qest:2013}.

McCamant and Ernst released FlowCheck~\cite{McCamant:2008:QIF:1375581.1375606}, a tool for security testing based on dynamic taint analysis. What FlowCheck
measures is the number of tainted bits, not an information-theoretic bound, so it is significantly different from our approach.
Another tool is described in \cite{Newsome:2009:MCC:1554339.1554349}, it is able to analyse large programs using the notion of channel capacity in the context of
dynamic taint analysis, while our approach
is based on verification techniques. In this sense, our work comes with stronger theoretical guarantees.

\chapter[Symbolic Execution as DPLL Modulo Theories]{Symbolic Execution as DPLL Modulo Theories}
\label{chap:SymExDPLL}
\newcommand{\theo}{\mathcal{T}}
\graphicspath{{chapter4/figs/}}
The previous chapter has introduced the Symbolic Quantitative Information Flow (SQIF) approach, which mimics the DPLL($\mathcal{T}$) algorithm. SQIF
was implemented on top of the Bounded Model Checker CBMC, used as a sub-routine, and can analyse programs written in C/C++. This chapter presents an alternative implementation for
the SQIF approach, using Symbolic Execution.

The implementation is based on a key observation that Symbolic Execution can be viewed as a variant of the DPLL($\mathcal{T}$) algorithm, 
or in other words, Symbolic Executors are SMT solvers. 

This view enables us to modify Symbolic PathFinder~\cite{Pasareanu:2013:ASE}, a Symbolic Executor for Java bytecode, into a QIF analysis tool, \textbf{jpf-qif}, with little effort.
The work in this chapter is the first to use Symbolic Execution for QIF analysis, and jpf-qif is the first QIF analysis tool for Java.

\section{Introduction}
Symbolic Execution (SE) \cite{King:1976:SEP:360248.360252} is now popular. It is increasingly used not only in academic 
settings but also in industry, such as in Microsoft, NASA, IBM and Fujitsu \cite{Cadar:2013:SES:2408776.2408795}. 
In the success of SE, the efficiency of SMT solvers \cite{DeMoura:2011:SMT:1995376.1995394} is a key factor. In fact, while SE was introduced more than three decades ago, 
it had not been made practical until research in SMT made significant advances \cite{Cadar:2013:SES:2408776.2408795}.

Recall that most state-of-the-art SMT solvers, e.g. \cite{DeMoura:2008:ZES:1792734.1792766,cav2008}, implement the DPLL($\theo$) algorithm \cite{Nieuwenhuis:2006:SSS:1217856.1217859}
which is an integration of two components as the following. The first component is a DPLL-based SAT solver,
to search on the \emph{Boolean skeleton} of the formula. The second component is a $\theo$-solver to check the consistency w.r.t. the theory $\theo$
of conjunctions of literals.
The path conditions generated by a Symbolic Executor, e.g. Symbolic PathFinder (SPF) \cite{Pasareanu:2010:SPS:1858996.1859035}, are also conjunctions of literals.
Therefore, when an SMT solver checks such a path condition, only the  $\mathcal{T}$-solver works on it, and the SAT component is not 
used\footnote{This claim is not for the decision procedure STP \cite{Ganesh:2007:DPB:1770351.1770421}, which converts Bit Vector formulas into 
propositional formulas and solves them with a SAT solver.}.

On the other hand, a \emph{classical} Symbolic Executor \cite{King:1976:SEP:360248.360252} can also be divided into two components. The first component, 
called \emph{Boolean Executor} hereafter, executes the instructions, and updates the path condition. The second component is a $\mathcal{T}$-solver (since the SAT solver is not used) to validate the consistency of the path condition. This thesis 
shows that a Boolean Executor does the same work as DPLL algorithm. Thus, SE is a variant of DPLL($\mathcal{T}$).
This view is important since it connects two communities and can give an insight for future research. 

\section{Illustration of DPLL(\texorpdfstring{$\mathcal{T}$}{})}
A complete formal description of the DPLL($\theo$) algorithm can be found in, e.g., \cite{Nieuwenhuis:2006:SSS:1217856.1217859}. Here we briefly 
recall some background via a running example as follows. 
\begin{align}
\varphi&:= 
  \!\begin{aligned}[t]
          &(\neg(x_0 > 5) \vee T_1) \wedge ((x_0 > 5)\vee T_2) \wedge (\neg(x_0 > 5) \vee (x_1 = x_0 + 1)) \wedge{}\\
          &(\neg(x_1 < 3) \vee T_3)  \wedge (\neg(x_1 < 3) \vee (x_2 = x_1 -1)) \wedge{} \\
          &((x_1<3) \vee T_4) \wedge ((x_1<3) \vee (y_1 = x_1 + 1))
  \end{aligned} \label{eq:ex}
\end{align}
$\varphi$ is a Linear Arithmetic formula. Boolean variables, $T_1\dots T_4$, are called Boolean atoms, and atomic formulas, e.g. $(x_0>5)$, are called theory
atoms or $\theo$-atoms. Any first-order formula $\varphi$ can be abstracted into a Boolean skeleton by replacing each $\theo$-atom in $\varphi$ with its 
Boolean abstraction. For the example above, we define new Boolean variables 
$G_1, G_2, A_1, A_2, A_3$ for the Boolean abstraction of $\theo$-atoms, and the abstraction can be expressed as:
\begin{align}
\mathcal{BA}&:= 
  \!\begin{aligned}[t]
              &G_1 = (x_0 > 5) \wedge G_2 = (x_1 < 3) \wedge{} \\
              &A_1 = (x_1 = x_0 + 1) \wedge A_2 = (x_2 = x_1 -1) \wedge A_3 = (y_1 = x_1 + 1) 
  \end{aligned} \label{eq:ba}
\end{align}
As the result, we obtain a formula $\varphi^P$ ($^P$ stands for propositional) as the Boolean skeleton of $\varphi$.
Obviously, $\varphi$ is logically equivalent to $\varphi^P \wedge \mathcal{BA}$.
\begin{align}
\varphi^P&:= 
  \!\begin{aligned}[t]
          &(\neg G_1 \vee T_1) \wedge (G_1 \vee T_2) \wedge (\neg G_1 \vee A_1)\wedge{}  \\
          &(\neg G_2 \vee T_3)  \wedge (\neg G_2 \vee A_2) \wedge{}\\
          &(G_2 \vee T_4) \wedge (G_2 \vee A_3)
  \end{aligned} \label{eq:exabs}
\end{align}

\begin{figure}[htp]
\centering
\fbox{
\begin{minipage}[b]{0.6\linewidth}
\begin{algorithmic}[0]
\Function{DPLL}{\texttt{BooleanFormula} $\varphi$}\{
\State $\mu$ = \textsc{True};
\texttt{status} = \texttt{propagate}($\varphi,\mu$);
\If {(\texttt{status} == \textsc{Sat})} \textbf{return} \textsc{Sat};
\ElsIf{(\texttt{status} == \textsc{UnSat})} \textbf{return} \textsc{UnSat};
\EndIf
\While {(\textsc{True})}\{
  \State $l$ = \texttt{decide}($\varphi$);
  \State $\mu = \mu \wedge l$;
  \State \texttt{status} = \texttt{propagate}($\varphi,\mu$);
  \If {(\texttt{status} == \textsc{Sat})} \textbf{return} \textsc{Sat};
  \ElsIf{(\texttt{status} == \textsc{UnSat})}
    \If {(\texttt{allStatesAreExplored}())}
      \State \Return \textsc{UnSat};
    \Else \texttt{ backtrack}($\varphi,\mu$);
    \EndIf  
  \EndIf 
\EndWhile
\EndFunction \} \} 
\end{algorithmic}
\end{minipage}
}
\caption{DPLL algorithm}
\label{DPLL}
\end{figure}

The DPLL($\theo$) algorithm is the integration of the DPLL algorithm with a $\theo$-solver. The DPLL algorithm searches on $\varphi^P$, returning a 
conjunction of Boolean literal $\mu^P$. Replacing all the new Boolean atoms, $G_i$ and $A_i$, in $\mu^P$
with their corresponding $\theo$-atoms, we obtain the conjunction $\mu$ in $\theo$.
The $\theo$-solver then checks whether $\mu$ is consistent with the theory $\theo$. Below is the illustration of DPLL($\theo$) on $\varphi$
(for the limit of space, only decision literals are shown in $\mu^P$):
\begin{align*}
&0\text{. } \mu^P = \texttt{True}      &\varphi^P &{} \\
&1\text{. } \mu^P = G_1                &\varphi^P  &= (\neg G_2 \vee T_3)  \wedge (\neg G_2 \vee A_2) \wedge (G_2 \vee T_4) \wedge (G_2 \vee A_3) \\
&2\text{. }\mu^P = G_1 \wedge G_2      &\varphi^P  &= \texttt{True} \text{ ; } \theo\text{-solver}(\mu) = \texttt{Inconsistent} \\
&3\text{. }\mu^P = G_1                 &\varphi^P  &= (\neg G_2 \vee T_3)  \wedge (\neg G_2 \vee A_2) \wedge (G_2 \vee T_4) \wedge (G_2 \vee A_3) \\
&4\text{. }\mu^P = G_1 \wedge \neg G_2 &\varphi^P  &= \texttt{True} \text{ ; } \theo\text{-solver}(\mu) = \texttt{Consistent}
\end{align*}
The DPLL algorithm tries to build a model using three main operations: \texttt{decide}, \texttt{propagate}, 
and \texttt{backtrack} \cite{DeMoura:2011:SMT:1995376.1995394}. The operation \texttt{decide} heuristically chooses a literal $l$ (which is an unassigned 
Boolean atom or its negation) for branching. The operation \texttt{propagate} then removes all the clauses containing $l$, and deletes all occurrences 
of $\neg l$ in the formula; this procedure is also called \emph{Boolean Constraint Propagation} (BCP). If after deleting a literal from a clause, the clause only
has only one literal left (\emph{unit clause}), BCP assigns this literal to \texttt{True}. If deleting a literal from a clause results in 
an empty clause, this is called a conflict. In this case, the DPLL procedure must \texttt{backtrack} and try a different branch value.

At step 1, $G_1$ is decided to be the branching literal, and the $\theo$-solver validates that $(x_0 > 5)$ is consistent. BCP removes the clause $(G_1 \vee T_2)$, 
and deletes all occurrences of $\neg G_1$. This results in two unit clauses $T_1$ and $A_1$, so they are assigned to \texttt{True},
which means $\mu^P = G_1 \wedge T_1 \wedge A_1$. Similarly, at step 2 $G_2$ is chosen, i.e. $\mu^P = G_1 \wedge T_1 \wedge A_1 \wedge G_2$. The $\theo$-solver 
checks the conjunction:
$\mu = (x_0 > 5)\wedge T_1 \wedge (x_1 < 3) \wedge (x_1 = x_0 + 1)$.
This is obviously inconsistent, thus DPLL($\theo$) backtracks and tries $\neg G_2$, which leads to a consistent model.

Note that DPLL($\theo$) refers to various procedures integrating DPLL and a $\theo$-solver. There are DPLL($\theo$) procedures with integration schemas different from what 
we have described here. The interested reader is pointed to \cite{jsat2007} for further references.
\section{Symbolic Execution as DPLL(\texorpdfstring{$\mathcal{T}$}{})}
\label{seDPLL}
Intuitively, a program can be encoded into a (first-order) formula whose models correspond to program traces. Symbolic Executors explore all program
traces w.r.t. the set of program conditions, therefore they can be viewed as SMT solvers that return all (partial) models w.r.t. 
a set of Boolean atoms. 

In this thesis we only consider bounded programs, since this is the class of programs that SE can analyse. This means every loop can be unwound into a sequence \texttt{if} 
statements. In order to encode a program into a formula, all program variables are renamed in the manner of Static Single Assignment form \cite{Cytron:1989:EMC:75277.75280}: each variable is assigned exactly 
once, and it is renamed into a new variable when being reassigned. In this way, assignments such as $x = x + 1$ will not be encoded into an unsatisfiable atomic formula.
Under these settings, a program \emph{P} can be modelled by a \emph{Symbolic Transition System} (STS) as follows:
$$
P \equiv (S, I, G, A , T)
$$
$S$ is the set of program states, $I \subseteq S$ is the set of initial states; each state in the STS models the computer memory at a program point. $G$ is 
the set of guards and $A$ is the set of actions; guards and actions are first-order formulas. 

An action models the effect of an instruction on the computer memory. Actions that do 
not update the computer memory (e.g. conditional jumps) are Boolean atoms, the others are $\theo$-atoms. $T \subseteq S \times G \times A \times S$ is the transition 
function, $t_{ij} = \langle s_i, g_{ij}, a_{ij}, s_j\rangle\in T$ models a transition from state $s_i$ to state $s_j$ by taking action $a_{ij}$ under 
the guard $g_{ij}$. 
After a transition $t_{ij}: s_i \to s_j$, the state $s_j$ is exactly as the state $s_i$ apart from the variable updated by the action $a_{ij}$.

Note that this STS models the program in more detail than the transition system in section \ref{sec:transit} that will be used in other chapters.

One way to encode a transition $t_{ij}$ into a first-order formula is to present it in the form: $t_{ij} \equiv g_{ij} \to a_{ij}$, or equally $t_{ij} \equiv \neg g_{ij} \vee a_{ij}$.
This encoding expresses that satisfying the guard $g_{ij}$ implies that the action $a_{ij}$ is performed. In this way, a program trace is defined as a sequence of transitions:
$$t_{01} \wedge t_{12} \wedge \dots \wedge t_{(k-1)k} = (\neg g_{01} \vee a_{01}) \wedge (\neg g_{12} \vee a_{12}) \dots \wedge (\neg g_{(k-1)k} \vee a_{(k-1)k}) $$
The semantics of the program is then defined as the set of all possible traces, or equally the set of all possible transitions, which can be represented 
as the following formula:
\begin{align}
\varphi = \bigwedge_{t_{ij} \in T} t_{ij} =  \bigwedge_{t_{ij} \in T} (\neg g_{ij} \vee a_{ij}) 
\label{equa:encode}
\end{align}

\begin{figure}[htp]
\begin{minipage}[b]{0.4\linewidth}
\begin{lstlisting}[language=C, frame=single,%xleftmargin=5.0ex,
basicstyle=\footnotesize\ttfamily,
numbers=left,stepnumber=0]
void test(int x, int y){
  if(x > 5){
    x++;
    if (x < 3) 
      x--;
    else 
      y = x + 1;
  }
}
\end{lstlisting}
\end{minipage}
\hspace{0.5cm}
\fbox{
\begin{minipage}[b]{0.45\linewidth}
\centering
\includegraphics[scale=0.45]{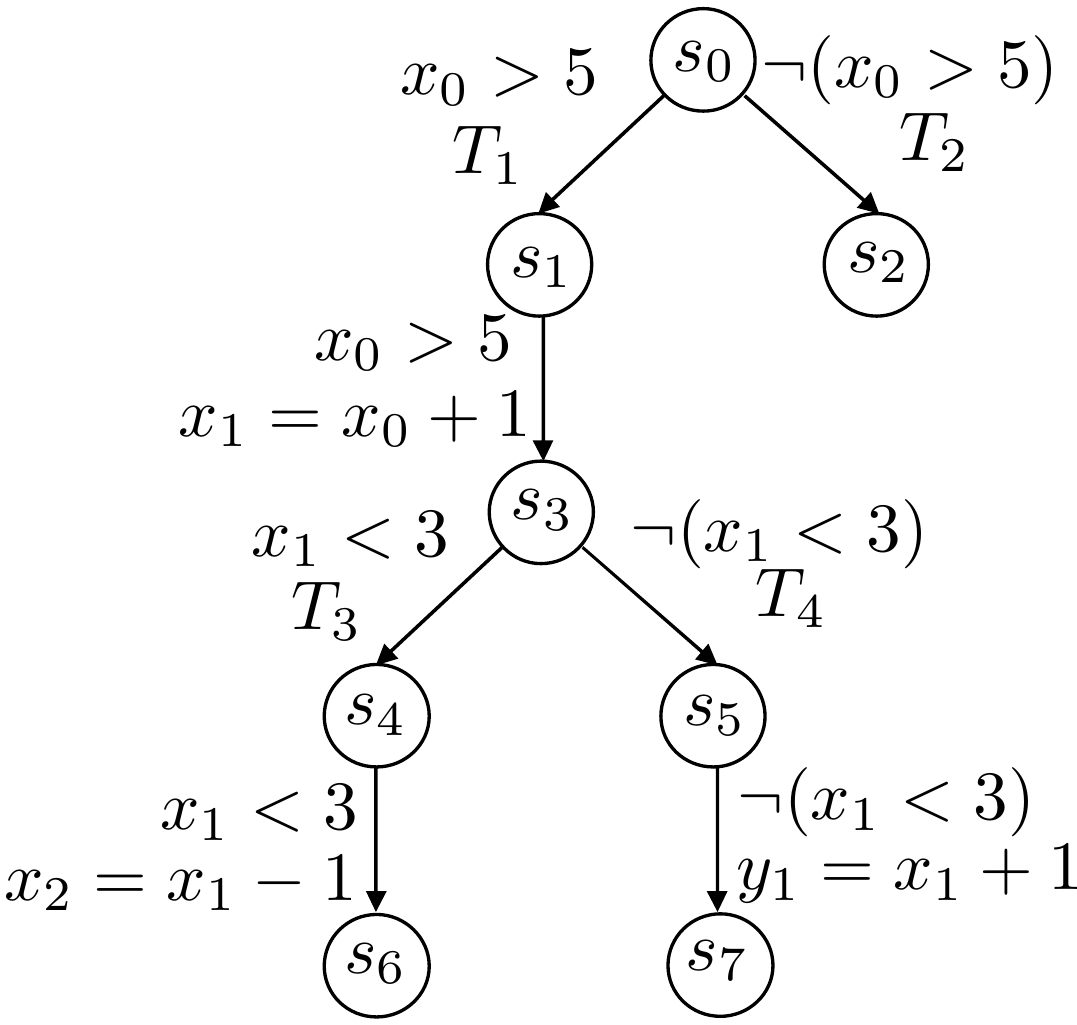}
\end{minipage}
}
\caption[A simple program and its associated STS]{A simple program and its associated STS. ``\texttt{if}$(x > 5)$'' is modelled by two transitions $\langle s_0, (x_0 > 5), T_1, s_1\rangle$
and $\langle s_0,\neg (x_0 > 5), T_2, s_2\rangle$; then ``\texttt{x++}'' is modelled by $\langle s_1, (x_0 > 5), x_1 = x_0 + 1, s_3\rangle$;
similarly for the rest of the program.}
\label{fig:ex}
\end{figure}

Figure \ref{fig:ex} depicts a simple example program and its associated STS. Encoding this STS following (\ref{equa:encode})
results in the formula (\ref{eq:ex}) that we have illustrated with DPLL($\theo$) in the previous section. We now illustrate this example with SE.

Similar to an SMT solver, a Symbolic Executor at a high level can be viewed as the integration of two components: a Boolean Executor (BE) to execute the instructions and a $\theo$-solver to check the feasibility of path conditions.
For example, SPF has a parameter \texttt{symbolic.dp} to customize which decision procedure to use. If we set this parameter with the option \texttt{no\_solver} 
then SPF solely works on the BE.

\begin{figure}[htp]
\centering
\fbox{
\begin{minipage}[b]{0.6\linewidth}
\begin{algorithmic}[0]
\Function{Executor}{\texttt{Program} $P$}\{
\State \texttt{PathCondition} pc = \textsc{True};
\State \texttt{InstructionPointer} i = \textsc{Null};
\State \texttt{update}($P,i$);
\If{($i$ == \textsc{Return})} \textbf{return};
\EndIf 
\While {(\textsc{True})}\{
  \State $l$ = \texttt{decide}($i$);
  \State $pc = pc \wedge l$;
      \State \texttt{update}($P,i$);
      \If{($i$ == \textsc{Return})}
	  \If {(\texttt{allStatesAreExplored}())}
	    \State \textbf{return};
	  \Else \texttt{ backtrack}($pc,i$);
	  \EndIf  
      \EndIf 
\EndWhile
\EndFunction \} \} 
\end{algorithmic}
\end{minipage}
}
\caption{A simplified Boolean Executor}
\label{SymEx}
\end{figure}

Figure \ref{SymEx} depicts a simplified procedure of a BE. This procedure can be described as trying to build \emph{all} path conditions using three main operations: \texttt{decide}, \texttt{update} and \texttt{backtrack}.
The operation \texttt{decide} chooses a literal $l$, a condition (or its negation) of an \texttt{if} statement, for branching, adding it to the
path condition. The operation \texttt{update}
then symbolically executes a block of statement, i.e. no branching statement presents, updating the computer memory. When the BE reaches the end of a symbolic path, it backtracks
to explore other paths. A Symbolic Executor, which is the integration of a BE and a $\theo$-solver, backtracks if the path condition is not satisfied.

Both DPLL and BE rely on Depth-First Search, they are similar in the way they \texttt{decide} and 
\texttt{backtrack}\footnote{We consider DPLL in its simplest form, without non-chronological backtracking.}. After choosing a literal, e.g. $(x_0 > 5)$, BE 
executes the block it guards, i.e. $T_1$ and $x_1 = x_0 + 1$. This is exactly the same as in DPLL: after choosing $g_{ij}$, for all the clauses 
$(\neg g_{ij} \vee a_{ij})$, BCP deletes $\neg g_{ij}$, assigning $a_{ij}$ to \texttt{True}.
Therefore, the operation \texttt{update} does the same work as BCP, we can view a BE as implementing the DPLL algorithm, and SE as DPLL($\theo$).

\section{SQIF by Symbolic Execution:}
\label{sec:sqif-se}
With the view of SE as \#DPLL($\mathcal{T}$), we are able to make a Symbolic Executor work as {\tt SymbolicQIF} with little effort.
The key idea here is to enumerate all concrete values from symbolic executions.

\begin{figure}[htp]
\begin{minipage}[b]{1\linewidth}
\centering
\includegraphics[width=0.6\textwidth]{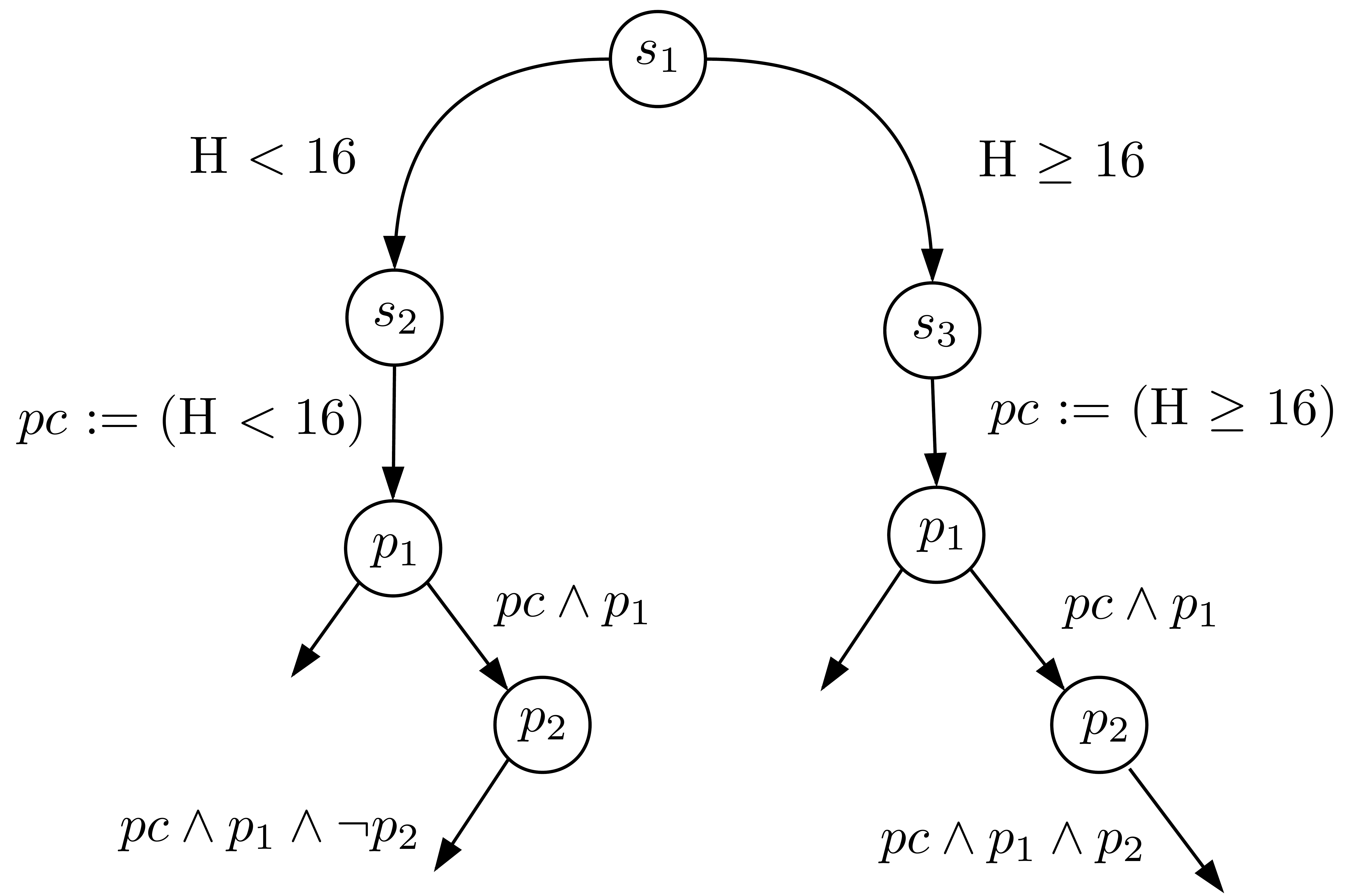}
\caption[Partial exploration path of SQIF-SE]{Partial exploration path of SQIF-SE for the data sanitization program from Figure \ref{sanitize}.}
\label{sqif-se}
\end{minipage}
\end{figure}

For a program $P$ that takes symbolic inputs $i_1, i_2,..i_{\alpha}$, and produces an output $O$, the result of running SE on $P$ is as
follows:
$$
  O = \left\{
  \begin{array}{l l}
    f_1(i_1,i_2..,i_{\alpha}) & \quad \text{if $pc_1$}\\
    f_2(i_1,i_2..,i_{\alpha}) & \quad \text{if $pc_2$}\\
    \dots & \quad \dots \\
    f_{\beta}(i_1,i_2..,i_{\alpha}) & \quad \text{if $pc_{\beta}$}
  \end{array} \right\}
$$
where $f_1,f_2,..f_{\beta}$ are formulas over symbolic inputs $i_1, i_2,..i_{\alpha}$. $pc_1,pc_2,..pc_{\beta}$ are the path conditions.
Notice $f_i$ expresses a symbolic final value for $O$, i.e. in terms of {\tt SymEx} instead of $f_i(i_1,i_2..,i_{\alpha}) $  we could write $\sigma_{i}(O)$ for $\sigma_i\in\varSigma $.
The following proposition was proved by King~\cite{King:1976:SEP:360248.360252}:
\begin{proposition}$$\forall i,j \in [1,{\beta}] \wedge i \neq j, \text{ } pc_i \wedge pc_j = \bot$$
which means that path conditions are mutually exclusive.
\end{proposition}
\begin{definition}
For a path condition $pc_i$ obtained from SE, the concretization set of $pc_i$, denoted $\mathcal{CS}(pc_i)$, is the set of all concrete values of output $O$
that can be reached by executing the program following $pc_i$.
\end{definition}
Consider again the illustrative example in Figure \ref{sanitize}, in which there are two path conditions: $pc_1 = H < 16$ and $pc_2 = H \geq 16$.
The corresponding concretization sets of these path conditions are: $\mathcal{CS}(pc_1) = [8..23]$ and $\mathcal{CS}(pc_2) = [24..2^{32}]$.
The set of all possible values of output $O$ is formed by the union of concretization sets of all paths, and thus:
$$N = \left| \bigcup_{i=1}^{{\beta}} \mathcal{CS}(pc_i)\right|$$
The set $\mathcal{CS}(pc_i)$ can be computed by inserting the code in Figure \ref{fig:instru}  at the end of the program and run SE: we add $M$ conditions, each one tests whether bit $b_i$ of the output $O$ is 0 or 1. These $M$ conditions test all the bits
of the output $O$. Exploring all possible combinations of these conditions leads to enumerating all possible values of $O$. We denote by SQIF-SE the implementation of SQIF using SE. A partial exploration path of SQIF-SE is described as in Figure \ref{sqif-se}.
SE as implemented by Symbolic PathFinder (SPF) returns a concrete values for each possible path. The number of distinct concrete values is the $N$
that we need to count.

SQIF-SE is implemented into a prototyping tool \textbf{jpf-qif} built on top of SPF. The tool works on Java programs.

\section{Soundness and Completeness}
SQIF-SE relies on a Symbolic Executor, and hence it is complete in programs with a bounded model
of runtime behaviour, which means programs have no recursion or unbounded loops. These are well-known issues in SE and handling
them is orthogonal to our work. SQIF-SE is also sound given a sound Symbolic Executor.

\section{Evaluation}
To evaluate \textbf{jpf-qif}, we compare it against our previous implementation \textbf{sqifc} and the \textbf{selfcomp} technique.
For case studies, we revisit the data sanitization program and the CRC programs in the previous chapter. Moreover, we consider the Tax program as follows.
\subsection{Tax Record}
Balliu et al. ~\cite{Balliu:2012:ESE:2354412.2355238} provided a very interesting case study of information flow security in Java programs  derived from the EU-funded FP7-project HATS.
The program contains 8 classes/interfaces and 267 LoC.
In our analysis we assume the year 2011-2012 basic tax rate in UK, which is applied for a person whose income does not exceed 35 thousand pounds per annum, is $F=20\%$\footnote{Since SPF can only handle conditions with integer values, we simplify the code by replacing $(income \times 20)/100$ with \emph{tax} as an integer.
The simplification we made does not change the secrecy, i.e. entropy, of \emph{income}, so it will not affect the result of our analysis.
}.
Thus, the tax is less than 7 thousands pound per annual. We assume that donations to charities is below the amount of tax to be paid. Obviously, one cannot pay more than what one earns.
Following~\cite{Balliu:2012:ESE:2354412.2355238} we are interested in leaks of a taxpayer's income and donations to a Tax checker.
\subsubsection{QIF vs. Declassification}
Balliu et al. considered two cases of declassification: the first one, called taxChecker1, is associated with the policy
``$income \times F \% + donation > payment$'', and the second one, called taxChecker2, is associated with the policy ``$income \times F \% + donation - payment$''.
They claimed that:
\emph{``The value declassified in the taxChecker1 case, resp. taxChecker2 case, is a lower bound, resp. upper bound, of the value revealed to the tax checker in the fixed tax rate variant.''}

We notice in this sentence the use of terms like  \emph{``value revealed''} and \emph{``bound''}: the authors
were trying to describe  \emph{quantitative} concepts. From the result of \textbf{jpf-qif}, we can give hence quantitative answers to these questions. In the case of \emph{taxChecker1}, the observable is
whether the payment is greater or smaller than the sum of the tax and donations, which means there are 2 possible outputs. This corresponds to the leak of 1 bit. In other words, the user policy or threshold $k = 1$.
Regarding the taxChecker2 case, under the assumptions listed above, the leakage is upper bounded by 4.86 bits obtained in 24.988 seconds.

\begin{figure*}[ht]
\centering
\begin{tabular}{|>{\centering}p{3.5cm}<{\centering}|>{\centering}p{0.9cm}<{\centering}|>{\centering}p{2cm}<{\centering}|>{\centering}p{2.5cm}<{\centering}|>{\centering}p{3cm}<{\centering}|}
\hline
\textbf{Case Study} & \textbf{LoC} &  \textbf{sqifc} time& \textbf{jpf-qif} time& \textbf{selfcomp} time\tabularnewline \hline
Data Sanitization &$ <10$   & 11.898 & 20.695 & timed out\tabularnewline \hline
CRC  (8)&$ <30 $  & 1.209 & 8.386 & 0.498\tabularnewline \hline
CRC (32) &$ $  & 8.657 & 9.357 & timed out \tabularnewline \hline
Tax Record & $267$   & - & 24.988 & - \tabularnewline \hline
\end{tabular}
\caption{Times in seconds, timeout is 30 minutes. ``-'' means inapplicable.}
\label{chap4:experiment}
\end{figure*}

Results of the analysis are shown in Figure \ref{chap4:experiment}. In general, 
\textbf{jpf-qif} is slower than \textbf{sqifc}, although they are different implementation of the same algorithm. It is not surprising, as Java is always
considered to be slower than C. Moreover, SPF is a virtual machine running on top of the Java Virtual Machine. Hence, there are more overheads
in the implementation using SPF.

\section{Discussion of related work}
The correspondence between Symbolic Execution and the DPLL($\mathcal{T}$) algorithm was first briefly mentioned in our previous work~\cite{Phan:2012:SQI:2382756.2382791}. In that paper, we described 
a preliminary version of the DPLL-based algorithm in the previous chapter (Figure~\ref{symcount}). However, at that time the tool sqifc is not yet 
available. Instead, we presented a quick implementation of the algorithm with SPF.

A year later, Brain et al. \cite{DBLP:conf/vmcai/BrainDHGK13} published an excellent paper showing the correspondence between DPLL($\mathcal{T}$) and Abstract Interpretation.
Although it is believed that Symbolic Execution is a case of Abstract Interpretation\footnote{\url{http://en.wikipedia.org/wiki/Symbolic_execution}}, we are not aware of any paper to discuss rigorously this relation.

\chapter[Concurrent Bounded Model Checking]{Concurrent Bounded Model Checking}
\label{chap:jpf-bmc}
\graphicspath{{chapter5/figs/}}
In the chapters \ref{chap:SQIF} and \ref{chap:SymExDPLL}, we have introduced two implementations of the Symbolic Quantitative Information Flow 
approach: the first one is \textbf{sqifc}, which employs the Bounded Model Checker CBMC; and the second one is \textbf{jpf-qif}, built on top of the
Symbolic PathFinder symbolic execution platform. A natural research question would be whether there is a relation between the two symbolic 
techniques: Bounded Model Checking and
Symbolic Execution.

This chapter studies this relation and introduces a methodology, based on Symbolic Execution, for Concurrent Bounded Model Checking.
In our approach, we translate a program into a formula in a disjunctive form, and this design enables concurrent verification: a main thread running 
a symbolic executor, without constraint solving, to build sub-formulas, and a set of worker threads running a decision procedure for satisfiability checks.

We have implemented this methodology in a tool  called JCBMC, the first bounded model checker for Java.
JCBMC is built as an extension of Java PathFinder, an open-source verification platform developed by NASA. 
JCBMC uses Symbolic PathFinder (SPF) for the symbolic execution, Z3 as the solver and implements concurrency with multi-threading.

For evaluation, we compare JCBMC against SPF and CBMC. The results of the experiments show that we can achieve significant advantages of performance over these two
state-of-the-art tools.

\section{Introduction}
Model checking techniques are often classified in two categories: explicit-state or symbolic, depending on how they process the states of the system.
While explicit-state model checking enumerates all possible states of the system explicitly, possibly on-the-fly \cite{jpf,spin}, symbolic model checking 
represents sets of states symbolically, and hence more efficiently, by using
Binary Decision Diagrams \cite{McMillan:1992:SMC:143233} or Boolean formulae \cite{Biere:1999:SMC:646483.691738}. SAT or SMT-based Bounded 
Model Checking (BMC)~\cite{Biere:1999:SMC:646483.691738} unwinds the transition relation of a program for a fixed number of steps $k$ and checks whether a 
property violation can occur in  $k$ or fewer steps. This bounded verification is reduced to a satisfiability check performed by a SAT or SMT solver. 
BMC is widely used in the hardware industry.

For software, the application of BMC for ANSI-C is embodied in  CBMC  \cite{Clarke:2003:BCC:775832.775928,ckl2004}, which has been 
successfully used for many practical applications. 
In CBMC a C program containing assertions is translated into a formula (in Static Single 
Assignment form) which is then fed to a SAT or SMT solver to check its satisfiability. A satisfying assignment indicates that an error was found.

Bounded model checking has not been explored so far for many other languages, including Java.  However, explicit-state model checking tools such as 
Java PathFinder (JPF) \cite{jpf} have been successfully used for the verification of many Java applications. Furthermore, there has been an explosion of symbolic execution \cite{King:1976:SEP:360248.360252} tools
that have been used successfully for test case generation and error detection in the context of many high level languages, for example \cite{Godefroid:2005:DDA:1065010.1065036,DBLP:conf/ndss/GodefroidLM08,Sen:2005:CCU:1081706.1081750,Cadar:2008:KUA:1855741.1855756}. In particular, 
relevant for the work reported here, Symbolic PathFinder (SPF) \cite{Pasareanu:2010:SPS:1858996.1859035} is a symbolic execution  tool built as an extension of JPF, 
that provides a symbolic analysis for Java programs involving multi-threading and complex data structures. 

In this thesis, we describe an alternative methodology for  BMC which is based on ``classical'' symbolic execution (SE) in the sense of King \cite{King:1976:SEP:360248.360252}.
Note that the way CBMC transforms a program into Static Single Assignment form can also be viewed as 
executing the program symbolically. However, this encoding is different from the SE of King
and we evaluate the two encodings as part of the work reported here.
Our methodology is not language specific and only relies on a  symbolic executor for that language and an SMT solver.

By using symbolic execution we obtain a  translation of a program and assertions into a disjunctive formula encoding the path conditions 
for each bounded (complete) path explored in the code. This suggests a simple concurrent verification strategy 
that relies on the observations that any subset of disjuncts in  a disjunction $F=F_1\vee \dots \vee F_n$ can be separately checked for satisfiability and whenever a subset is found to 
be satisfiable the satisfiability task can be stopped. Hence the verification of disjunctive formulas is naturally parallelizable.

We have implemented this methodology in a tool called JCBMC, which stands for {\bf J}ava {\bf C}oncurrent {\bf B}ounded {\bf M}odel {\bf C}hecker. The tool is built on top of SPF, and uses it to generate the disjunctive formula from the code (constraint solving is turned off in SPF itself) and while generating the formula it sends sub-formulas to multiple worker threads for satisfiability checking. JCBMC handles programs with multi-threading and recursive input data structures  and relies on a standard SMT solver, namely Z3~\cite{Z3} for solving the constraints. Other solvers can easily be incorporated. One can even use different solvers for solving different path constraints in parallel.

Although JCBMC is only a prototype, and concurrency is implemented by multi-threading but not parallelized yet, its performance, compared with existing tools, i.e. Symbolic PathFinder and CBMC, is remarkable. 
We summarize our contributions as follows:
\begin{itemize}
\item A methodology for concurrent bounded model checking that is based on ``classical'' symbolic execution and it is naturally parallelizable.
\item The methodology is language independent and supports assume-guarantee reasoning.
\item A tool JCBMC, a concurrent bounded model checker for Java. 
\item Experiments to show effectiveness of the tool for verification of programs with multi-threading and data structures.
\item Comparisons with bounded model checking and ``classical'' symbolic execution, as embodied by CBMC and SPF respectively.
\end{itemize}

\section{Illustrative example}
\label{sec:example}
We illustrate our approach using the simple example program in Figure \ref{fig:ex2}. We want to check if the assertion in line 9 is valid for all possible inputs. Note that an analysis using an explicit state model checker such as  JPF would not be feasible, as this would involve enumerating all the possible inputs to the program.

\begin{figure}[htp]
\begin{minipage}[b]{0.38\linewidth}
\begin{lstlisting}[language=C, frame=none,
xleftmargin=3ex,basicstyle=\footnotesize\ttfamily,
numbers=left,stepnumber=1, morekeywords={assert} ]
void test(int x, int y){
  if(x > 5){
    x++;
    if (x < 3) 
      x--;
    else 
      y = x;
  }
  assert (x < 10 );
}
\end{lstlisting}
\caption{A simple example}
\label{fig:ex2}
\end{minipage}
\hspace{-0.5cm}
\begin{minipage}[b]{0.6\linewidth}
\begin{tabular}{ l l l }
  1. & $pc$ = $\top$; & $\sigma = \{x \mapsto x_0, y \mapsto y_0\}$ \\
  2. & $pc = (x_0 > 5)$; & $\sigma = \{x \mapsto x_0, y \mapsto y_0\}$ \\
  3. & $pc = (x_0 > 5)$; & $\sigma = \{x \mapsto x_0 + 1, y \mapsto y_0\}$ \\
  4. & $pc = (x_0 > 5)$; & $\sigma = \{x \mapsto x_0 + 1, y \mapsto y_0\}$ \\
  6. & $pc = (x_0 > 5)$; & $\sigma = \{x \mapsto x_0 + 1, y \mapsto y_0\}$ \\
  7. & $pc = (x_0 > 5)$; & $\sigma = \{x \mapsto x_0 + 1, y \mapsto x_0+1\}$
\end{tabular}
\caption[Symbolic Execution for the path: (1,2,3,4,6,7)]{SE for the path: (1,2,3,4,6,7)}
\label{fig:se-ex}
\end{minipage}
\end{figure}

Figure \ref{fig:se-ex} illustrates a part of classical SE  with constraints solving on the program following the path $(1,2,3,4,6,7)$. This analysis could be performed using e.g. SPF.
Instead of concrete values, SE takes the symbols $x_0$ and $y_0$ as 
inputs and executes them just like concrete values. It also keeps track of the path condition $pc$ which consists of the conditions true along that path and the symbolic environment $\sigma$ which maps
variables into expressions over the input symbols $x_0$, $y_0$. Typically, whenever the $pc$ is updated, SE checks the satisfiability of $pc$ using an off-the-shelf solver.

Initially, $pc$ is true, and $\sigma$ maps inputs to theirs symbols. When SE reaches line 2, it updates $pc$ as $(x_0 > 5)$ since this is the condition 
to reach line 3 where $\sigma$ becomes $x \mapsto x_0 + 1$. In line 4, the condition $(x < 3)$ is translated in $\sigma$ to $c \equiv (x_0 + 1 < 3)$.
At this point SE calls an SMT solver, and detects that $pc \vdash \neg c$ because  $(x_0 > 5) \vdash \neg (x_0 + 1 < 3)$, therefore it jumps to line 6 with $pc$ unchanged.

In our approach for BMC, SE plays the role of generating the formula which encodes the program behaviour and the property to be checked. The satisfiability of the resulting 
formula will be checked separately by an SMT solver. Therefore, we execute SE \emph{without invoking constraint solving} whenever $pc$ is updated, and we postpone checking the $pc$ until the end of the execution path. In this way, we can save the execution time of calling the solver, but
the trade-off is that infeasible paths are also included. 
However, this will not affect the soundness of the analysis, since constraint solving is performed later.
When SE reaches line 9 following the path $\{1,2,3,4,6,7\}$, we have:
$$pc = (x_0 > 5) \wedge \neg (x_0 + 1 < 3)$$
Here we reach the property $\mathcal{P}$ to verify, which is $(x < 10)$.
We denote by $\mathcal{P}|_{\sigma}$ the evaluation of $\mathcal{P}$ in the symbolic environment $\sigma$. At this point, $\sigma$ maps $x$ to $x_0 + 1$, which leads to the following: 
$$\mathcal{P}|_{\sigma} = (x_0 + 1 < 10)$$
The property $\mathcal{P}$ is violated in this path if we can find a model for 
$$pc \wedge \neg \mathcal{P}|_{\sigma} = (x_0 > 5)  \wedge \neg (x_0 + 1 < 3) \wedge \neg (x_0 + 1 < 10)$$ 
Setting $x_0=11$ will provide such a model.
The whole formula generated by our approach for the code in Figure \ref{fig:ex2} is:
\begin{eqnarray*}
((x_0 > 5)  \wedge \neg (x_0 + 1 < 3) \wedge  \neg (x_0 + 1 < 10)) &\vee & \\
 ((x_0 > 5)  \wedge  (x_0 + 1 < 3) \wedge  \neg(x_0  < 10)) &\vee& \\
 (\neg(x_0 > 5)  \wedge \neg(x_0  < 10)) & &
\end{eqnarray*}
In general, we use SE to explore all possible symbolic paths up to a certain length, and then encode the program together with the property to check into a formula of the form:
$$\bigvee_{i=0}^{M} (pc_i \wedge \neg \mathcal{P}|_{\sigma_i})$$
where $N$ is the number of paths that may trigger the error. This form allows us to divide the formula into blocks of $D$ disjunctions:
$$\bigvee_{i=0}^{D-1} (pc_i \wedge \neg \mathcal{P}|_{\sigma_i}) \vee \bigvee_{i=D}^{2D-1} (pc_i \wedge \neg \mathcal{P}|_{\sigma_i})\dots\vee\bigvee_{i=(k-1)D}^{kD-1} (pc_i \wedge \neg \mathcal{P}|_{\sigma_i}) \vee\bigvee_{i=kD}^{M}  (pc_i \wedge \neg \mathcal{P}|_{\sigma_i}) $$
In this way, we can solve the formula concurrently using several threads, each one solving a single block. A model of a single block is also a model of the formula, therefore the procedure stops when any of the threads
find out a model. In JCBMC, after the main thread generates a sub-formula and passes it to a worker thread, it moves on to generate the next sub-formula, while the worker thread solves the given sub-formula concurrently.

\section{Concurrent Bounded Model Checking}
\label{sec:BMC-SE}
\begin{figure}[htp]
\centering
\includegraphics[scale=0.5]{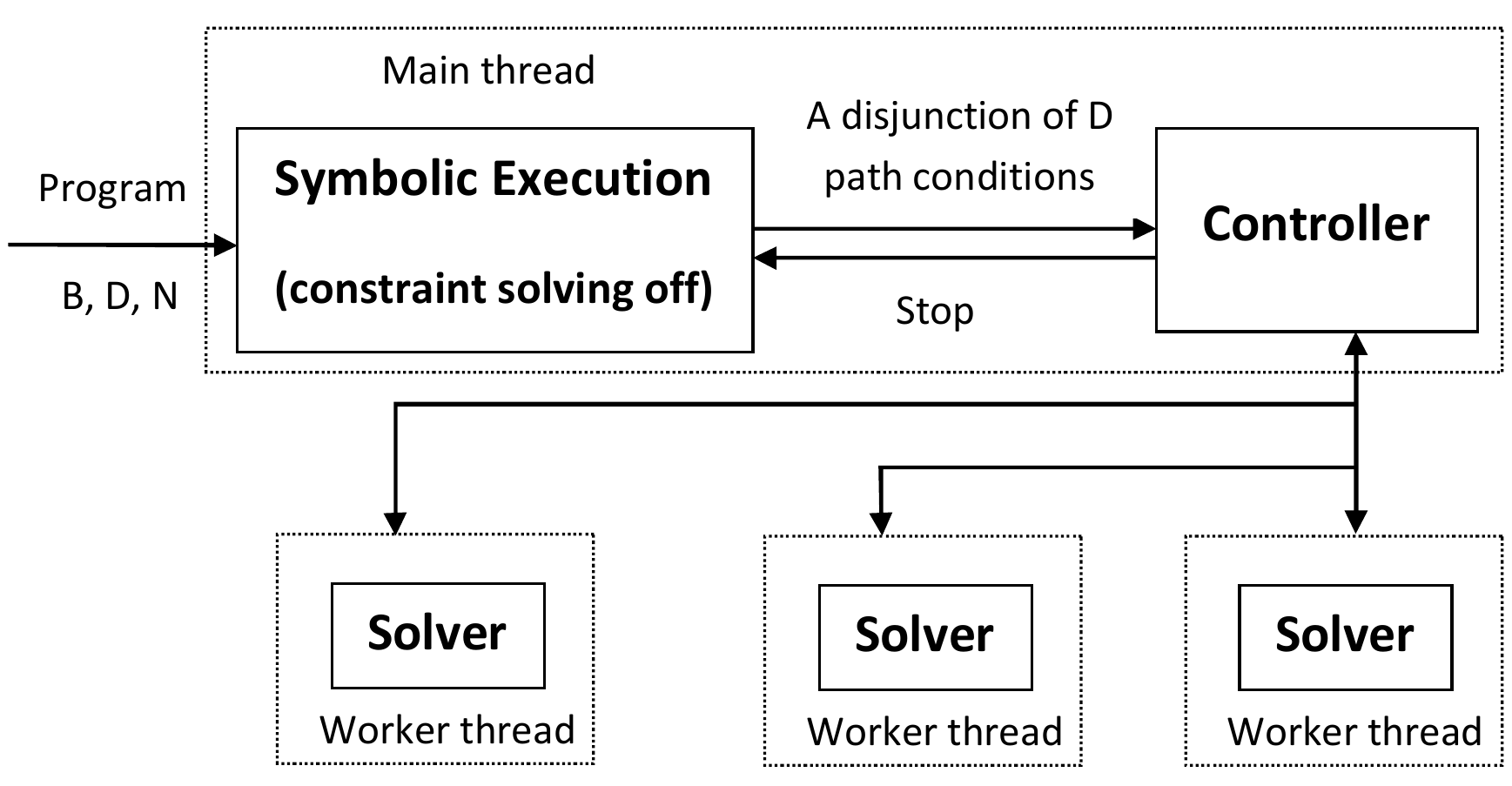}
\caption{Concurrent Bounded Model Checking architecture}
\label{fig:JCBMC}
\end{figure}

Our method for concurrent bounded model checking is illustrated in Figure \ref{fig:JCBMC}. The inputs to the method are: a program under test, a  property to verify and three parameters -- $B$ is the search bound, $N$ is the number of  workers and $D$ is the number of disjuncts to give to one worker. The goal is to check if the property holds in the program, up to exploration bound $B$.

\subsection{Bounded Model Checking by Symbolic Execution}
The program under test is analysed using ``classical'' bounded symbolic execution with constraint solving turned off. This means that whenever a path condition is updated, we do not check its satisfiability, but rather continue the exploration. As a result, symbolic execution may explore infeasible paths, which will be checked later using constraint solving. Our approach can be used for the bounded verification of safety properties, which we assume have been reduced to checking assertions embedded in the code. 
Furthermore, our method supports both {\em assume} and {\em assert} statements to enable {\em assume-guarantee style} verification. The assumed conditions are simply added to the path conditions during the symbolic execution.

The result of SE is a disjunction of path conditions, encoding constraints on the inputs to follow those paths, up to the pre-specified search bound. From among these paths, only the ones that may lead to assert violations are selected for solving. This is achieved by the {\em controller} which collects sets of $D$ violating path conditions and sends them for solving to parallel worker threads, using off-the-shelf solvers. The workers start solving as soon as they receive the disjunctive formulas, which may happen while the symbolic execution is still exploring the program. The verification terminates as soon as one of the threads finds a satisfying assignment, in which case an error is reported, or when all the disjunctions are found to be un-satisfiable, in which case the assertion holds (no error) up to the given bound.  Note that if the symbolic execution discovers no potentially violating paths (i.e. the error is unreachable), then no solving will be performed. 

\subsection{Comparing our approach with BMC and SE}
Compared with classical BMC we use an explicit enumeration of paths, while BMC uses an implicit enumeration of paths. Although at first glance the implicit encoding should be better our experiments, even with sequential JCBMC, show that this is not the case. Furthermore, the explicit enumeration is easily parallelizable, with simple and natural load balancing for different threads. Crucially our approach stops as soon as a path leading to an error is found to be satisfiable, while with classical BMC, all the program needs to be explored.

Compared with classical symbolic execution: we solve only in the end. So obviously the price to be paid is the exploration of infeasible paths. On the other hand, again, it is naturally parallelizable and constraint solving, which is one of the bottlenecks in SE, can be done in parallel, even with different solvers, with little coordination, if any, needed. 



 In the following we describe in more detail our method. We start with a description of a sequential approach, to clarify how we use symbolic execution to built a disjunctive formula of the path conditions. Solving this formula happens after the symbolic execution, sequentially.

\subsection{Sequential Verification}

Our approach of employing SE for BMC is based on a simple observation. Suppose $s_k$ is an error state in the  transition system $P$.
To determine the reachability of $s_k$ from the initial state $s_0$, BMC builds a series of transitions $s_0 \to s_1\to..\to s_{k}$, resulting in the formula in (\ref{equa:constraints}) of Section \ref{Sec:BMC-CBMC}.
On the other hand, SE builds the path condition $pc$ which is a first order formula also characterising  reachability of $s_k$ from $s_0$. Therefore, both condition (\ref{equa:constraints}) and $pc$ characterize all the inputs that reach the error.

Suppose we want to prove formally that the program \emph{P} satisfies, within given bounds, a property $\mathcal{P}$ represented by a set of assertions. 
This means whenever
the program reaches an assertion point, the assertion needs to be valid, which means:
$\bigwedge (pc_i \rightarrow \mathcal{P}|_{\sigma_i})$.
This can be verified by checking the satisfiability of
$\bigvee (pc_i \wedge \neg\mathcal{P}|_{\sigma_i})$.
\begin{proof}
First notice one formula is the negation of the other one, i.e.
$$\bigwedge (pc_i \rightarrow \mathcal{P}|_{\sigma_i}) = \bigwedge (\neg pc_i \vee \mathcal{P}|_{\sigma_i}) = \bigwedge \neg( pc_i \wedge \neg\mathcal{P}|_{\sigma_i}) = \neg\bigvee ( pc_i \wedge \neg\mathcal{P}|_{\sigma_i})$$  
hence if $\bigvee (pc_i \wedge \neg\mathcal{P}|_{\sigma_i})$ is satisfiable there exists an $i$ such that $ pc_i \wedge \neg\mathcal{P}|_{\sigma_i}$ 
is satisfiable i.e. an initial state leading to the path $pc_i$ and not satisfying the property $\mathcal{P}|_{\sigma_i}$. 
On the other hand if $\bigvee (pc_i \wedge \neg\mathcal{P}|_{\sigma_i})$ if not satisfiable then no disjunct $ pc_i \wedge \neg\mathcal{P}|_{\sigma_i}$ 
is satisfiable hence  there exists no initial state leading to an execution path satisfying the assertion.
\end{proof}

The algorithm in Figure \ref{symexBMC} shows how to build the formula $\Gamma = \bigvee (pc_i \wedge \neg\mathcal{P}|_{\sigma_i})$ for the GC language described in the 
previous section. In essence, the algorithm runs classical SE with no constraint solving for checking $pc$ satisfiability.
\begin{figure}[t]
\centering
\begin{minipage}[b]{0.65\linewidth}
\begin{algorithmic}[1]
\Function{SymExBMC}{$\sigma$, $pc$, $l$, $i$}
\If{($i > B$) } \label{ifbound:start}
        \State \textbf{return}
\EndIf \label{ifbound:end}
\State Extract statement $s$ at $l$ \label{while:start}
\While{($s$ is not an \textbf{if}-statement $\wedge$ $l \neq EOF$)}
    \If{($\text{\emph{isSAT}} = \top$) } \label{ifsat:start}
        \State \textbf{return}
    \EndIf \label{ifsat:end}
    \If{($s$ is `\textbf{assume} c')} 
        \State $pc \leftarrow pc \wedge c|_{\sigma}$ \label{line:assume}
    \ElsIf{($s$ is `\textbf{assert} c')}
           \State Process($pc \wedge \neg c|_{\sigma}$) \label{line:assert}
    \Else
        \State Execute the assignment $s$, update $\sigma$
    \EndIf
    \State $l \leftarrow {\tt next}(l)$
    \State{Extract statement $s$ at $l$}
\EndWhile \label{while:end}
\If{($l = EOF$)}
        \State \textbf{return}
\EndIf
\State Extract $\{c,l_{\top},l_{\bot}\}$ from \textbf{if}-statement
\State $i \leftarrow i+1$
\State $pc_1 \leftarrow pc \wedge c|_{\sigma}$ 
\State SymExBMC($\sigma$, $pc_1$, $l_{\top}$, $i$)
\State $pc_2 \leftarrow pc \wedge \neg c|_{\sigma}$
\State SymExBMC($\sigma$, $pc_2$, $l_{\bot}$, $i$)
\EndFunction
\end{algorithmic}
\caption{Formula generation for BMC}
\label{symexBMC}
\end{minipage}
\begin{minipage}[b]{0.33\linewidth}
\centering
\begin{algorithmic}
\Function{Process}{$\gamma$}
\State $\Gamma \leftarrow \Gamma \vee \gamma$ 
\EndFunction
\end{algorithmic}
\caption{Process error paths}
\end{minipage}
\end{figure}
A statement of GC is determined by its location $l$, and the function {\tt next}(l) 
returns the location of the next statement. An if-statement consists of a condition $c$, the location $l_{\top}$ if $c$ is true, and the location $l_{\bot}$ 
if $c$ is false. 
In the recursive procedure {\tt SymExBMC} as well as the function {\tt Process}, all parameters
are passed by value. In {\tt SymExBMC}, the checking from line \ref{ifsat:start} to line \ref{ifsat:end} is only used in concurrent mode. In sequential mode, \emph{isSAT} is set to false,
and it is not changed in the whole procedure. $\Gamma$ is declared as a global variable, initialised to false (denoted by $\bot$). $k$ is global constant, defining the bound of BMC. 

Similar to standard SE, at the beginning the symbolic environment $\sigma$ maps program inputs to symbols, the path condition $pc$ is initialised to 
true (denoted by $\top$). The depth $i$ of the recursion is initialised to 0. The procedure  {\tt SymExBMC} starts by ensuring the
search depth $i$ not to reach the bound $B$ (line \ref{ifbound:start} to line \ref{ifbound:end}). As from line \ref{while:start} to line \ref{while:end}, it 
symbolically executes a basic block, i.e. without branching statement, of the program. The basic block ends by an if-statement or when the location $l$ 
reaches the end of the source file ($l = EOF$). Assumptions and assertions are evaluated by the current symbolic environment (line \ref{line:assume}, \ref{line:assert}).
When there is an assignment, the symbolic environment updates the mapping for the variable in the left hand side by the evaluation of the right hand side.
At the end of the block, if there is an if-statement at the next location,  {\tt SymExBMC} is recursively called for both the \emph{then} path and the \emph{else} 
path. The condition is added to the path conditions without any checking of path feasibility.

\subsection{Concurrent Verification}
\label{sec:concur-BMC}

The simple algorithm presented in the previous sections will essentially enumerate all the possible feasible and infeasible paths through a program (up to a given bound), collect the path conditions for each path into a formula in disjunctive form and then invoke a constraint solver to check the satisfiability, all at once. The disadvantage of the approach is that it needs to enumerate all the possible paths through the program, which can quickly become  expensive, especially if multi-threading is also considered. We therefore propose a 
 concurrent algorithm that parallelises {\tt SymExBMC}  by delegating the satisfiability check of the disjuncts in $\Gamma$ to worker threads. 
%
The  concurrency of the algorithm relies on two parameters: the number of concurrent workers available and the number of disjuncts sent to each worker: the optimal choice is architecture and SMT solver dependent. In our experiments we found 200 disjuncts to be a reasonable choice.

The main function is in Figure~\ref{Main}. It initialises $\Gamma$, $pc$, $\sigma$ exactly the same as in sequential mode of {\tt SymExBMC}. Here, $isSAT$ is a boolean variable shared between the threads,
and can be modified (set to true) by them. $d$ is also a global variable of the main thread only to keep the number of current disjuncts. After initializing, the function {\tt SymExBMC} is called. 
The main difference between sequential and concurrent mode is the function {\tt Process} in Figure \ref{ConBoM} and the checking from line \ref{ifsat:start} to line \ref{ifsat:end} in {\tt SymExBMC}.
In sequential mode, \emph{isSAT} is always false, so  {\tt SymExBMC} keeps building the formula until it reaches EOF. In concurrent mode, when 
the number of disjuncts in $\Gamma$ reaches a bound $B$, {\tt Process} sends $\Gamma$ for satisfiability check to a worker thread. Crucially this worker thread can run in parallel to any other running thread as they run completely independent tasks.
 
Whenever a worker thread finds a model for its own $\Gamma$ it sets the shared variable $isSAT$ to true which will return control to the main thread and end the computation with 
{\tt Verification failed}. If no thread sets  $isSAT$ to true then  ${\tt SymExBMC}$  will eventually terminate and the remaining disjuncts (whose number is hence less than $D$) 
are sent for satisfiability check to a final thread. {\tt Verification successful} is returned only if no thread  has set  $isSAT$ to true during the computation.
 
\begin{figure}[t]
\fbox{
\begin{minipage}[b]{0.56\linewidth}
\centering
\begin{algorithmic}
\State $\Gamma$, \emph{isSAT} $\leftarrow\bot$; $pc \leftarrow\top$; $d,i \leftarrow 0$
\State $l \leftarrow$ first statement
\State SymExBMC($\sigma$, $pc$, $l$, $i$)
\If{($i > 0$ $\wedge$ \emph{isSAT} = $\bot$)}
        \State Execute Run($\Gamma$, \emph{isSAT}) in worker thread
\EndIf
\If{(\emph{isSAT} = $\bot$)}
     \State \textbf{Return} Verification successful
\Else     
     \State \textbf{Return} Verification failed
\EndIf
\end{algorithmic}
\caption{Main thread}
\label{Main}
\end{minipage}
}
\fbox{
\begin{minipage}[b]{0.35\linewidth}
\centering
\begin{algorithmic}
\Function{Run}{$\Gamma$, \emph{isSAT}}
\State Run SMT-Solver on $\Gamma$
\If{($\Gamma$ has a model)}
        \State \emph{isSAT} = $\top$
\EndIf
\EndFunction
\end{algorithmic}
\caption{Worker thread}
\end{minipage}
}
\end{figure}

\begin{figure}[t]
\centering
\fbox{
\begin{minipage}[b]{0.65\linewidth}
\begin{algorithmic}
\Function{Process}{$\gamma$}
	\State $\Gamma \leftarrow \Gamma \vee \gamma$
	\State $d \leftarrow d+1$
	\If{($d \ge D$)} 
	    \State Execute Run($\Gamma$, \emph{isSAT}) in worker thread
	    \State $\Gamma \leftarrow \bot$; $d \leftarrow 0$
  \EndIf
\EndFunction
\end{algorithmic}
\caption{Process error paths for Concurrent BMC}
\label{ConBoM}
\end{minipage}
}
\end{figure}

\subsubsection{Implementation}
\label{sec:JCBMC}
Our prototype tool, JCBMC, has been implemented following the Observer Design Pattern~\cite{Gamma:1995:DPE:186897}.  SPF executes symbolically the Java bytecode  program and acts as 
the subject. The Controller acts as the observer, it waits for SPF to have generated a sub-formula $\Gamma$ with $D$ disjuncts  and then sends it to  an available  worker thread. $D$ is a user chosen parameter of JCBMC. The worker thread executes 
{\tt Run}($\Gamma$, \emph{isSAT}) which first writes $\Gamma$ into a file in SMT2 format \cite{smt2}, then calls the SMT solver Z3 to check for satisfiability.
JCBMC creates a thread pool of N workers; current architecture doesn't support parallelism but only 
 multi-threading.

JCBMC is built on top of SPF and as an extension of the JPF platform, therefore it inherits all the power of JPF and SPF.

%
%

\section{Evaluation}
\label{sec:cases}
Our evaluation comprises cases studies  to compare JCBMC with SPF (with default configuration) and case studies 
to compare JCBMC with CBMC\footnote{To compare both tools with the same solver in the experiments CBMC will be called with option --smt2, and we 
will use Z3 for satisfiability checks.
}. To compare with CBMC we have considered C code whose Java translation is almost literal. 

An important reminder is that the current implementation of JCBMC is multi-threaded but not yet parallel and we expect a parallel implementation to have a significant advantage over the current one. 

By JSBMC we denote the sequential implementation of JCBMC where a single thread is used.
Experiments are run on a machine equipped with dual Xeon(R) E5-2670 CPUs. The results are shown in Tables~\ref{fig:all} and \ref{fig:java}.
Unless otherwise specified times are in seconds, $x$m$y$ means $x$ minutes and $y$ seconds, ``timed out'' is one hour and x denotes a memory hit\footnote{A memory hit is a ``run out of memory'' problem. This can be addressed by a different memory manager in JPF or with a direct implementation of {\tt SymExBMC}.}. 
The source code for the examples can be found at: \url{https://github.com/qsphan/jpf-bmc}.
\subsection{Comparing with CBMC and SPF}

\subsubsection{Bubble Sort}
We consider the classical  bubble sort algorithm, which has already been studied in the BMC community \cite{svcomp14,Armando:2009:BMC:1501670.1501673}. Here, differently from
\cite{Armando:2009:BMC:1501670.1501673}, we consider the more challenging symbolic version where the values of the array are non-deterministically chosen.
We consider both the verification of the assertion ``the elements of the array are ordered after bubble sort'' and its negation 
``the elements of the array are not ordered after bubble sort''. We analyse a program implementing bubble sort. It will hence contain no bugs for the positive assertion and will be buggy for the negation. Results are shown in Fig \ref{fig:all}. 
We notice that while CBMC is better for the positive assertion, JCBMC outperforms the other tools for the negative assertion and is capable
of find a counterexample for array sizes of a higher order of magnitude.

\begin{figure}[t]
\centering
\begin{tabular}{|c|c|c|c|c|c|} 
  \hline
   & \textbf{SPF}     & \textbf{JSBMC}  & \textbf{CBMC} & \textbf{JCBMC} (10) & \textbf{JCBMC} (200)   \\ \hline
 \multicolumn{1}{|c|} {Array size} & \multicolumn{5}{c|}{Bubble sort with assertion negated} \\ \hline
  5          & 4.517 &  2.174     &  0.460    & 1.097   & 1.338 \\ \hline
  6          & 5.622 &  12.604     & 0.817     & 1.160  & 1.389 \\ \hline
 15         & 36.194	&  x        & 56m34.033     & 1.195  & 1.948  \\ \hline
  30         & 4m32.790		& x        & timed out     & 1.387  &  2.905  \\ \hline
  100          & 	timed out	&  x        &  timed out    & 4.944  & 34.697  \\ \hline
  & \multicolumn{5}{c|}{Verification of bubble sort} \\ \hline
     5       &      6m19.222    &   3.712    &  7.171    &  4.193             & 3.622  \\ \hline
     6       &      timed out     &   26.293    &  37.816    &  29.512              & 21.834  \\ \hline
     7       &      x  &   x  &  5m22.641    &  x             & x \\ \hline
     8       &      x    &   x    &  timed out    &  x             & x \\ \hline
\multicolumn{1}{|c}{}& \multicolumn{5}{c|}{Sum of array} \\ \hline
  unsafe & 1.403  & 12.671 & 1m5.738  & 1.576 & 2.479  \\ \hline
  safe & {failed} & 12.030 & 2.252  & 9.466  & 10.614   \\ \hline
\end{tabular}
\caption[Comparing performance JCBMC against SPF and CBMC]{Performance of all tools. JCBMC(10) and JCBMC(200) mean JCBMC is run with the parameter D as 10 and 200 respectively. ``failed'' refers to SPF failing to solve the constraints using the integrated solver.
}
\label{fig:all}
\end{figure}

\subsubsection{Sum of array}
We consider the array case studies taken from \cite{svcomp14}, in particular \emph{sum\_array\_safe.c} for verification and \emph{sum\_array\_unsafe.c} for refutation.
The array size is set to 1000. Results show both SPF and JCBMC outperform CBMC for the unsafe version, while CBMC has a slight advantage for the safe version. 
\subsection[Comparing with SPF]{Comparing with SPF (Java code)}
The following examples consist of substantial Java code which is not naturally translatable in C; we hence compare JCBMC only with SPF. Notice
JCBMC and SPF are both extensions of JPF: in the case all inputs are concrete they both reduce to JPF-core hence their performance is identical. Hence   we only consider programs with symbolic inputs. 

\subsubsection{Flap controller}
This case study is shipped with the distribution of SPF. It is a multi-threaded program modelling a simplified flap controller on an aircraft. It contains 3 classes, and 80 lines of 
code. This example demonstrates handling of multi-threading.

\subsubsection{Red Black Tree}
This is another example from the SPF distributions (3474 LOC in one class). We check for consistency of the tree after performing {\tt put}, {\tt remove}, {\tt get} and {\tt firstKey} symbolically.  Results show that both JSBMC and JCBMC significantly outperform SPF.

\subsubsection{MER Arbiter}
The MER Arbiter models a component of the flight software for NASA
JPL's Mars Exploration Rovers (MER).  The MER
Arbiter has been modelled in Simulink/Stateflow and it was
automatically translated into Java using the Polyglot
framework and analyzed with SPF~\cite{Balasubramanian:2013:PSA:2450387.2450440}. The configuration for our analysis
involved two users and five resources. 
The example has 268 classes,
553 methods, 4697 lines of code (including the Java Polyglot execution
framework) but only approx. 50 classes are relevant.  We analyse the code with and without the error (see~\cite{Balasubramanian:2013:PSA:2450387.2450440}). The performances of SPF, JSBMC and JCBMC are comparable, with SPF slightly better.

\begin{figure}[t]
\centering
\begin{tabular}{|l|c|c|c|c|c|} 
  \hline
  Tool & \textbf{SPF}     & \textbf{JSBMC}  & \textbf{JCBMC} (10) & \textbf{JCBMC} (200)  \\ \hline
  Flap controller (unsafe) & 1.141   & 2.899           & 0.948  & 1.370  \\ \hline
  Red-black tree (safe) & 53.602   & 3.942  & 3.267 & 2.774    \\ \hline
  MER Arbiter (unsafe) & 5.275   & 8.111  & 7.479 & 7.579    \\ \hline
  MER Arbiter (safe) & 47.065   & 59.145  & 57.740 & 58.886    \\ \hline

\end{tabular}
\caption[Comparing performance JCBMC against SPF]{Performance on Flap controller, Red-black tree and MER Arbiter. JCBMC(10) and JCBMC(200) mean JCBMC is run with the parameter D as 10 and 200 respectively.}
\label{fig:java}
\end{figure}

\subsubsection{Discussion of experiments} Comparing with CBMC, JCBMC scores better in finding counterexamples than in verifying their absence; this is consistent with its design because a counterexample corresponds to a worker thread finding a model of the formula. Compared with SPF, JCBMC can be much better (see bubble sort or red black tree) but can also be comparable or slightly worse (see MER Arbiter and Flap Controller results). The reason for the latter is that the cost of generating path conditions  dominates the cost of solving them. Similarly, SPF failed to generate formulas for bubble sort for sizes 7 and higher. Furthermore, an error path (e.g. in MER) may occur at the beginning of the SE exploration, and it is therefore discovered quickly by SPF, while JCBMC still needs to generate the pre-specified number $D$ of error paths before solving them. The results suggest one direction for future work, namely to investigate improving the cost of SE-based path generation (see last section).
 
\section{Discussion of related work}
\label{relwork}
Related approaches on parallelising BMC~\cite{Abraham:2006:PSS:1757571.1757595,DBLP:journals/corr/abs-0912-2552} 
address parallel solving of the conjunctive formula that is built for BMC and aim at performing solving at different bounds, where 
some clauses are shared to enable more efficient SAT solving.
In contrast we aim to solve the formulas generated with SE for the same bound, which are naturally disjoint resulting in a simpler parallel algorithm. Furthermore our 
work aims at verifying programs written in high-level languages such as Java and it is not 
clear how the previous work, performed in the context of finite state automata, would be applicable.
Also related is the work on parallel SAT and SMT solving~\cite{Sinz01pasat,Pamira4022225,Wintersteiger:2009:CPA:1575060.1575124}, which 
can be seen complementing the work presented here, in the sense that we can use e.g. the parallel version of Z3~\cite{Wintersteiger:2009:CPA:1575060.1575124} in each of the workers to further speed up our proposed approach.

PKIND \cite{DBLP:journals/corr/abs-1111-0372} is a parallel model checker for Lustre that uses k-induction. PKIND  runs in parallel the different tasks 
involved in performing the induction, e.g. the base step, the induction step and also the generation of auxiliary invariants used for verification. 
Therefore PKIND performs the parallel work at a higher level of granularity than JCBMC. It would be interesting to investigate if we can replace the parallel tasks in PKIND with our own version of SE-based bounded verification, which in turn is parallelized at the level of 
granularity of symbolic paths. 

Parallel model checking has been investigated in the context of explicit-state~\cite{BBCR10,Palmer02partialorder,Burns:2012:PMC:2364065.2364083,Holzmann:2007:DME:1314033.1314051,Stern:1997:PMV:647766.736020,Jabbar:2006:PED:2146228.2146244}
and symbolic~\cite{Nalla_parallelbounded,DBLP:conf/ecai/KwiatkowskaLQ10} exploration. The latter 
 were done in the context of verification using Binary Decision Diagrams, and hence are very different from ours. These approaches concentrate 
on partitioning the state space to be explored in parallel and on dealing with the communication overhead between 
parallel workers. In contrast, in our approach the workers perform the solving independently, with no communication between them.



In previous work we have developed a framework for performing parallel symbolic execution in SPF~\cite{Staats:2010:PSE:1831708.1831732}. We used a set of pre-conditions to partition
the symbolic execution tree to distribute its processing. These pre-conditions were computed \emph{a priori}, using a ``shallow'' symbolic execution up to a small 
exploration bound, to {\em statically} compute the different partitions of the input space with no communication overhead.  Other approaches to parallel 
symbolic execution \cite{King,Ciortea:2010:CST:1713254.1713257,Sarfraz} operate primarily by {\em dynamically} partitioning the symbolic execution tree 
for load balancing, which may result in better use of computational resources but also in more communication overhead. All these approaches were done in the 
context of ``classical"~\cite{Staats:2010:PSE:1831708.1831732,King} or dynamic~\cite{Ciortea:2010:CST:1713254.1713257,Sarfraz} symbolic execution, using constraint solving during path generation. In contrast the approach we advocate here has a clear 
separation between path generation and constraint solving, allowing us to easily achieve load balancing between workers, with little communication 
overhead. It would be interesting to compare experimentally and to also combine the techniques in JCBMC and the parallel version of SPF and we plan to do that 
in future work.

\chapter[Quantifying Information Leaks using Reliability Analysis]{Quantifying Information Leaks using Reliability Analysis}
\label{chap:qilura}
\graphicspath{{chapter6/figs/}}
\lstset{language=C,frame=single,numbers=none,
basicstyle=\ttfamily
}
In chapter \ref{chap:SymExDPLL}, we have presented the first technique to use Symbolic Execution for Quantitative Information Flow analysis. 
There, Symbolic Execution has been used for both exploring the program and counting the models. In this chapter, we explore an alternative approach 
which uses Symbolic Execution only for exploring the program, leaving the task of counting models for a reliability analysis tool.

This chapter can be divided in two parts. The first part studies the \emph{qualitative} aspect of information flow analysis. Its contribution
is a novel and practical approach for self-composition using Symbolic Execution. 

In the second part of the chapter, we show the relation between reliability analysis and Quantitative Information Flow. Exploiting this relation, 
we combine our new self-composition technique with a Symbolic Execution-based reliability analysis tool to
quantify information leaks in Java bytecode.
\section{Introduction}
Recall that in section \ref{sec:selfcomp}, we have introduced the theorem proving approach to non-interference, in which non-interference is logically formulated by \emph{self-composition}.
As far as we are aware, this has been the only approach for qualitative analysis that returns neither false positives nor false negatives. We also
quoted the claim of Terauchi and Aiken~\cite{Terauchi:2005:SIF:2156802.2156828} that self-composition was impractical and that it would be unlikely
to expect any future advance. The main limitations of self-composition, as pointed out by Terauchi and Aiken, come from the symmetry and redundancy of the self-composed program, which
lead to some partial-correctness conditions that hold between \code{P} and \code{P}$_1$. To find these conditions is crucial for the effectiveness of the analysis, 
however, finding them is in general impractical. He we present a \emph{practical} implementation for self-composition. 

\subsection{Qualitative analysis}
The idea of self-composition is to have a copy \code{P}$_1$ of the program \code{P} to compare with itself. The approach can be divided into two steps:
the first step is to compose the program with a copy of itself; the second one is to perform analysis on the self-composed program. Our approach is to delay self-composing to the second step: first,
we perform analysis on the original program with Symbolic Execution; second, we self-compose the result of the analysis to get the formula of self-composition.

We expand the idea of comparing the program \code{P} with it copy \code{P}$_1$ into comparing all pairs of executions $\rho$ of \code{P} and $\rho_1$ of \code{P}$_1$.
Since it is impossible to enumerate all possible executions, we use Symbolic Execution to synthesize the symbolic paths that represents a set of concrete executions,
and perform comparison on these symbolic paths, which we formulate as \emph{path-equivalence}.

The delay of self-composing after performing the analysis is the main novelty of our approach. In this way, we could avoid the symmetry and redundancy of the self-composed program.
Moreover, the symbolic paths synthesized by Symbolic Execution are presented by first-order theories, just as the generated formula of self-composition. The validity of this formula can
be automatically and efficiently checked by powerful SMT solvers.

\subsection{Quantitative analysis}
Traditional self-composition technique can only tell if a program leaks information. On the contrary, we can refine our Symbolic Execution-based self-composition
into a more fine-grained analysis that can decide if a path of the program leaks information. We then quantify the leaks for each symbolic path using
a reliability analysis tool.

Our approach is implemented into an automated tool, called QILURA (which stands for
\textbf{Q}uantify \textbf{I}nformation \textbf{L}eaks \textbf{U}sing
\textbf{R}eliability \textbf{A}nalysis). Given a
program, and inputs labeled as {\em high} and {\em low}, QILURA
computes an upper bound on the maximum number of bits that
the program can leak to a public observer. Our implementation is done
in the context of Java bytecode programs and the
SPF~\cite{Pasareanu:2010:SPS:1858996.1859035} symbolic execution
engine, extended for reliability
analysis~\cite{Filieri:2013:RAS:2486788.2486870}. However, the work is
general and can be applied in the context of any programming language
for which a symbolic execution tool exists.

\begin{figure}[htp]
\centering
\includegraphics[scale=0.6]{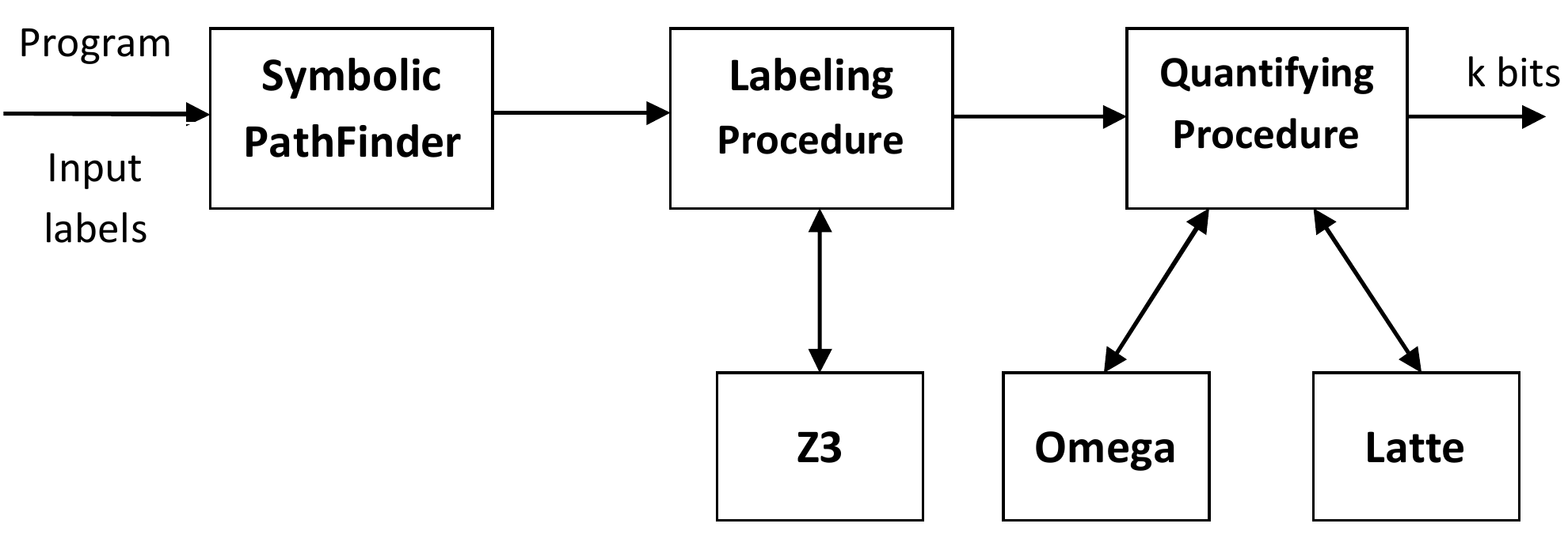}
\caption{Architecture of QILURA}
\label{fig:tool}
\end{figure}

At a high level, the architecture of QILURA is depicted in
Figure \ref{fig:tool}. The user labels the inputs of the
program with \emph{high} and \emph{low}. The program is then
passed to SPF to collect all possible symbolic paths. 
The \emph{Labeling Procedure}, using a fine-grained self-composition \cite{Barthe:2004:SIF:1009380.1009669}, classifies all the paths into three
categories: \texttt{clean}, \texttt{direct} and
\texttt{indirect}. The procedure uses z3~\cite{DeMoura:2008:ZES:1792734.1792766} for satisfiability checking of self-composition condition.

Finally, the \emph{Quantifying Procedure} uses Barvinok model counting techniques \cite{latte}
over the symbolic constraints (simplified using Omega \cite{omega})
collected by SPF to count the number of inputs that follow paths labeled with ``direct'' 
and provides an upper bound of $k$ bit on the leaks.
\section{Preliminaries}
This section reformulates Symbolic Execution and provides some background on the new Symbolic Execution-based reliability analysis framework of Filieri et al.~\cite{Filieri:2013:RAS:2486788.2486870}.
\subsection{Symbolic Execution} 
In chapter \ref{chap:SymExDPLL}, we described Symbolic Execution on a Symbolic Transition System which modelled in detail the transition
of a program with guards and actions. These details are not necessary in this chapter and are not captured in the
transition system in section \ref{sec:transit}, that we use to describe concrete execution. Therefore, we reformulate Symbolic Execution in a more
coarse-grained transition system that only takes in to account the source and target states of a transition. A program \code{P} is modelled as follows:
$$P = (S^*, I^*, F^*, T^*)$$
where $S^*$ is the set of symbolic states; each $s^* \in S^*$ represents a set of concrete states $s \in S$.
$I^* \subseteq S^*$ is the set of initial symbolic states; $F^* \subseteq S^*$ is the set of final symbolic states; and
$T^* \subseteq S^* \times S^*$ is the transition function.
A symbolic path (symbolic trace) of the program \code{P} is represented by a sequence of symbolic states:
$$\rho^* = s^*_0s^*_1..s^*_k$$ 
such that $s^*_0 \in I^*, s^*_k \in F^*$ and $\langle s^*_i,s^*_{i+1}\rangle\in T^*$ for all $i \in \{0, \dots,k-1\}$. 
The symbolic semantics of \code{P} is then defined as the set of all 
symbolic paths $\mathcal{R}^*$, which is also called as the \emph{symbolic execution tree}. Likewise, each $\rho^* \in \mathcal{R}^*$ represents a 
set of traces $\rho \in \mathcal{R}$.

We denote by $X|_y$ the value of the variable $X$ at the state $y$. After symbolically executing the program \code{P} with initial input symbols $H = \alpha, L = \beta$, 
for each $s^*_i \in F^*$, i.e. each leaf of the symbolic execution tree, we have a symbolic formula for the value of the output \code{O} in the symbolic environment:
$$O|_{s^*_i} = f_i(\alpha,\beta)$$
Another product of SE is the path condition $pc_i \equiv c_i(\alpha,\beta)$ for $s^*_i$ to be reachable. Each $pc_i$ corresponds to a symbolic path $\rho^*_i$. The following theorem was also proved by King \cite{King:1976:SEP:360248.360252}:
\begin{theorem}$$\forall i,j \in [1,n] \wedge i \neq j. pc_i \wedge pc_j = \bot$$
\end{theorem}
We define the function \emph{path} such that: $$\text{\emph{path}}(\rho^*_i) = pc_i 
$$
The output \code{O} can be considered as a result of the following function:
\begin{equation}
O = \left\{
  \begin{array}{l l}
    f_1(\alpha,\beta) & \quad \text{if $c_1(\alpha,\beta) $}\\
    f_2(\alpha,\beta) & \quad \text{if $c_2(\alpha,\beta)$}\\
    \dots & \quad \dots \\
    f_n(\alpha,\beta) & \quad \text{if $c_n(\alpha,\beta)$}\\
  \end{array} \right\}
\label{equa:se}
\end{equation}
Or the following always holds:
\begin{corollary}
\label{selem}
$$
\forall i \in [1,n]. c_i(\alpha,\beta)\rightarrow O =  f_i(\alpha,\beta) 
$$
\end{corollary}
$f_i$ and $c_i$ are in general combination of first-order theories, e.g. \emph{linear arithmetic}, \emph{bit vector} and so on. SE tools make use of off-the-shelf SMT solvers
to check the satisfiability of $c_i$, and eliminate unreachable paths (which may appear in the control flow graph).

\subsection{Reliability Analysis}
Reliability analysis \cite{Cheung:1980:USR:1313319.1313511} aims to compute the probability that a program successfully accomplishes its task
without errors.
Most previous work perform reliability analysis at early stages of design, on a architectural abstraction of the program, and thus they are not
applicable to source code. 

In \cite{Filieri:2013:RAS:2486788.2486870}, Filieri et al. introduced the first approach that can compute the reliability
of program from Java bytecode. Their approach is to use SE to enumerate each of the symbolic paths (and its path condition $pc_i$).
The symbolic path is then labelled as: (i)\textbf{T} if the program accomplishes the task; (ii) \textbf{F} if the program reaches an error state;
(iii) \textbf{G} if we cannot decide because the path is not fully explored (G stands for grey).

From the path condition $pc_i$, Filieri et al. use the tool Latte~\cite{latte} to compute efficiently the number of inputs $\#(pc_i)$ that satisfies the 
path condition $pc$. The reliability of the program, i.e. the probability that the program accomplishes its task, is then computed as:
$$\mathcal{R} = \frac{\Sigma \#(pc^T)}{\Sigma \#(pc^T)+\Sigma \#(pc^F)+\Sigma \#(pc^G)}$$

For QILURA we do not compute probabilities but use directly the counts over the computed symbolic constraints.

\section{Self-composition by Symbolic Execution}
To avoid the limitation of the theorem proving approach, we need to reformulate the self-composition formula into a simpler logic which does not contain the program \code{P}. This is made
possible by using the trace semantics of programs. 
\subsection{Self-composition as path-equivalence}
Given a program \code{P} that takes secret input \code{H}, public input \code{L} and producing public output \code{O}; \code{P}$_1$ is the same program as \code{P}, with all variables renamed: \code{H} as \code{H}$_1$, \code{L} as \code{L}$_1$ and \code{O} as \code{O}$_1$.
The trace semantics of \code{P} and \code{P}$_1$ are $\mathcal{R}$ and $\mathcal{R}_1$ respectively.

\begin{definition}[trace-equivalence] The program \code{P} satisfies non-interference if:
\begin{equation}
\forall \rho \in \mathcal{R}, \rho_1 \in \mathcal{R}_1. \code{L}|_{init(\rho)} = L_1|_{init(\rho_1)} \rightarrow O|_{fin(\rho)} = O_1|_{fin(\rho_1)}
\label{traceself}
\end{equation}
\end{definition}
It is stated similarly to the Hoare triple in (\ref{scHoare}): for all possible pairs of traces $\rho$ of \code{P}, and $\rho_1$ of \code{P}$_1$: if $L = L_1$ at the initial states, then $O = O_1$ at the
final states. At this point, we have a formulation of self-composition that does not involve the programs \code{P} and \code{P}$_1$.

However, even with simple programs, it is impossible to compute all the traces. Our solution is to use trace-equivalence with SE. Recall that each symbolic path represents a set of traces,
and it is possible to build a complete symbolic execution tree (here we only consider bounded programs). 
Following Corollary \ref{selem}, trace-equivalence in the context of SE is redefined as follows:
\begin{definition}[path-equivalence] The program \code{P} satisfies non-interference if and only if for all $\rho^* \in \mathcal{R}^*$ and for all $\rho^*_1 \in \mathcal{R}^*_1$, the following equation holds:
\begin{equation}
 (L|_{init(\rho^*)} = L_1|_{init(\rho^*_1)}) \wedge path(\rho^*) \wedge path(\rho^*_1) \rightarrow (O|_{fin(\rho^*)} = O_1|_{fin(\rho^*_1)}) 
\label{pathself}
\end{equation}
\end{definition}
In this way, we have an SMT formula, i.e. a combination of first-order theories.
This is the key novelty of our approach, since the formulation of self-composition in first-order theories enables us to solve it efficiently using 
off-the-shelf SMT solvers.
\subsection{Path-equivalence generation}
\label{sec:path-equiv}
Suppose \code{P} is symbolically executed with $H = \alpha, L = \beta$. To simplify the formula, we choose the input symbols for \code{P}$_1$ as $H_1 = \alpha_1, L_1 = \beta$ so
that $L|_{init(\rho^*)} = L_1|_{init(\rho^*_1)}$ is automatically satisfied. That means:
$$(H|_{init(\rho^*)} = \alpha) \wedge (L|_{init(\rho^*)} = \beta) \wedge (H_1|_{init(\rho^*_1)} = \alpha_1) \wedge  (L_1|_{init(\rho^*_1)} = \beta)$$
Given the result of SE is a function of the output \code{O} as in (\ref{equa:se}), the path-equivalence in (\ref{pathself}) can be rewritten as:
$$PE \equiv DF \wedge IF$$
where:
\begin{align}
DF &\equiv \bigwedge_{i=1}^{n} c_i(\alpha,\beta) \wedge c_i(\alpha_1,\beta) \rightarrow (f_i(\alpha,\beta) = f_i(\alpha_1,\beta)) \nonumber\\
IF &\equiv \bigwedge_{i=1}^{n-1} \bigwedge_{j=i+1}^{n} c_i(\alpha,\beta) \wedge c_j(\alpha_1,\beta) \rightarrow (f_i(\alpha,\beta) = f_j(\alpha_1,\beta)) \nonumber
\end{align}
\emph{DF} checks the path-equivalence when both \code{P} and \code{P}$_1$ follow the same symbolic path, and thus it guarantees the absence of direct flows.
On the other hand, \emph{IF} checks the path-equivalence when \code{P} and \code{P}$_1$ follow different symbolic paths, and it guarantees the absence of implicit flows. 

\subsection{Examples}
We illustrate the approach with some toy examples. Here we assume the same setting as above: a program \code{P} with confidential input \code{H}, public input \code{L}, and output \code{O}.
SE executes \code{P} with input symbols $H = \alpha$ and $L = \beta$.

\subsubsection{Implicit flow}
Consider the password checking program:
\begin{figure}[htp]
\centering
\begin{minipage}[b]{0.25\linewidth}
\begin{lstlisting}
if (H == L)
    O = true;
else
    O = false;
\end{lstlisting}
\end{minipage}
\end{figure}
By SE, we have: 
$$
  O = \left\{
  \begin{array}{l l}
    true & \quad \text{if $\alpha = \beta$}\\
    false & \quad \text{if $\alpha \neq \beta$}\\
  \end{array} \right\}
$$
\emph{DF} and \emph{DF} are generated as follows:
\begin{align}
DF &\equiv (\alpha = \beta \wedge \alpha_1 = \beta \rightarrow true = true) \wedge (\alpha \neq \beta \wedge \alpha_1 \neq \beta \rightarrow false = false)\nonumber\\ 
IF &\equiv \alpha = \beta \wedge \alpha_1 \neq \beta \rightarrow true = false\nonumber
\end{align}
It is trivial to prove that \emph{DF} is valid and \emph{IF} is invalid, and thus the program violates non-interference and leaks information via implicit flows.
\subsubsection{No flow}
Consider the modified version of the password checking procedure in Listing \ref{secpass}. 

\begin{figure}[htp]
\centering
\begin{minipage}[b]{0.25\linewidth}
\begin{lstlisting}
if (H == L)
    O = false;
else
    O = false;
\end{lstlisting}
\end{minipage}
\end{figure}
By SE, we have: 
$$
  O = \left\{
  \begin{array}{l l}
    false & \quad \text{if $\alpha = \beta$}\\
    false & \quad \text{if $\alpha \neq \beta$}\\
  \end{array} \right\}
$$
\emph{DF} and \emph{IF} are generated as follows:
\begin{align}
DF &\equiv (\alpha = \beta \wedge \alpha_1 = \beta \rightarrow false = false) \wedge (\alpha \neq \beta \wedge \alpha_1 \neq \beta \rightarrow false = false)\nonumber\\ 
IF &\equiv \alpha = \beta \wedge \alpha_1 \neq \beta \rightarrow false = false\nonumber
\end{align}
It is trivial to prove that both \emph{DF} and \emph{IF} are valid, and thus the program satisfies non-interference. Note that this is the case that type systems,
taint analysis would decide as violating non-interference.
\subsubsection{No confidential data involved.} Consider the password checking program, with a small modification to exclude the confidential data in its computation, i.e.
to make it secure.
\begin{figure}[htp]
\centering
\begin{minipage}[b]{0.24\linewidth}
\begin{lstlisting}
if (L == 3)
    O = true;
else
    O = false;
\end{lstlisting}
\end{minipage}
\end{figure}

Similarly we have:
$$
  O = \left\{
  \begin{array}{l l}
    true & \quad \text{if $\beta$ = 3}\\
    false & \quad \text{if $\neg(\beta = 3)$}\\
  \end{array} \right\}
$$
\emph{DF} and \emph{IF} are derived as:
\begin{align}
DF &\equiv (\beta = 3 \wedge \beta = 3 \rightarrow true = true) \wedge (\neg(\beta = 3) \wedge \neg(\beta = 3) \rightarrow false = false)\nonumber\\ 
IF &\equiv \beta = 3 \wedge \neg(\beta = 3) \rightarrow true = false\nonumber
\end{align}
Both \emph{DF} and \emph{IF} are valid, which confirms the intuition that the program is secure.
\subsubsection{Both implicit and explicit flows}
Consider again the data sanitization program:
\begin{figure}[htp]
\centering
\begin{minipage}[b]{0.24\linewidth}
\begin{lstlisting}
if (H < 16)
    O = H + L;
else
    O = L;
\end{lstlisting}
\end{minipage}
\end{figure}

The summaries and path conditions returned by SE are as follows:
$$
  O = \left\{
  \begin{array}{l l}
    \alpha + \beta & \quad \text{if $\alpha < 16$}\\
    \beta & \quad \text{if $\neg(\alpha < 16)$}\\
  \end{array} \right\}
$$
\emph{DF} and \emph{DF} are generated similarly:
\begin{align}
DF &\equiv (\alpha < 16 \wedge \alpha_1 < 16 \rightarrow \alpha + \beta = \alpha_1 + \beta) \wedge (\neg(\alpha <16) \wedge \neg(\alpha_1 <16) \rightarrow \beta = \beta)\nonumber\\ 
IF &\equiv \alpha < 16 \wedge \neg(\alpha_1 <16) \rightarrow \alpha + \beta = \beta\nonumber
\end{align}
It is easy to find counterexamples to make \emph{DF} and \emph{IF} invalid, for example: ($\alpha = 1; \alpha_1=2$) for \emph{DF} and ($\alpha = 1; \alpha_1=17$) for \emph{IF}. So the program leaks via both implicit and explicit flows.

\subsection{Optimization}
After SE collects the symbolic paths and path conditions as in (\ref{equa:se}), there are three possible cases for a $\rho^*\in \mathcal{R}^*$:
\begin{itemize}
 \item $O|_{\text{\emph{fin}}(\rho^*)}$ and $path(\rho^*)$ does not contain $\alpha$: $\rho_s$ can be classified as ``secure'', and is excluded in 
computing \emph{DF} and \emph{IF}. With this optimization, programs like the one in Figure \ref{secpass} can be classified as secure without
computing anything.
\item $O|_{\text{\emph{fin}}(\rho^*)}$ does not contain $\alpha$; $path(\rho^*)$ contains $\alpha$: in this case, computing \emph{DF} for this path will result a formula
$C(\beta) \rightarrow true$ which is valid for any $C$. Therefore, $\rho^*$ is excluded in computing \emph{DF}.
 \item $O|_{\text{\emph{fin}}(\rho^*)}$ contains $\alpha$: we can conclude that the program is insecure, leaking information via direct flow. This is very similar
to taint analysis, however SE is more precise, since it can detect cases such as $H \times 0$ or $H - H$ and so on.
\end{itemize}
These three optimizations are sufficient to eliminate the computation of \emph{DF}, and simplify the formula to be validated. 

In a previous paper~\cite{Terauchi:2005:SIF:2156802.2156828}, Terauchi and Aiken presented an interesting program that computes Fibonacci numbers, and containing confidential data. 
Applying the self-composition technique for this program turned out to be very tricky because of the symmetry and redundancy of the self-composed program, and the state-of-the-art safety analysis tool BLAST failed to terminate.
With our Symbolic Execution-based approach and these optimizations, the case study becomes trivial, and can be checked quickly without computation of \emph{DF} and \emph{IF}.

\section[Quantifying Information Leaks using Reliability Analysis]
{Quantifying Information Leaks by combining Self-composition and Reliability Analysis}
At a high level, QILURA performs a two-step analysis. First, SE is run to collect all symbolic paths of the program (up to a user-specified depth), then each path is assigned a label: (i) \textbf{clean}: if it leaks no information, (ii) \textbf{direct}: if it leaks information via direct flow, and 
(iii) \textbf{indirect}:
if it leaks information via indirect flow.
Secondly, a model counter for symbolic paths from~\cite{Filieri:2013:RAS:2486788.2486870} is used to count the number of possible inputs that go to ``direct'' paths, and
compute an upper bound on the leakage. QILURA is available at: {\url{https://github.com/qif/jpf-qilura}}.

\subsection{Fine-grained self-composition}
Checking the satisfiability of the path-equivalence condition $PE \equiv DF \wedge IF$ in section \ref{sec:path-equiv} can only 
decide whether a program leaks information. However, we can refine it to a path-level analysis of leakage.  

We assume the same settings as in section \ref{sec:path-equiv}: we symbolically execute the program \code{P} with the input symbols: $\code{H} = \alpha, \code{L} = \beta$,
and the program \code{P}$_1$ with the input symbols $\code{H}_1 = \alpha_1, \code{L}_1 = \beta$. Based on the path-equivalence condition for
direct flow and indirect flow, we define the self-composition condition for each symbolic path as follows.

\begin{definition} Given a path $\rho^*_i$ such that $path(\rho^*_i) = c_i(\alpha,\beta)$ and $O|_{fin(\rho^*_i)} = f_i(\alpha,\beta)$, $\rho^*_i$
does not leak information via direct flow if and only if the following condition holds:
 $$c_i(\alpha,\beta) \wedge c_i(\alpha_1,\beta) \rightarrow (f_i(\alpha,\beta) = f_i(\alpha_1,\beta))$$
\end{definition}
Applying the \emph{material implication} rule on the condition above results in:
$$\neg (c_i(\alpha,\beta) \wedge c_i(\alpha_1,\beta)) \vee (f_i(\alpha,\beta) = f_i(\alpha_1,\beta)) $$
In case $\rho^*_i$ leaks information via direct flow, the condition above is violated, which means its negation is satisfiable:
$$\neg(\neg (c_i(\alpha,\beta) \wedge c_i(\alpha_1,\beta)) \vee (f_i(\alpha,\beta) = f_i(\alpha_1,\beta))) $$
This formula can be simplified using De Morgan's law as:
\begin{equation}
c_i(\alpha,\beta) \wedge c_i(\alpha_1,\beta) \wedge \neg (f_i(\alpha,\beta) = f_i(\alpha_1,\beta)) \label{equa:DF} 
\end{equation}
\begin{definition} Given two symbolic paths $\rho^*_i$ and $\rho^*_j$ such that
$path(\rho^*_i) = c_i(\alpha,\beta)$, $O|_{fin(\rho^*_i)} = f_i(\alpha,\beta)$, $path(\rho^*_j) = c_j(\alpha,\beta)$ and $O|_{fin(\rho^*_j)} = f_j(\alpha,\beta)$,
$\rho^*_i$ and $\rho^*_j$ do not leak information via indirect flow if and only if the following condition holds:
 $$ c_i(\alpha,\beta) \wedge c_j(\alpha_1,\beta) \rightarrow (f_i(\alpha,\beta) = f_j(\alpha_1,\beta))$$
\end{definition}
Similar to the derivation above, in case $\rho^*_i$ and $\rho^*_j$ leaks information via indirect flow, the following formula is satisfiable.
\begin{equation}
 c_i(\alpha,\beta) \wedge c_j(\alpha_1,\beta) \wedge \neg(f_i(\alpha,\beta) = f_j(\alpha_1,\beta)) \label{equa:IF}
\end{equation}
Based on the conditions in (\ref{equa:DF}) and (\ref{equa:IF}), we implement a procedure to label all symbolic paths
as in Figure \ref{fig:labeling}. The function \texttt{isSAT} is implemented by calling the SMT solver z3~\cite{DeMoura:2008:ZES:1792734.1792766}.

\begin{figure}[htp]
\centering
\fbox{%
\begin{minipage}[b]{0.75\linewidth}
\begin{algorithmic}[0]
\ForAll{$\rho_i$} \{
  \State label[i] $\leftarrow$ \texttt{clean}
  \State $\varphi \leftarrow c_i(\alpha,\beta) \wedge c_i(\alpha_1,\beta) \wedge \neg (f_i(\alpha,\beta) = f_i(\alpha_1,\beta))$
  \If{(\texttt{isSAT}($\varphi$))}
    label[i] $\leftarrow$ \texttt{direct}
  \EndIf
\EndFor \}
\For{$i$ = 1 \textbf{to} $n-1$}
  \For{$ j= i+1$ \textbf{to} $n$} \{
    \State $\varphi \leftarrow c_i(\alpha,\beta) \wedge c_j(\alpha_1,\beta) \wedge \neg(f_i(\alpha,\beta) = f_j(\alpha_1,\beta))$
    \If{(\texttt{isSAT}($\varphi$))} \{
      \If{(label[i] = \texttt{clean})}
	label[i] $\leftarrow$ \texttt{indirect}
      \EndIf
      \If{(label[j] = \texttt{clean})}
	label[j] $\leftarrow$ \texttt{indirect}
      \EndIf
    \EndIf \}
  \EndFor \}
\EndFor
\end{algorithmic}
\end{minipage}
}
\caption{Fine-grained self-composition}
\label{fig:labeling}
\end{figure}

The algorithm of this procedure is straightforward. At the beginning, all paths are labeled as being \texttt{clean}. The procedure then searches for direct flow by checking the condition (\ref{equa:DF}) for all paths.
It then searches for indirect flow by checking the condition (\ref{equa:IF}) on all possible pairs of symbolic paths.
\subsection{Model Counting for Symbolic Paths}
For a symbolic path $\rho$, let \#$in(\rho)$ and \#$out(\rho)$ denote the number of concrete inputs and outputs of $\rho$ respectively.
Obviously $\#in(\rho_i)$ is $\#(pc_i)$ computed in \cite{Filieri:2013:RAS:2486788.2486870}. After being labeled, all paths are classified into three categories: clean, direct and indirect. So the channel capacity is bounded by:
$$CC(P) \leq \log_2(S\#out(\rho_{c}) + S\#out(\rho_{i}) + S\#out(\rho_{d}))$$
where $\rho_{c}$ is the clean path, $\rho_{i}$ is the indirect path, and $\rho_{d}$ is the indirect path.
\begin{itemize}
 \item Since clean paths are not interfered by the confidential input we can replace $S\#out(\rho_{c})$ with 1.
 \item An indirect path only reveals that the program follows
that path, its output is not interfered, and each path has one output. Thus, $S\#out(\rho_{i})$ is just the number of indirect paths.
 \item We hence only need to compute $S\#out(\rho_{d})$.
\end{itemize}

A deterministic program can be viewed as a function that maps each input to 
exactly one output (denotational semantics).
Therefore, the number of inputs is always greater than or equal to the number of possible outputs. This means $\#in(\rho) \geq \#out(\rho)$,
and $S\#in(\rho_{d}) \geq S\#out(\rho_{d})$.

By using the model counting engine for symbolic paths in \cite{Filieri:2013:RAS:2486788.2486870}, we can compute $S\#in(\rho_{d})$, and hence compute an upper bound of channel capacity $CC(P)$.

\section{Evaluation}
Automated QIF analysis is notoriously hard. To the best of our knowledge, the only tool for QIF analysis of Java bytecode is our own work jpf-qif \cite{Phan:2012:SQI:2382756.2382791} which uses SE for QIF analysis, but no model counting. Instead jpf-qif adds the conditions for testing each bit of the output at the end of the program, hence exploring all these conditions using SPF. We compare jpf-qif with QILURA below.

We also compare with BitPattern~\cite{Meng:2011:CBI:2166956.2166957}, which computes an upper bound on channel capacity by exploring the relations between every pair of bits of the output. 
In more recent work~\cite{Meng:2013:FTB}, BitPattern was improved using new heuristics. We compare QILURA with (the improved) BitPattern on several case studies taken from~\cite{Meng:2011:CBI:2166956.2166957,Meng:2013:FTB}.

\begin{figure}[htp]
\centering
\begin{minipage}[b]{0.24\linewidth}
\begin{lstlisting}
if (H > 999){ 
  O = -1; 
} 
O = H; 
O = O - H;
\end{lstlisting}
\end{minipage}
\caption{No flow}
\end{figure}

Moreover, we consider a special case when the program does not leak any information to assess the effectiveness and precision of our technique in such a corner case.
The program does not leak information because the output \code{O} is always 0 regardless of the value of the secret \code{H}. 
However, the assignment $O = H$ and the condition $H > 999$
make the program be rejected by other qualitative information-flow techniques, e.g. the ones based on type systems or taint analysis. 
 
\subsection{Results and discussions}
Figure \ref{chap6:experiment} summaries our experiment, we take the time from the faster version of BitPattern in \cite{Meng:2013:FTB}. Note that in both \cite{Meng:2011:CBI:2166956.2166957} and \cite{Meng:2013:FTB},
the authors manually transform the programs into bit vector predicates, so there will be extra time if they automate this process.

\begin{figure}[htp]
\small
\begin{tabular}{|l|c|c|c|c|c|c|c|}
\hline
\multirow{2}{*}{\textbf{Case Study}} & \multicolumn{2}{c|}{\textbf{jpf-qif}} & \multicolumn{2}{c|}{\textbf{QILURA}} & \multicolumn{2}{c|}{\textbf{BitPattern}} \tabularnewline \cline{2-7}
		                     &  Capacity   & Time                    &	Bound             &  Time      &	Bound             &  Time    \tabularnewline \hline
No Flow                              &  0          & 2.304                   &    0                     &  0.790     &    -                     &   -      \tabularnewline \hline
Sanity check, base =\texttt{0x00001000}  &  4      & 45.324                  &   4.09                   &  1.066     &   4                      &   0.036  \tabularnewline \hline
Sanity check, base =\texttt{0x7ffffffa}  &  4      & 35.346                  &   4.09                   &  1.049     &   4.59                   &   0.203  \tabularnewline \hline
Implicit Flow                        &  2.81       & 0.897                   &    3                     &  0.796     &   3                      &   0.011  \tabularnewline \hline
Electronic Purse                     &  2          & 1.169                   &   2.32                   &  0.854     &   2                      &   0.157  \tabularnewline \hline
Ten random outputs                   & 3.32        & 1.050                   &   3.32                   &  0.814     &   18.645                 &   0.224  \tabularnewline \hline
\end{tabular}
\caption[Performance of all tools]{Capacity and bounds are in bits, times are in seconds. ``-'' means ``not reported''.}
\label{chap6:experiment}
\end{figure}

Comparing with jpf-qif, QILURA has both advantage and disadvantage.
As shown in Figure \ref{chap6:experiment}, thanks to the model counting tool Latte, QILURA is much faster than jpf-qif while the upper bounds it 
computed only deviate to a small extent from the exact channel capacities. 

This increase in performance comes with a price, Latte can only count models
of a system of linear integer inequalities $A\bar{x} \geqslant \bar{b}$. For this reason, QILURA cannot analyse the case studies of CRC and Tax Record 
in chapter \ref{chap:SymExDPLL}, which have complicated constraints. The limitation due to using Latte is shared with previous work~\cite{Backes:2009:ADQ:1607723.1608130,Klebanov12a} which
were also demonstrated with toy examples.

The BitPattern technique can also compute rather tight upper bounds in most of the cases. However, by analysing the relations of pairs of bits, the technique
is vulnerable when possible values of the output are not in a specific range, as shown in the last case study.

\section{Discussion of related work}
Self-composition was first introduced by Darvas et al. \cite{Darvas:2005:TPA:2154040.2154072} who expressed it in a dynamic logic and proved information flow properties 
for Java CARD programs. Their approach is not automated, requiring users to provide loop invariants, induction hypotheses and so on. Barthe et al. \cite{Barthe:2004:SIF:1009380.1009669}
then coined the term ``self-composition'' and investigated its theoretical aspects, extending the problem to non-deterministic and termination-sensitive cases.

Terauchi and Aiken \cite{Terauchi:2005:SIF:2156802.2156828} found that self-composition was problematic, since the self-composed programs contains symmetry and redundancy.
They proposed a type-directed transformation for a simple imperative language to deal with the problem. Milushev et al. \cite{Milushev:2012:NVS:2366649.2366659} implemented this
type-directed transformation and used \emph{Dynamic Symbolic Execution} (also known as \emph{concolic testing}) as a program analysis tool for non-interference. 

To our knowledge, our technique is unique in that it only 
performs analysis on the original program, rather than the self-composed program, the idea of self-composition is shown in the way we rename the symbolic formula, not in the analysis stage.

Backes et al. \cite{Backes:2009:ADQ:1607723.1608130} describe how to use the model checker ARMC and Latte for QIF analysis. Their technique is very 
precise but also extremely expensive: it involves input counting to compute the pre-image of the observables; in contrast our input counting is used 
for counting the observable. The work of Backes et al. is extended in \cite{Klebanov12a}, which uses the interactive theorem prover KeY instead of ARMC, and requiring significant 
user effort. Moreover, this work is based on ``\emph{classical}'' self-composition, and as Terauchi and Aiken \cite{Terauchi:2005:SIF:2156802.2156828} have pointed out,
it is unlikely practical. Of course, both \cite{Backes:2009:ADQ:1607723.1608130} and \cite{Klebanov12a} were demonstrated with toy examples.

The only technique that can precisely determine if a program leaks information is self-composition \cite{Barthe:2004:SIF:1009380.1009669}. 
QILURA also uses self-composition with the key difference that it is able to determine if a single symbolic path leaks information.

\chapter[A solver for Model Counting Modulo Theories]{A solver for Model Counting Modulo Theories}
\label{chap:allsmt}
Recall that in Section \ref{chap3:method}, we have casted the problem of QIF analysis into the \#SMT problem and proposed two \#SMT-based approaches to QIF, which can be summarized by
the figure below.
\begin{center}
\begin{tikzpicture}
\node at (0,0) {\code{P}};
\draw [<->] (.7,0) -- (5.2,0);
\node at (5.9,0) {$\varphi_P$};
\node at (0,-.8) {QIF};
\node at (5.9,-.8) {\#SMT};
\node at (0,-1.6) {Formal methods};
\node at (5.9,-1.6) {DPLL($\mathcal{T}$)};
\end{tikzpicture}
\end{center}

So far we have investigated the approach on the left-hand side: we directly analyse the program \code{P}, and use formal methods, specifically Bounded Model Checking and Symbolic Execution, in a way that mimics a \#SMT solver for QIF analysis.

This chapter investigates our second approach on the right-hand side: it demonstrates how to build from the program \code{P} a formula $\varphi_P$ with the two properties described in section \ref{chap3:method}, and how to build a \#SMT solver to count 
the models of $\varphi_P$.

Moreover, we study a variant of the \#SMT problem: the \emph{All-Solution Satisfiability Modulo Theories} (All-SMT) problem. All-SMT is only different from \#SMT in that 
instead of counting the number of model, it asks for the enumeration of all models. We show that our algorithms for \#SMT can also be used for 
All-SMT, and we propose the use of an All-SMT solver in new application domains: Bounded Model Checking, automated test generation and reliability analysis.

\section{Quantifying Information Leaks using a \#SMT solver}
We assume the setting in our attacker model in Figure \ref{attackermodel}: a program \code{P}
that takes secret input \code{H}, public input \code{L} and producing public output \code{O}.
Our analysis consists of two steps as the following.
\begin{itemize}
 \item The first step is to build a first-order formula from the program \code{P} a formula $\varphi_P$ with the following properties: (i) $\varphi_P$ contains a set of Boolean variables 
$V_I := \{p_1,p_2,..,p_M\}$; (ii) $p_i = \top$ if and only if $b_i$ is 1, and $p_i = \bot$ if and only if $b_i = 0$.
 \item The second step is to use a \#SMT solver to count the number of 
models of $\varphi_P$ with respect to the set $V_I$.
\end{itemize}
We will show how to build a \#SMT solver later in the next section. At the moment, we assume that there is such a solver for our QIF analysis.

\subsection{Illustrative Example}
To demonstrate the approach, let us consider again the previously used illustrative example in Figure~\ref{sanitize}, which can be encoded into a first-order formula 
as in Figure~\ref{chap7:ex}.

\begin{figure}[htp]
\centering
\begin{minipage}[t]{0.22\linewidth}
\centering
\begin{lstlisting}
L = 8;
if (H < 16)
  O = H + L;
else
  O = L;
\end{lstlisting}
\end{minipage}
\begin{minipage}[t]{0.41\linewidth}
\centering
\vspace{-0.1cm}
\fbox{
$ \begin{array}{ll}
(L_1 = 8) &\wedge \\
(G_0 = H_0 < 16) &\wedge \\
(O_1 = H_0 + L_1) &\wedge \\
(O_2 = L_1) &\wedge \\
(O_3 = g ? O_1 : \code{O}_3) &
  \end{array}
$
}
\end{minipage}
\caption{A simple program encoded into a first-order formula}
\label{chap7:ex}
\end{figure}

The program is transformed into Static Single Assignment (SSA) form \cite{Cytron:1989:EMC:75277.75280}: variables are renamed when they are reassigned. At the beginning, assuming that the variables
\code{H}, \code{L} and \code{O} take the value $H_0$, $L_0$ and $O_0$ respectively after being declared. Then, $L_1$ is the value of the variable \code{L} after being
reassigned (as 8), and similarly for the rest of the program.

We then need to build the set of boolean variable $V_I := \{p_1,p_2,..,p_{32}\}$. In chapter~\ref{chap:SQIF}, we directly analysed the program, and thus hence the set $V_I$ by
instrumenting the source code of the program \code{P} with the block of code in Figure \ref{fig:instru}. This block of code used bitwise operators to extract each bit of the output \code{O}.
Here, we analyse the formula (encoded from the program), and hence build the set $V_I$ by instrumenting the formula. 
Moreover, the formula needs to be in a first-order theory $\mathcal{T}$ that supports 
bitwise operators. Fortunately, the model checker CBMC can automatedly transform a C program into a formula in the theory of bit vector QF\_AUFBV \cite{smt2} which satisfies
this requirement.

The formula in Figure \ref{chap7:ex} can be easily expressed in QF\_AUFBV with $O_3$ declared as a 32-bit vector.
We instrument the formula by adding a set of Boolean variables $V_I = \{p_1, p_1,\dots p_{32}\}$, each one tests the value of a bit of $O_3$. For
example:
\begin{center} 
   \texttt{(assert (= (= \#b1 ((\_ extract 0 0) $O_3$)) $p_1$))}
\end{center}
This statement in SMT-LIB v2 format~\cite{smt2} extracts the first bit of $O_3$ (all bits from position 0 to position 0), then comparing if this bit is equal to 1 (\#b1). The Boolean variable $p_1$ is 
asserted to be the truth value of this comparison. Similar settings are applied for the rest of Boolean variables $p_2, p_3,\dots, p_{32}$. 

At this point we have built a formula $\varphi_P$ that characterizes the behaviour of the program \code{P}, and contains a set $V_I$ of Boolean 
variables, each one represents a bit of the output \code{O} of the program \code{P}. By using a \#SMT solver to count the number of models of $\varphi_P$ with
respect to the set $V_I$, which is also the number of possible values of the output \code{O}, we can conclude the maximum leakage of the program 
\code{P} as per definition \ref{DEF:QIF}.

\subsection{Program transformation with CBMC}
In our two-step analysis, the second step is automated with a \#SMT solver, we only need to automate the first step: building a formula $\varphi_P$ from 
the program \code{P}.

\begin{figure}[htp]
\centering
\fbox{
\begin{minipage}{0.75\linewidth}
\texttt{$\mathcal{C}(\textnormal{``}$if(c) $I_1$ else $I_2\textnormal{''},g) := \mathcal{C}(I_1,g \wedge \rho(c)) \wedge \mathcal{C}(I_2,g \wedge \neg\rho(c))$\\
$\mathcal{P}(\textnormal{``}$if(c) $I_1$ else $I_2\textnormal{''},g) := \mathcal{P}(I_1,g \wedge \rho(c)) \wedge \mathcal{P}(I_2,g \wedge \neg\rho(c))$\\
$\mathcal{C}(\textnormal{``} I_1 ; I_2\textnormal{''},g) := \mathcal{C}(I_1,g) \wedge \mathcal{C}(I_2,g)$\\
$\mathcal{P}(\textnormal{``} I_1 ; I_2\textnormal{''},g) := \mathcal{P}(I_1,g) \wedge \mathcal{P}(I_2,g)$\\
$\mathcal{P}(\textnormal{``}$assert(a)$\textnormal{''},g) := g \rightarrow \rho(a)$\\
$\mathcal{C}(\textnormal{``}$v = e$\textnormal{''},g) := (v_{\alpha} = (g? \rho(e): v_{\alpha - 1}))$
}
\end{minipage}
}
\caption{The two functions $\mathcal{C}(\code{P},g)$ and $\mathcal{P}(\code{P},g)$ for program transformation. $\rho(c)$ is expression $c$ after being renamed as per SSA form; $v_{\alpha - 1}$ and 
$v_{\alpha}$ are value of $v$ before and after the assignment respectively.}
\label{chap7:trans}
\end{figure}

The model checker CBMC transforms a program \code{P} and a guard $g$ into a logical formula using two functions: $\mathcal{C}(\code{P},g)$ transforms the program constraints, 
and $\mathcal{P}(\code{P},g)$ transforms the program specification, namely assertions. At the beginning, the guard $g$ is initialised to $\top$.
Both functions are defined by induction on the syntax of program as in Figure \ref{chap7:trans} (interested readers are pointed to 
\cite{Clarke:2003:BCC:775832.775928} for full details).

In its default settings, CBMC transforms the program and its specification into a propositional formula. However, it also has the option \texttt{$--$smt2} (still experimental) to 
transform the program and specification into a QF\_AUFBV formula in SMT-LIB v2 format. Hence, we use leverage this option to build the formula $\varphi_P$.

Recall (section \ref{Sec:BMC-CBMC}) that the formula generated by CBMC is in the form $\mathcal{C} \wedge \neg \mathcal{P}$, where $\mathcal{C}$ is the program constraints,
and $\mathcal{P}$ is the program specification, i.e. assertions. Since we only assess the security of a program when it is free from errors, we only need the program constraints $\mathcal{C}$. However,
without a specification, the aggressive program slicing in CBMC can decide immediately that the program does not violate any specification, and do not generate 
any formula. This is actually a strong feature of CBMC, however it prevents us from getting a formula $\mathcal{C}$.

A simple solution for the problem above is to add a fake error, ``\texttt{assert(0);}'', at the end of the program. Hence, the generated formula is $\mathcal{C} \wedge \neg \bot$, or simply $\mathcal{C}$.

\subsection{Formula instrumentation}
As we have demonstrated with the example, we need to instrument the formula generated by CBMC with a set of Boolean variables $V_I$.

In the SSA form, a variable is renamed when it is reassigned. For example, the output \code{O} is named $O_0$ when initialised. When it is assigned,
it is renamed to $O_1$, and so on.
The index is incremented, and \code{O} keeps the final value after final assignment. Therefore, we build a simple parser to locate the variable $O_k$ renamed from the output \code{O}, which has the maximal index $k$.

The declaration of the set of Boolean variables $V_I$, and theirs binding with the bits of $O_k$ can be appended to the end of the formula.
The whole formula instrumentation procedure is implemented in a simple Java program.
\section{All-Solution Satisfiability Modulo Theories}
The SMT solver MathSAT, from version 4 \cite{cav2008}, provides a functionality, called All-SMT, that given a formula $\varphi$ and a set $V_I$ of \emph{important} Boolean variables,
MathSAT in All-SMT mode computes all models of $\varphi$ with respect to the set $V_I$. 

In this thesis, we extend the All-SMT of MathSAT with a set $V_R$ of \emph{relevant}, possibly non-Boolean, variables. The extended All-SMT($\varphi,V_I,V_R)$ problem is to
compute all models of $\varphi$ with respect to the set $V_I$ and the models includes value assignment for variables in $V_R$. We show how this All-SMT problem can be used to analyse the availability, reliability and security of programs:
\begin{itemize}
 \item \emph{Bounded Model Checking} \cite{Biere:1999:SMC:646483.691738}: SMT-based Bounded Model Checking can only return a single error traces, the user has to fix the error, then run the model checker again for other error traces. This is because
 SMT solvers can only return one model. Combining Bounded Model Checking with an All-SMT solver, we can compute multiple counterexamples in one run of the model checker.
 \item \emph{Automated Test Generation}: an All-SMT solver can be combined with either a Symbolic Executor or a Bounded Model Checker for test input generation. Although traditional Symbolic 
 Execution with an SMT solver is capable of generating test inputs, it needs to make hundreds or thousands of calls to the SMT solver. In our approach, the Symbolic Execution tool
 needs to make only one call to the All-SMT solver for any programs.
 \item \emph{Reliability analysis}: we can build a reliability analysis tool by combining an All-SMT solver with a Symbolic Executor to enumerate all path conditions of the program,
 then using the Barvinok model counting technique \cite{latte} to compute the number of inputs that go into each symbolic path. In this way,
 we can compute the reliability of the program, i.e. the probability that the program successfully accomplishes its task without errors.
\end{itemize}
\subsection{Multiple-counterexamples for BMC}
Recall that (section \ref{Sec:BMC-CBMC}), Bounded Model Checking (BMC) transforms the program and its specification into a formula, then solving the
resulting formula with a SAT or SMT solver.

Since a SAT or SMT solver can only return a single model, state-of-the-art SAT-based or SMT-based Bounded Model Checker can only return a single error trace per run. The user has to fix the error and 
run the model checker again to find more error traces. On the other hand, All-SMT solver can return all models w.r.t. a set of Boolean variable, it can be
exploited to find multiple counterexamples for BMC.

\begin{figure}[htp]
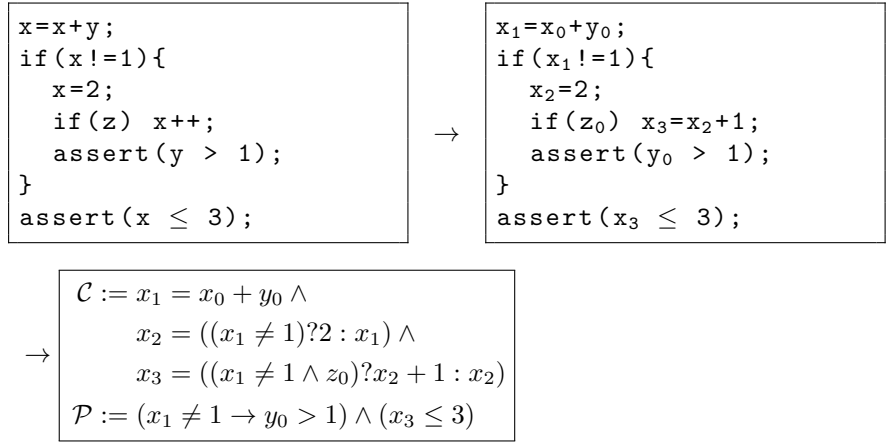

\begin{lrbox}{\leftlisting}
\begin{minipage}{0.33\linewidth}
\begin{qsplisting}
x=x+y;
if(x!=1){
  x=2; 
  if(z) x++; 
  assert(y > 1);
}
assert(x $\le$ 3);
\end{qsplisting}
\end{minipage}
\end{lrbox}
\begin{lrbox}{\rightlisting}
\begin{minipage}{0.33\linewidth}
\begin{qsplisting}
x$_\mt{1}$=x$_\mt{0}$+y$_\mt{0}$;
if(x$_\mt{1}$!=1){
  x$_\mt{2}$=2; 
  if(z$_\mt{0}$) x$_\mt{3}$=x$_\mt{2}$+1;
  assert(y$_\mt{0}$ > 1); 
}
assert(x$_\mt{3}$ $\le$ 3);
\end{qsplisting}
\end{minipage}
\end{lrbox}

\footnotesize{
\begin{align*}
& \usebox{\leftlisting} \quad\to\quad \usebox{\rightlisting} \\
&\to\boxed{
\begin{aligned}
\mathcal{C}&:= 
  \!\begin{aligned}[t]
  x_1 &= x_0 + y_0 \wedge{} \\ 
  x_2 &= ((x_1 \neq 1)?2:x_1) \wedge{} \\ 
  x_3 &= ((x_1 \neq 1 \wedge z_0)?x_2+1:x_2)
  \end{aligned}
\\
\mathcal{P}&:= (x_1 \neq 1 \rightarrow y_0 > 1) \wedge (x_3\le 3)
\end{aligned}
}
\end{align*}
}
\caption[Encoding a program into a logical formula]{Example, modified from \cite{Clarke:2003:BCC:775832.775928}: the program is transformed into Static Single Assignment form, and then encoded into a logical formula}
\label{fig:cbmc}
\end{figure}
To illustrate our approach, we reconsider an example from \cite{Clarke:2003:BCC:775832.775928}, which was used to illustrate CBMC \cite{ckl2004}. 
The example is shown in Figure \ref{fig:cbmc}, we have modified it, adding the assertion \texttt{(y > 1)}, so that the program $P$ contains more 
than one error. At the first step, the program is transformed into SSA form. 

As shown in Figure \ref{fig:cbmc}, applying $\mathcal{C}(\code{P},g)$ and $\mathcal{P}(\code{P},g)$ in the SSA program results in the set of guards $(x_1 \neq 1)$ and $(z_0 \neq 0)$.
We denote Boolean variable $g_1$ and $g_2$ such that $g_1 := \mathcal{BA}(x_1 \neq 1)$ and $g_2 := \mathcal{BA}(z_0 \neq 0)$.

A model of the formula $\mathcal{C} \wedge \neg \mathcal{P}$ will correspond to a trace of the program that violates the specification $\mathcal{P}$.
By asking an All-SMT solver to return all models of $\varphi = \mathcal{C} \wedge \neg \mathcal{P}$ with respect to the set of Boolean abstraction of variables in the guards,
i.e. $V_I = \{g_1,g_2\}$, we can get a set of all models, each one corresponds to an error trace.

Note that each error trace represents a set of concrete execution, to get the inputs for just one representative concrete execution that triggers the error,  we set them as the relevant variables to be included in the models,
which means $V_R = \{x_0,y_0,z_0\}$.
In this case hence $V_I $ represents the guards of the program and  $V_R $ the inputs.

\subsection{Automated Test Generation}
This section shows how an All-SMT solver can be used in two different approaches for Automated Test Generation (ATG), namely Bounded Model Checking and 
Symbolic Execution.
\subsubsection{ATG using Bounded Model Checking}
We use the same trick as in the previous section. The goal is to compute all models of a formula, each one corresponds to a program traces in the program.
Take an example as in Figure \ref{symex}. Different from the one in Figure \ref{fig:cbmc}, the program contains no error, so it will go through CBMC without any solver being called.
CBMC will not generate a formula either.

\begin{figure}[htp]
\centering
\begin{minipage}[t]{0.4\linewidth}
\centering
\begin{lstlisting}[]
void foo(int x, int y){
  if(x > 5){
    x++;
    if (x < 3) 
      x--;
    else 
      y = x + 1;
  }
  return;
}
\end{lstlisting}
\end{minipage}
\begin{minipage}[t]{0.41\linewidth}
\centering
\vspace{-0.1cm}
\fbox{
$ \begin{array}{ll}
(g_1 = x_1 > 5) &\wedge \\
(x_2 = 1 + x_1) &\wedge \\
(g_2 = x_2 < 3) &\wedge \\
(x_3 = -1 + x_2)&\wedge \\
(x_4 = x_2)     &\wedge \\
(y_2 = 1 + x_4) &\wedge \\
(x_5 = g_2 ? x_3 : x_4)  &\wedge \\
(y_3 = g_2 ? y_1 : y_2)  &\wedge  \\
(x_6 = \neg g_1 ? x_1 : x_5)&\wedge \\
(y_4 = \neg g_1 ? y_1 : y_3) &
  \end{array}
$
}
\end{minipage}
\caption{A simple program encoded into a formula}
\label{symex}
\end{figure}

In order to generate test inputs that cover all program paths (to a given bound), similar to the previous section, we append ``\texttt{assert (0)};'' as a fake error at the end of the program.
Since this error is reachable by all program paths, CBMC will include all the paths into the formula.

The box in the right in Figure \ref{symex} shows the formula encoded by CBMC. We run the All-SMT solver on the formula with $V_I = \{g_1,g_2\}$,
$V_R = \{x_1,y_1\}$. The All-SMT solver will return a set of solutions, each one contains value assignments for $x_1$ and $y_1$, which can be used as test input for the function \texttt{foo}.
\subsubsection{ATG using Symbolic Execution}
Recall that Symbolic Execution (SE) executes programs on 
unspecified inputs, by using symbolic inputs instead of concrete data. For each executed program path, a path condition $pc$ is built which represents the 
condition on the inputs for the execution to follow that path, according to the branching conditions in the code. 
In classical SE, the 
satisfiability of the path condition is checked at every branching point, using off-the-shelf solvers. In this way only feasible program paths are 
explored. Test generation is performed by solving the path conditions.

Here we propose another approach for SE using All-SMT solver. We use SE with the \emph{constraint solver turning off},
to compute the set of all possible program paths: $pc_1, pc_2, \dots pc_M$. Since there is no constraint solving, a $pc_i$ can be infeasible. Hence, the program under test 
can be viewed as corresponding to the following formula: \[\varphi := pc_1 \vee pc_2 \dots \vee pc_M\]
To illustrate, let us consider again the program in Figure \ref{symex}. The program can be viewed as corresponding to the formula:
\begin{equation}
\label{equa:symex}
             \begin{aligned}[t]
 \varphi \text{ } := \text{ }  &((x > 5) \wedge (x + 1 < 3)) &\vee {} \\ 
	     & ((x > 5) \wedge \neg (x + 1 < 3)) &\vee {} \\ 
	     &\neg (x > 5)  &
  \end{aligned}
\end{equation}
Notice that the path {\normalsize{$(x > 5) \wedge (x + 1 < 3)$}} is infeasible, but it is still included in the formula, since we do not check the constraint at each branching point.
Applying Boolean abstraction on $\varphi$ leads to:
\begin{center}
          $\mathcal{BA} := ((C_1 = (x > 5)) \wedge (C_2 =(x+1 <3)))$
\end{center}
We use an All-SMT solver on $\varphi \wedge \mathcal{BA}$ with $V_I = \{C_1,C_2\}$ and $V_R = \{x,y\}$.
The set of models returned by the All-SMT solver is the set of feasible paths, and the evaluation of relevant variables can be used as test inputs for the program.

Also for ATG  $V_I $ represents the guards of the program and  $V_R $ the inputs.

\subsection{Reliability analysis}
This section introduces an alternative implementation for the approach in \cite{Filieri:2013:RAS:2486788.2486870} by using our All-SMT-based SE
instead of classical SE. The improvement is that we only need to make only one call to the All-SMT solver to explore all feasible paths.

\begin{figure}[htp]
\centering
\begin{minipage}[t]{0.4\linewidth}
\begin{lstlisting}[]
void foo(int x, y){
  if(x > 5){
    x++;
    if (x < 3) 
      x--;
    else { 
      y = x + 1;
      assert false;
    }
  }
  return;
}
\end{lstlisting}
\end{minipage}
\end{figure}

Let us consider again the previous example with only one difference: we add an error for the path $x \geq 3$.
Similar to the previous section, we use SE with the constraint solver turning off, to encode the program into a logical formula $\varphi$
as in (\ref{equa:symex}). Moreover, the two paths $((x > 5) \wedge (x + 1 < 3))$ and $\neg (x > 5)$ are labelled with \textbf{T},
as in these two paths the program finishes normally. On the other hand, the path $((x > 5) \wedge \neg (x + 1 < 3))$ is labelled with \textbf{F} since an error
is reachable in this path.

Similar to the previous section, using an All-SMT solver on $\varphi \wedge \mathcal{BA}$ with $V_I = \{C_1,C_2\}$ and $V_R = \{x,y\}$ will eliminate
the infeasible path $((x > 5) \wedge (x + 1 < 3))$. We then can use the Latte tool to count the models for each paths, and compute the reliability of the program.

\section{Algorithms for \#SMT and All-SMT solver}
\label{sec:algo}
Obviously, there is no off-the-shelf solver for our new problem \#SMT. The closet to a \#SMT solver is the functionality All-SMT of MathSAT.
However, MathSAT does not support model generation for relevant variables in All-SMT mode. Moreover, when using MathSAT for our analysis, MathSAT 
returns incorrect number of models in several benchmarks (we will show later in section \ref{sec:impl}). 
For these reasons, we have developed a lightweight approach to implement an All-SMT/\#SMT front-end for SMT solvers. 
We build our algorithms from a number of APIs provided by the SMT solver, which we list below.

\vspace{.6cm}
\begin{center}
\begin{tabular}{@{}ll} 
API                               & Description 	\\ \midrule
\api{Assert}($\varphi$)           & Assert formula $\varphi$ into the solver.       \\
\api{Check}()		          & Check consistency of all assertions.            \\
\api{Model}()		          & Get model of the last \api{Check}.    \\
\api{Eval}(\stmt{t})		  & Evaluate expression \texttt{t} in current model.  \\
\api{Push}()		          & Create a backtracking point.                    \\
\api{Pop}(\stmt{n}$=1$)		  & Backtracks \texttt{n} backtracking points.        \\
\end{tabular}
\end{center}
\vspace{.6cm}

A key feature of SMT solvers for our algorithms is that of being \emph{incremental} and \emph{backtrackable}.
The following example shows a sequence of API calls and their effects to the solver.

\vspace{.6cm}
\begin{center}
\begin{tabular}{lllll} 
\api{Assert}($\varphi_1$); \api{Check}();    &      & $\varphi_1$                                       && $\Rightarrow$ \stmt{SAT} \\
\api{Push}();                                &      & $\varphi_1$                                       &&  \\
\api{Assert}($\varphi_2$); \api{Check}();    &      & $\varphi_1 \wedge \varphi_2$                      && $\Rightarrow$ \stmt{SAT} \\
\api{Push}();                                &      & $\varphi_1 \wedge \varphi_2$                      &&  \\
\api{Assert}($\varphi_3$); \api{Check}();    &      & $\varphi_1 \wedge \varphi_2 \wedge \varphi_3$     && $\Rightarrow$ \stmt{UNSAT} \\
\api{Pop}(2);                                &      & $\varphi_1$                                       &&  \\
\api{Assert}($\varphi_4$); \api{Check}();    &      & $\varphi_1 \wedge \varphi_4$                      && $\Rightarrow$ \stmt{SAT} \\
\end{tabular}
\end{center}
\vspace{.6cm}

It is possible for an incremental SMT solver to add additional assertions to the original formula. Moreover, when \api{Check} is being called several times,
the solver can remember its computation from one call to the other. Thus, when being called to
check $\varphi_1 \wedge \varphi_2$ after checking $\varphi_1$,  it avoids restarting the computation from scratch by restarting
the computation from the previous status.
Backtrackable means that the solver is able to undo steps, using \api{Push} and \api{Pop}, and returns to a previous status on the stack in an efficient manner.

Both z3 \cite{DeMoura:2008:ZES:1792734.1792766} and MathSAT \cite{cav2008} provide similar APIs to interact with the solver in incremental mode.
Beside the APIs, we develop a function \stmt{filter}(\stmt{m}, $V_I, V_R$) that given a model of the formula $\varphi$, and a set of important Boolean
variable $V_I$, and the set of relevant variables $V_R$, the function will return a subset \stmt{m}$_{ir}$ of \stmt{m} that only contains literals from $V_I$ and $V_R$.
This function will be used in both algorithms.
\subsection{Blocking clauses method}
A straightforward approach for \#SMT is to add clauses that prevent the solver from finding the same solution again. 

\begin{figure}[htp]
\centering
\fbox{
\begin{minipage}[b]{0.55\linewidth}
\begin{algorithmic}[0]
\Function{\stmt{All-BC}}{$\varphi, V_I, V_R$} \{
\State \stmt{N} $\leftarrow$ 0;  $\varPsi \leftarrow \epsilon$;
\State \api{Assert}($\varphi$);
\While{(\api{Check}() $=$ \stmt{SAT})} \{
  \State \stmt{N} $\leftarrow$ \stmt{N} + 1;
  \State \stmt{m} $\leftarrow$ \api{Model}($\varphi$); 
  \State \stmt{m}$_{ir}\leftarrow$ \stmt{filter}(\stmt{m}, $V_I, V_R$);
  \State $\varPsi \leftarrow \varPsi \cup {}$\{\stmt{m}$_{ir}$\};
  \State \stmt{block} $\leftarrow$ \stmt{FALSE};
  \ForAll{$p_i \in V_I$} \{
      \State \stmt{block} $\leftarrow$ \stmt{block} $\vee$ ($p_i \neq $ \api{Eval}($p_i$));
  \EndFor \}
  \State \api{Assert} (\stmt{block});
\EndWhile \}
\State \Return \stmt{N},  $\varPsi$;
\EndFunction \}
\end{algorithmic}
\end{minipage}
}
\caption{Blocking clauses \#SMT}
\label{blockingclause}
\end{figure}

The pseudo-code for the blocking clauses method is shown in Figure \ref{blockingclause}. Every time the solver discovers a solution \texttt{m} of $\varphi$ such that \texttt{m} $= l_0 \wedge l_1 \wedge {} \dots \wedge l_n \wedge \dots$, in which only $l_0,l_1\dots l_n$ are literals of $p_0,p_1\dots p_n$ in $V_I$. The negation of $l_1 \wedge l_2 \wedge {} \dots \wedge l_n$ 
would be, by De Morgan's law, as follows:
\[\text{\stmt{block}} = \neg l_0 \vee \neg l_1 \vee \dots \vee \neg l_n\]
A literal $l_i$ in the model can be viewed as a mapping $p_i$ to \{\stmt{TRUE}, \stmt{FALSE}\}, thus the negation $\neg l_i$ is $p_i \neq $ \api{Eval}($p_i$).
By adding this clause to the formula, by \api{Assert}(\stmt{block}), a solution with $l_0 \wedge l_1 \wedge {} \dots \wedge l_n$ will not be discovered again. This procedure repeats until 
no other solution is found. At that point, we have enumerated all the solution of $\varphi$ with respect to $V_I$. All solutions are stored in $\varPsi$, and \stmt{N} $= |\varPsi|$ is the result for
the corresponding \#SMT problem.

The blocking clauses method is straightforward and it is simple to implement. However, adding a large number of blocking clauses
will consume a large amount of memory. Moreover, increasing number of clauses also means that the \emph{Boolean Constraint Propagation} procedure is slowed down.
Despite these inefficiencies, the blocking clauses method can be used to verify the results of other techniques.
\subsection{Depth-first search}
To address the inefficiencies of adding a large number of clauses, we introduce an alternative method which avoids re-discovering solutions using depth-first search (DFS).

We divide the set of variables of $\varphi$ into two sets: $V_I$ is the set of important Boolean variables, and $V_U$ is the set of unimportant, possibly non-Boolean, variables ($V_R \subseteq V_U$).
Hence, the formula $\varphi$ can be viewed as a function:
\[V_I \times V_U \rightarrow \{\texttt{TRUE}, \texttt{FALSE}\}\] 
Our \#SMT procedure 
is the integration of two components: the first component is a simple SAT solver to enumerate all possible partial truth assignments $\mu_I$ of $V_I$; the second component
is the SMT solver to check the consistency of $\varphi \wedge \mu_I$.

\begin{figure}[ht]
\centering
\fbox{
\begin{minipage}[b]{0.55\linewidth}
\begin{algorithmic}[0]
\Function{\stmt{All-DFS}}{$\varphi,V_I,V_R$} \{
\State \stmt{N} $\leftarrow$ 0; $\varPsi \leftarrow \epsilon$; 
\State \api{Assert}($\varphi$); 
\If {(\api{Check}() $\neq$ \stmt{SAT})} \Return \stmt{N}, $\varPsi$; \EndIf
\State \stmt{depth} $\leftarrow$ 0; \stmt{finished} $\leftarrow$ \stmt{FALSE};
\While {(\stmt{finished} $=$ \stmt{FALSE})}\{
  \State $l \leftarrow$ \stmt{choose\_literal}($V_I$);
  \State \api{Push}();
  \State \api{Assert}($l$); \stmt{depth} $\leftarrow$ \stmt{depth} + 1;
  \If {(\api{Check}() $=$ \stmt{SAT})} \{
     \If {(\stmt{depth} $=$ $|V_I|$)} \{
         \State \stmt{N} $\leftarrow$ \stmt{N} + 1;
         \State \stmt{m} $\leftarrow$ \api{Model}($\varphi$); 
	 \State \stmt{m}$_{ir}$ $\leftarrow$ \stmt{filter}(\stmt{m}, $V_I, V_R$);
         \State $\varPsi \leftarrow \varPsi \cup {}$\{\stmt{m}$_{ir}$\};
         \State \stmt{backtrack}();
     \EndIf \} 
    \} \Else \stmt{ backtrack}(); 
  \EndIf 
\EndWhile \}
\State \Return \stmt{N}, $\varPsi$;
\EndFunction \}
\end{algorithmic}
\end{minipage}
}
\caption{Depth-first search \#SMT}
\label{dfs}
\end{figure}

The pseudo-code for DFS-based \#SMT is depicted in Figure \ref{dfs}. The method \texttt{choose\_literal} chooses the next states to explore 
from $V_I$ in a DFS manner, and the variable \texttt{depth} keeps the number of important variables that has been chosen. That means,
\texttt{choose\_literal} will select a literal $V_I[\texttt{depth}]$ or $\neg V_I[\texttt{depth}]$. This literal is ``\emph{pushed}'' to the formula as a unit clauses.
Recall that the plain DPLL algorithm \cite{Davis:1962:MPT:368273.368557} is a depth-first search combining with
the BCP procedure. Here we do not perform BCP, however by adding all literals of $\mu_I$ as unit clauses to $\varphi$,
we force the SMT solver to perform BCP on those literals.

When all important variables has been assigned a truth value, i.e. $\texttt{depth} = |V_I|$, and the formula in the solver is consistent, then the search
has found a model. It then \emph{backtracks} to find another one. The method \texttt{backtrack} implements a simple chronological backtracking, it ``\emph{pops}'' the unit clauses and 
sets the variable \texttt{finished} to \texttt{TRUE} when all states are explored. It is also called when the formula in the solver is inconsistent.

Compare to the blocking clauses method, the DFS-based method is much more efficient in term of memory usage. The blocking clauses method 
needs to add $N$ blocking clauses to find all models while the DFS adds maximum $|V_I|$ of unit clauses. The memory efficiency leads to timing efficiency
when there are a large number of models.
\subsection{Implementation}
We have implemented both of the methods discussed above in a prototype tool, called \textbf{aZ3}. The tool is built in Java, using the APIs provided by the SMT 
solver z3 \cite{DeMoura:2008:ZES:1792734.1792766}. aZ3 supports standard SMT-LIB v2 with two additional commands: the first one is \texttt{check-allsat}, similar to MathSAT,
 to specify the list of important variables, and the second one is \texttt{allsat-relevant} to specify the list of relevant variables.

We have also implemented a QIF analyzer, called \textbf{sqifc++}, which uses CBMC to encode a program into a formula, then invoking aZ3
to compute channel capacity. 
\section{Evaluation}
\label{sec:impl}
We make two experiments: the first one is to compare our \#SMT solver aZ3 against MathSAT modified for \#SMT; the second one is to compare the QIF approach using a \#SMT 
solver with the one in chapter~\ref{chap:SQIF}, which uses formal methods.

The benchmarks, the aZ3 solver, and the wrapper of MathSAT for \#SMT can be found at: {\url{http://www.eecs.qmul.ac.uk/~qsp30/test/allsmt.tar.gz}}.
\subsection{Evaluation of \#SMT solvers}
%
\begin{figure}[ht]
\centering
\begin{tabular}{|l l|c|c|c|c|c|c|c|}
\hline
\multicolumn{2}{|c|}{\multirow{2}{*}{\textbf{Benchmark}}} & Expected  & \multicolumn{2}{c|}{\textbf{MathSAT 5}} &\multicolumn{2}{c|}{\textbf{aZ3}}          \tabularnewline
 &  &   N          		      				&  		                N        &      Time   &  \textbf{BC} time  &  \textbf{DFS} time  \tabularnewline \hline
\multicolumn{1}{|l|}{\multirow{8}{*}{\begin{sideways} \texttt{QF\_LIA}  
\end{sideways}}} &   Example in Figure \ref{fig:cbmc}  &              2           &    2   &     0.007   &         0.021      &    0.013        \tabularnewline \cline{2-7}
\multicolumn{1}{|l|}{} & Example in Figure \ref{symex}  &              3           &    3   &     0.005   &         0.008      &    0.007        \tabularnewline \cline{2-7}
\multicolumn{1}{|l|}{} & Flap controller \cite{Filieri:2013:RAS:2486788.2486870}  &              5           &    5   &     0.031   &         0.020      &    0.012        \tabularnewline \cline{2-7}
\multicolumn{1}{|l|}{} & Red-black tree \cite{spflink} &              31           &    31   &     0.016   &         0.054      &    0.073        \tabularnewline \cline{2-7}
\multicolumn{1}{|l|}{} & Bubble sort \cite{svcomp14} &              541           &    541   &     0.136   &         1.850      &    2.069        \tabularnewline \cline{2-7}
\multicolumn{1}{|l|}{} & Array false \cite{svcomp14} &              1370           &    1370   &     0.037   &         3.008      &    2.650        \tabularnewline \cline{2-7}
\multicolumn{1}{|l|}{} & Sum array false \cite{svcomp14} &              1024           &    1024   &     0.026   &         0.899      &    0.792        \tabularnewline \cline{2-7}
\multicolumn{1}{|l|}{} & Linear search false \cite{svcomp14} &              1024           &   1024   &     0.028   &         0.899      &    0.604        \tabularnewline \cline{2-7}
\hline
\multicolumn{1}{|l|}{\multirow{9}{*}{\begin{sideways} \texttt{QF\_AUFBV} 
\end{sideways}}}       & Data sanitization \cite{Meng:2011:CBI:2166956.2166957} &              16           &    16   &     0.008   &         0.035      &    0.086        \tabularnewline \cline{2-7}
\multicolumn{1}{|l|}{} & Implicit flow \cite{Meng:2011:CBI:2166956.2166957}& 7               &    7        &     0.012   &       0.029        &    0.049           \tabularnewline \cline{2-7}
\multicolumn{1}{|l|}{} & Population count \cite{Meng:2011:CBI:2166956.2166957}& 33           &    \textbf{71}       &     0.012   &       0.074        &    0.398           \tabularnewline \cline{2-7}
\multicolumn{1}{|l|}{} & Mix and duplicate \cite{Meng:2011:CBI:2166956.2166957} &             65536           &    \textbf{162087}   &     4.648   &         -          &    136.947         \tabularnewline \cline{2-7}
\multicolumn{1}{|l|}{} & Masked copy \cite{Meng:2011:CBI:2166956.2166957} &              65536           &    65536   &     1.319   &         -          &    18.630         \tabularnewline \cline{2-7}
\multicolumn{1}{|l|}{} & Sum query \cite{Meng:2011:CBI:2166956.2166957} &              28           &    \textbf{64}   &     0.010   &         0.055          &    0.133        \tabularnewline \cline{2-7}
\multicolumn{1}{|l|}{} & Ten random outputs \cite{Meng:2011:CBI:2166956.2166957} &              10           &    10   &     0.014   &         0.038          &    0.093        \tabularnewline \cline{2-7}
\multicolumn{1}{|l|}{} & CRC (8) \cite{Phan:2014:AMC} &              8               &    \textbf{12}       &     0.018   &       0.041        &    0.099           \tabularnewline \cline{2-7}
\multicolumn{1}{|l|}{} & CRC (32) \cite{Phan:2014:AMC} &              32              &    \textbf{36}       &     0.019   &       0.075        &    0.325           \tabularnewline \hline
\end{tabular}
\caption{N is the number of models. \textbf{BC} time and \textbf{DFS} time are the time of aZ3 using the blocking clauses method and depth-first search-based method respectively. Times are in seconds. ``-'' means ``timed out in 1 hour''. Notice that for both aZ3 implementations the number of models is Expected N.}
\label{chap7:experiment}
\end{figure}
In order to evaluate aZ3 and MathSAT, we create two set of benchmarks.
The first group of benchmarks are formulas in QF\_LIA (integer linear arithmetic)~\cite{smt2}. These benchmarks are used to evaluate All-SMT solvers
in the context of test input generation. The formulas are generated using Symbolic PathFinder~\cite{Pasareanu:2010:SPS:1858996.1859035} (SPF).
SPF has a parameter, \texttt{symbolic.dp}, to set the constraint solver for it. If this parameter set to \texttt{no\_solver}, the tool will
run without constraint solving. 

The architecture of SPF enables us to attach a ``listener'' to it. When SPF executes a program, the listener collects the path conditions, and outputs them to a QF\_LIA formula.
The models of the integer variables can be used as test inputs for the original programs.

The second group of benchmarks that we considered are formulas in QF\_AUFBV~\cite{smt2} (bit vector with array). The source of these benchmarks are programs
in the QIF literature, mostly re-collected in \cite{Meng:2011:CBI:2166956.2166957}. We use CBMC with the option \texttt{$--$smt2} to transform the programs
into QF\_AUFBV formulas, and instrument the resulting formulas to make them \#SMT problems. There are no relevant variables in these benchmarks.

\subsubsection{Discussion of evaluation} Figure \ref{chap7:experiment} summaries our experiments with the two solvers aZ3 and MathSAT 5.2.11 on the benchmarks.
In order to compare with MathSAT, we commented out the relevant variables in the QF\_LIA benchmarks.
As shown in the figure 
MathSAT is faster than aZ3. This is not surprised, since we build the tool from the front-end, while the All-SMT functionality of MathSAT is built from the back-end, making use of the internal data structure.

However, MathSAT returns incorrect models in several benchmarks. Especially, in the benchmark ``Mix and duplicate'' MathSAT is significantly faster
than aZ3, but it is also extremely imprecise at the same time. Note that benchmarks in QF\_AUFBV are derived from the QIF literature, and their number of
models were already reported in other papers. For example ``Mix and duplicate'' was reported in \cite{Newsome:2009:MCC:1554339.1554349} and \cite{Meng:2011:CBI:2166956.2166957} to have $2^{16}$ models.

The blocking clauses methods is comparable, or even faster than the DFS-based method when the number of models is small. However, for the benchmarks
with $2^{16}$ models, adding $2^{16}$ blocking clauses is obviously not efficient in both time and memory. As a result, the method failed to provide the answer for such
benchmarks. On the other hand, the DFS-based method was still able to provide the answer in a reasonable time.

\subsection{Evaluation of QIF analysers} 
\label{chap7:eval}
Figure \ref{chap7:qif} compares the performance of sqifc++ against the tool sqifc in chapter \ref{chap:SQIF}. The results show that 
sqifc++ is much more efficient. The reason is that, sqifc makes several calls to CBMC, and for each call CBMC has to transform the program into
a formula, then calling a SAT/SMT solver to check the formula. On the other hand, sqifc++ transform the program only once, and its search is also more
efficient.

\begin{figure}[htp]
\centering
\begin{tabular}{|l|c|c|c|c|c|}
\hline
{\multirow{2}{*}{\textbf{Benchmark}}} & \multirow{2}{*}{Leaks} & \textbf{sqifc}  &\multicolumn{3}{c|}{\textbf{sqifc$++$} time} \tabularnewline 
                                      &        & time  &         \textbf{CBMC} time   &  \textbf{aZ3} time & \ Total \tabularnewline \hline
Data sanitization \cite{Meng:2011:CBI:2166956.2166957}                        & 4       & 11.898   &       0.165      &    0.086     & 0.251   \tabularnewline \cline{2-6}
Implicit flow \cite{Meng:2011:CBI:2166956.2166957}                            & 2.81    & 5.033   &       0.169      &    0.049     & 0.218   \tabularnewline \cline{2-6}
Population count \cite{Meng:2011:CBI:2166956.2166957}                         & 5.04    & 17.278  &       0.162      &    0.398     & 0.560   \tabularnewline \cline{2-6}
Mix and duplicate \cite{Meng:2011:CBI:2166956.2166957}                        & 16      & - &       0.154      &    136.947   & 137.101 \tabularnewline \cline{2-6}
Masked copy \cite{Meng:2011:CBI:2166956.2166957}                              & 16      & - &       0.175      &    18.630    & 18.805  \tabularnewline \cline{2-6}
Sum query \cite{Meng:2011:CBI:2166956.2166957}                                & 4.81    &64.557   &       0.162      &    0.133     & 0.295   \tabularnewline \cline{2-6}
Ten random outputs \cite{Meng:2011:CBI:2166956.2166957}                       & 3.32    &64.202   &       0.160      &    0.093     & 0.253   \tabularnewline \cline{2-6}
CRC (8) \cite{Phan:2014:AMC}                                                  & 3       &2.551    &       0.184      &    0.099     & 0.283   \tabularnewline \cline{2-6}
CRC (32) \cite{Phan:2014:AMC}                                                 & 5       &7.755    &       0.193      &    0.325     & 0.518   \tabularnewline \hline
\end{tabular}
\caption{Comparing the new approach with the sqifc tool in chapter \ref{chap:SQIF}. Leaks are in bits. aZ3 runs with the DFS-based algorithm. 
Times are in seconds, ``-'' means timeout in one hour. Total time of sqifc++ is the sum of CBMC time and aZ3 time.}
\label{chap7:qif}
\end{figure}

However, sqifc++ relies on CBMC for program transformation, and this functionality of CBMC (\texttt{$--$smt2}) is still experimental.
We found that CBMC generates incorrect formulas for several case studies in chapter \ref{chap:SQIF}, therefore we could not use sqifc++ to analyse,
for example, the dining cryptos case study.
\section{Discussion of related work}
\subsection{Quantitative Information Flow}
We have just compared the two prototype tools sqifc++ and sqifc in the previous section. 
Moreover, in chapter \ref{chap:SQIF}, we already compared our \#SMT-based approach with other work in QIF literature, so we do not repeat the discussion here.
\subsection{Multiple-counterexamples for BMC} 
The most relevant work to ours is that of Bhargavan et al. \cite{Bhargavan:2002:VFA:506201.506203} embodied in the Verisim testing tool for network 
protocols. 
When an error trace is found to violate the specification, which is an extended LTL formula $\phi$, 
Verisim uses a technique, called \emph{tuning}, to replace $\phi$ with $\varphi$ that ignores the violation. Tuning is not fully automatic.

Another   technique introduced by Ball et al. \cite{Ball:2003:SCL:604131.604140} is embodied in the SLAM tool-kit. The algorithm uses a model checker as a 
sub-routine. When the model checker finds an error trace, SLAM localizes the error cause, modifying the source code with a \texttt{halt} statement at the
error cause. The model checker is then invoked again, and the \texttt{halt} statements instruct the model checker to stop exploring paths at the previously found error causes.
This procedure is very expensive, it requires comparing the error trace with all correct traces to localize the error, and requires to run the model checker
several times. Our work is much simpler, and faster but the error traces we compute can come from the same causes.

\subsection{Automated Test Generation}
The closest to our work is
FShell \cite{Holzer:2008:FST:1427782.1427809}, which also uses CBMC for automated test generation. FShell transforms the program under test into a 
CNF formula, and solves it using an incremental SAT solver. Every time the SAT solver finds a solution representing a symbolic path, FShell adds a blocking clauses
to prevent that path from being explored again. As our experiments have shown, the blocking clauses method is suffered from rapid space growth.

Classical Symbolic Execution also uses SMT solvers to check the satisfiability of path condition. The SMT solver is called whenever a conditional statement
is executed, hence it may be called hundreds or thousands of times. In our approach, the symbolic executor
makes only one call to the All-SMT solver.

\subsection{Reliability analysis}
Our approach to reliability analysis is based on the paper of Filieri et al. \cite{Filieri:2013:RAS:2486788.2486870} that uses classical 
Symbolic Execution and Barvinok model counting tool. We extend the approach using our new All-SMT-based Symbolic
Execution instead of classical Symbolic Execution. The main difference is the same as in the case of test generation, our approach only makes one call
to the All-SMT solver in the whole analysis.

\subsection{All Solutions SAT Modulo Theories}
As we have discussed throughout the chapter, MathSAT is the only SMT solver that supports All-SMT. Its algorithm is briefly described 
in \cite{allsmtpatent}, and it has been used to compute predicate abstraction in \cite{Cavada:2007:CPA:1333874.1334136} and 
\cite{DBLP:conf/fmcad/CimattiDJR09}. Our experiment results show that MathSAT is imprecise in several benchmarks. However, we also notice that the imprecisions seem to be limited in QF\_AUFBV benchmarks,
while the authors performed experiments with QF\_LRA formulas in \cite{Cavada:2007:CPA:1333874.1334136}. 

Also in the context of predicate abstraction, Lahiri et al. \cite{Lahiri:2006:STF:2135909.2135964} have proposed several techniques, which use the SMT solver Barcelogic
 to generate the set of all satisfying assignments over a set of predicates. However, we are not able to include Barcelogic in our experiments, since the solver provided to us by the author
 does not support All-SMT.

A principal difference between the work mentioned above and the one in this chapter is that we implemented from the front-end of an SMT solver. For this reason, our 
implementation is slower than MathSAT. On the good side, our approach is applicable to implement even in closed-source SMT solvers that do not support All-SMT but provide 
similar APIs.

\chapter[Conclusions]{Conclusions}
\label{chap:conclusions}
\section{Summary}
This thesis introduces a new research problem, Model Counting Modulo Theories or \#SMT, and presenting a \#SMT-based approach to quantification of information leaks. 
Although our implementations are far from being optimised, they drastically outperform the existing technique based on self-composition, e.g.
reducing the time of analysing some programs from Linux kernel from some hours to a few seconds. 
Our approach is applicable to programs with difficult data structures including pointers, and to Java bytecode.

On the theoretical side, this thesis makes the original contributions by discovering the relations among different research areas: (i) it casts the QIF problem
into the \#SMT problem; (ii) it shows the correspondence between Symbolic Execution and the DPLL($\mathcal{T}$) algorithm; (iii) it explores the relation
between Symbolic Execution and Bounded Model Checking; (iv) finally, it exploits the connection between QIF analysis and Reliability analysis.

On the application side, this thesis is the first to use Symbolic Execution to quantify information leaks. It is also the first to use classical Symbolic Execution
for Bounded Model Checking. Beside, it proposes the use of an All-SMT solver for multiple-counterexamples in Bounded Model Checking, for automated
test generation, and for reliability analysis.

On the practical side, this thesis has developed several tools in both C/C++ and Java: sqifc, jpf-qif and QILURA, for the quantification of information leaks in software.
It also demonstrated the use of these tools to analyse vulnerabilities from the National Vulnerability Database of the US government, 
and anonymity protocols.

For the verification community, important contributions of this thesis include the development of JCBMC, a Concurrent Bounded Model Checker for Java, and the All-SMT solver aZ3.
\section{Future Research}
In the previous chapter, we have proposed the use of an All-SMT solver for multiple-counterexamples in Bounded Model Checking, for automated
test generation and for reliability analysis. An immediate direction would be to implement these ideas into automated tools, and to perform experiments on standard benchmarks.
Some other possible directions for investigation are the following.

\subsubsection{Fault localization}
Another interesting avenue of further research would be to the All-SMT solver with CBMC to localize error causes using similar idea in \cite{Ball:2003:SCL:604131.604140}.
Models of $\varphi_1 = \mathcal{C} \wedge \mathcal{P}$ correspond to correct traces that satisfy the specification, and models of $\varphi_2 = \mathcal{C} \wedge \neg\mathcal{P}$
correspond to error traces that violate the specification. Using an All-SMT solver, we can compute the sets of all models of $\varphi_1$ and $\varphi_2$ with respect to 
the set of guards. Comparing the two sets of models, we can localize the transitions that only appear in error traces.

\subsubsection{Concurrent Bounded Model Checking}
A first improvement on this direction would be to upgrade its concurrency from single CPU multi-threading to true parallelism
 and to perform obvious optimisations. Another improvement would be to replace Symbolic PathFinder with a lighter weight tool, or a parallel version, to reduce 
 the cost of generating path conditions. It will also be interesting to implement the methodology for other languages (C, Python) and investigate 
 how to use Symbolic Execution for IC3 style verification.

\subsubsection{Statistical analysis for QIF}
The QILURA tool is still just a prototype. A possible improvement for it would be to use approximate exploration techniques to replace the exact, complete exploration presented in this thesis. In this way, for 
the tool can be used with increased scalability, but with formal statistical guarantees on the results.


\renewcommand{\bibname}{References}
\bibliography{papers}        
\bibliographystyle{plain}  

\end{document}